\documentclass[acmtocl,acmnow]{acmtrans2m}

\usepackage{prooftree}
\usepackage[dvips]{graphicx}
\usepackage{latexsym}
\usepackage{amssymb}
\usepackage{color}

\newcommand {\linea} {\vspace{0.1cm} \begin{footnotesize} \noindent ------------------------------------------------------------------------------------------------------------------------------ \end{footnotesize}}
\newcommand {\lineacorta} {\vspace{0.1cm} \begin{footnotesize} \noindent ----------------------------------------------------------------------------------------------------------------------- \end{footnotesize}}
\newcommand {\tratteggio} {\vspace{0.1cm} \begin{center}  \ldots\ldots\ldots\ldots\ldots\ldots\ldots\ldots\ldots\ldots\ldots\ldots\ldots\ldots\ldots\ldots\ldots\ldots\ldots\ldots\ldots \end{center} \vspace{0.3cm}}

\newcommand {\Cu} {{\bf C}}
\newcommand {\Ra} {{\bf R}}
\newcommand {\Cl} {{\bf CL}}
\newcommand {\Gammas} {\Gamma^{\cond^{\pm}},\Gamma^{\bbox^{\freccia}}}

\newcommand {\freccia} {\downarrow}
\newcommand {\imp} {\rightarrow}
\newcommand {\cond} {\mathrel{{\scriptstyle\mid\!\sim}}}
\newcommand {\ent} {\cond}
\newcommand {\orr} {\vee}
\newcommand {\andd} {\land}
\newcommand {\prova} {\vdash}
\newcommand {\nott} {\lnot}
\newcommand {\perogni} {\forall}
\newcommand {\esiste} {\exists}
\newcommand {\sx} {\langle}
\newcommand {\dx} {\rangle}

\newcommand {\appartiene} {\in}
\newcommand {\emme} {\mathcal{M}}

\newcommand {\elle} {\mathcal{L}}
\newcommand {\modello} {\models}

\newcommand {\unione} {\cup}
\newcommand {\intersezione} {\cap}

\newcommand {\tc} {\mid}
\newcommand {\vuoto} {\emptyset}
\newcommand {\WW} {\mathcal{W}}

\newcommand {\diverso} {\neq}

\newcommand {\bbox} {\square}

\newcommand{\be}{\begin{enumerate}}
\newcommand{\ee}{\end{enumerate}}

\newcommand{\hide}[1]{}

\newcommand{\lan}{{\cal L}}

\newcommand{\irule}[3]
{\prooftree{#1}\justifies{#2}\using{\:#3}\endprooftree}

\hide{

{\qed \end{trivlist}}

{\qed \end{trivlist}}

\newenvironment{definition}
{\begin{defi} \rm}{\qed \end{defi}}

\newenvironment{example}
{\begin{exa} \rm}{\qed \end{exa}}

\newenvironment{remark}
{\begin{rem} \rm}{\end{rem}}

}

\def \cases{\left \{\begin{array}{l}}
\def \endcases{\end{array}\right .}

\newcommand {\Pe} {{\bf P}}

\newcommand {\sse} {\leftrightarrow}

\newcommand {\bes} {\begin{description}}
\newcommand{\ens} {\end{description}}

\newcommand {\beq} {\begin{quote}}
\newcommand {\enq} {\end{quote}}
\newcommand {\bit} {\begin{itemize}}
\newcommand {\enit} {\end{itemize}}

\newcommand {\calcoloP} {{\bf \mathcal{T}P}}
\newcommand {\calcoloPterminante} {{\bf \mathcal{T}P^{T}}}
\newcommand {\calcoloCL} {{\bf \mathcal{T}CL}}
\newcommand {\calcoloC} {{\bf \mathcal{T}C}}
\newcommand {\calcoloCtent} {{\bf \mathcal{T}C}^{\natural}}
\newcommand {\calcoloCLterminante} {{\bf \mathcal{T}CL^{T}}}
\newcommand {\calcoloR} {{\bf \mathcal{T}R}}
\newcommand {\calcoloRterminante} {{\bf \mathcal{T}R^{T}}}
\newcommand {\provafatto}[1]{\begin{flushright}$\bbox$ \emph{Fact #1}\end{flushright}}
\newcommand {\Gammam}[2]{\Gamma^{M}_{#1 \imp #2}}
\newcommand {\ellep}[1]{\elle^{\Gamma_0}_{\bbox^{#1}}}

\newenvironment{ultimissimi}{\color{black}}{\color{black}} 
\newenvironment{rosso}{\color{black}}{\color{black}} 
\newenvironment{blu}{\color{black}}{\color{black}} 

\newtheorem{theorem}{Theorem}[section]
\newtheorem{lemma}[theorem]{Lemma}
\newtheorem{corollary}[theorem]{Corollary}
\newtheorem{proposition}[theorem]{Proposition}
\newtheorem{definition}[theorem]{Definition}

\newtheorem{fact}[theorem]{Fact}

\newenvironment{provaposu}{\noindent \emph{Proof.}}{\begin{flushright}$\blacksquare$\end{flushright}}
\newenvironment{provaallettore}{\noindent \emph{Proof.}}{\vspace{0.5cm}}
\newenvironment{pf*}{\noindent}{\vspace{0.5cm}}
\newenvironment{sketch}{\noindent \emph{Proof sketch.}}{\begin{flushright}$\blacksquare$\end{flushright}}

\acmVolume{V}
\acmNumber{N}
\acmYear{YY}

\title{Analytic Tableaux Calculi for KLM Logics of Nonmonotonic Reasoning}
\author{LAURA GIORDANO \\ Dipartimento di Informatica - Universit\`a del
Piemonte Orientale "A. Avogadro" \and VALENTINA GLIOZZI \\
 Universit\`a degli Studi
di Torino \and NICOLA OLIVETTI \\
LSIS - UMR CNRS 6168 Universit\'e Paul C\'ezanne (Aix-Marseille 3)
\and GIAN LUCA POZZATO \\ Universit\`a degli Studi di Torino}

\begin{abstract}
We present tableau calculi for some logics of
\begin{blu}nonmonotonic\end{blu} reasoning, as defined by Kraus, Lehmann and
Magidor. We give a tableau proof procedure for \begin{blu}all KLM
logics, namely preferential, loop-cumulative, cumulative and
rational logics\end{blu}. Our calculi are obtained by introducing
suitable modalities to interpret conditional assertions. We
provide a decision procedure for the logics considered, and we
study their complexity.
\end{abstract}



\begin{document}

\begin{bottomstuff}
Authors' addresses: L. Giordano, Dipartimento di Informatica -
Universit\`a del Piemonte Orientale "A. Avogadro" - via Bellini
25/G -
 15100 Alessandria - Italy - e-mail: {\tt laura@mfn.unipmn.it}. \\
 V. Gliozzi, Dipartimento di Informatica - Universit\`a degli Studi di
Torino, corso Svizzera
185 - 10149 Turin - Italy, e-mail: {\tt gliozzi@di.unito.it}.\\
 N. Olivetti, LSIS - UMR CNRS 6168
Universit\'e Paul C\'ezanne (Aix-Marseille 3) Avenue Escadrille
Normandie-Niemen 13397 Marseille Cedex 20 - France - e-mail: {\tt
nicola.olivetti@univ.u-3mrs.fr - nicola.olivetti@lsis.org}.\\ G.L.
Pozzato, Dipartimento di Informatica - Universit\`a degli Studi di
Torino, corso Svizzera 185 - 10149 Turin - Italy, e-mail: {\tt
pozzato@di.unito.it}.
\end{bottomstuff}

\maketitle

\section{Introduction}
In the early 90s \cite{KrausLehmannMagidor:90} Kraus, Lehmann and
Magidor (from now on KLM) proposed a formalization of nonmonotonic
reasoning that was early recognized as a landmark. Their work
stemmed from  two sources: the theory of nonmonotonic consequence
relations initiated by Gabbay \cite{gabbay} and the preferential
semantics proposed by Shoham \cite{shoham} as a generalization of
Circumscription. Their work led to a classification of
nonmonotonic consequence relations, determining a  hierarchy of
stronger and stronger systems. The so called \emph{KLM properties}
have been widely accepted as the ``conservative core'' of default
reasoning. The role of KLM logics is similar to the role of AGM
postulates in Belief Revision \cite{gardenfors}: they give a set
of postulates for default reasoning that any concrete reasoning
mechanism should satisfy.

According to the KLM framework, defeasible knowledge is
represented by a  (finite) set of nonmonotonic conditionals or
assertions of the form

\[A \ent B\]

\noindent whose reading is \emph{normally (or typically) the $A$'s
are $B$'s}. The operator ``$\ent$'' is nonmonotonic, in the sense
that $A \ent B$ does not imply $A \land C \ent B$. For instance, a
knowledge base $K$ may contain the following set of conditionals:
\[adult \ent worker,  adult \ent taxpayer, student \ent adult,
student \ent \neg worker,\] \[student \ent \neg taxpayer, retired
\ent adult, retired \ent \neg worker\] whose meaning is that
adults typically work, adults typically pay taxes, students are
typically adults, but they typically do not work, nor do they pay
taxes, and so on. Observe that if $\ent$ were interpreted as
classical (or intuitionistic) implication, we simply would get
$student \ent \bot$, $retired \ent \bot$, i.e. typically there are
not students, nor retired people, thereby obtaining a trivial
knowledge base. One can derive new conditional assertions from the
knowledge base by means of a set of inference rules.

In KLM framework, the set of adopted inference rules defines some
fundamental types of inference systems, namely, from the weakest
to the strongest: Cumulative (\Cu), Loop-Cumulative (\Cl),
Preferential (\Pe) and Rational (\Ra) logic. All these systems
allow one to infer new assertions from a given knowledge base $K$
without incurring in the trivializing conclusions of classical
logic: regarding our example, in none of them, one can infer
$student \ent worker$ or $retired \ent worker$. In cumulative
logics (both \Cu \ and \Cl) one can infer $adult \land student
\ent \neg worker$ (giving preference to more specific
information), in Preferential logic \Pe \ one can also infer that
$adult \ent \neg retired$ (i.e. typical adults are not retired).
In the rational case \Ra, if one further knows that $\nott (adult
\ent \lnot married)$ (i.e. it is not the case that adults are
typically unmarried), one can also infer that $adult \land married
\ent worker$.

From a semantic point of view, to each logic (\Cu, \Cl, \Pe, \Ra)
corresponds one kind of models, namely possible-world structures
equipped with a preference relation among worlds or states.  More
precisely, for \Pe \ we have models with a preference relation (an
irreflexive and transitive relation) on worlds.  For the stronger
\Ra \ the preference relation is further assumed to be {\em
modular}. For the weaker logic  \Cl, the transitive and
irreflexive preference relation is defined on {\em states}, where
a state can be identified, intuitively, with a set of worlds. In
the weakest case of \Cu, the preference relation is on states, as
for \Cl, but it is no longer assumed to be transitive. In all
cases, the meaning of a conditional assertion $A \ent B$ is that
$B$ holds in the {\em most preferred} worlds/states where $A$
holds.

In KLM framework the operator ``$\ent$'' is considered as a
meta-language operator, rather than as a connective in the object
language. However, it has been readily observed that KLM systems
\Pe \  and \Ra  \ coincide to a large extent with the flat (i.e.
unnested) fragments of well-known conditional logics, once we
interpret the operator ``$\ent$'' as a binary connective
\cite{CroccoLamarre:92,Boutilier:94,Katsuno-Sato:91}.

A recent result by Halpern and Friedman
\cite{halpern-plausibility} has shown that preferential and
rational logic are natural and general systems: surprisingly
enough, the axiom system of preferential (likewise of rational
logic) is complete with respect to a wide spectrum of semantics,
from ranked models, to parametrized probabilistic structures,
$\epsilon$-semantics and possibilistic structures. The reason is
that all these structures are examples of {\em plausibility
structures}  and the truth in them is captured by the axioms of
preferential (or rational) logic. These results, and their
extensions to the first order setting \cite{halpern-first-order}
are the source of a renewed  interest in KLM framework. A
considerable amount of research in the area has then concentrated
in developing concrete mechanisms for plausible reasoning in
accordance with KLM systems (\Pe \ and \Ra \ mostly). These
mechanisms are defined by exploiting a variety of models of
reasoning under uncertainty (ranked models, belief functions,
possibilistic logic, etc.
\cite{dubois-prade,nicola5,nicola6,pearl,nicola1,nicola2}) that
provide, as we remarked, alternative semantics to KLM systems.
These mechanisms are based on the restriction of the semantics to
preferred classes of models of KLM logics; this is also the case
of Lehmann's notion of rational closure introduced in
\cite{whatdoes} (not to be confused with the logic \Ra). More
recent research has also explored the integration of KLM framework
with paraconsistent logics \cite{nicola3}. Finally, there has been
some recent investigation on the relation between KLM systems and
decision-theory \cite{nicola7,nicola8}.

Even if KLM was born as an inferential approach to nonmonotonic
reasoning, curiously enough, there has not been much investigation
on deductive mechanisms for these logics. In short, the state of
the art is as follows:

\begin{itemize}
\item Lehmann and Magidor \cite{whatdoes} have proved
that validity in \Pe \ is  {\bf coNP}-complete. Their decision
procedure for \Pe \ is more a theoretical tool than a  practical
algorithm, as it requires to guess sets of indexes and
propositional evaluations. They have also provided another
procedure for \Pe \ that exploits its reduction to \Ra. However,
the reduction of \Pe \ to \Ra \ breaks down if boolean
combinations of conditionals are allowed, indeed it is exactly
when such combinations are allowed that  the difference between
\Pe \ and \Ra \ arises.

\item A tableau proof  procedure  for \Cu \ has been given in
\cite{Governatori:02}. Their tableau procedure is fairly
complicated; it uses labels and a complex unification mechanism.
Moreover, the authors show how to extend the system to
Loop-Cumulative logic \Cl \ and discuss some ways to extend it to
the other logics.

\hide{ Moreover, the extension to the other logics of the KLM
family is problematic. IL REFEREE SOTTOLINEA: }

\item In \cite{GGOSTableaux2003} and \cite{Berlino} some labelled tableaux
calculi have been defined for the conditional logic {\bf CE} and
its main extensions, including {\bf CV}. The   flat fragment (i.e.
without nested conditionals)  of   {\bf CE} and of  {\bf CV}
corresponds   respectively to {\bf P} and to {\bf R}. These
calculi however need a fairly complicated loop-checking mechanism
to ensure termination. It is not clear if they match complexity
bounds and if they can be adapted in a simple way to \Cl \ and to
\Cu.

\item Finally, decidability of \Pe \ and \Ra \ has  also been
obtained by interpreting them into standard modal logics, as it is
done by Boutilier \cite{Boutilier:94}. However, his mapping is not
very direct and natural, as we discuss below.

\item To the best of our knowledge, for  \Cl \ no decision
procedure and complexity bound was known before the present work.
\end{itemize}

In this work we \begin{blu}introduce tableau procedures for all
KLM logics, starting with the preferential logic \Pe. Our approach
is based on a novel interpretation of \Pe \ into modal logics. As
a difference with previous approaches (e.g. Crocco and Lamarre
\cite{CroccoLamarre:92} and Boutillier \cite{Boutilier:94}), that
take S4 as the modal counterpart of \Pe, we consider here
G\"odel-L\"ob modal logic of provability G (see for instance
\cite{hughes}). Our tableau method provides a sort of run-time
translation of \Pe \ into modal logic G.\end{blu}

The idea is simply to interpret the preference relation as an
accessibility relation: a conditional $A \ent B$ holds in a model
if  $B$ is true in all minimal $A$-worlds $w$ (i.e. worlds in
which $A$ holds and that are minimal). An $A$-world $w$ is a
minimal $A$-world if all smaller worlds are not $A$-worlds. The
relation with modal logic G is motivated by the fact that we
assume, following KLM, the so-called {\em smoothness condition},
which is related to the well-known \emph{limit assumption}. This
condition ensures that minimal $A$-worlds exist whenever there are
$A$-worlds, by preventing infinitely descending chains of worlds.
This condition therefore corresponds to the finite-chain condition
on the accessibility relation (as in modal logic G). Therefore,
our interpretation of conditionals is different from the one
proposed by Boutilier, who rejects the smoothness condition and
then gives a less natural (and more complicated) interpretation of
\Pe \ into modal logic S4.

As a further difference with previous approaches, we do not give a
formal translation of \Pe \ into G. Rather, we directly provide a
tableau calculus for \Pe. One can notice some similarities between
some of the rules for \Pe \ and some of the rules for G. This is
due to the correspondence between the semantics of the two logics.
For deductive purposes, we believe that our approach is more
direct, intuitive, and efficient than translating \Pe \ into G and
then using a calculus for G.

We are able to extend our approach to the cases of \Cl \ and  \Cu
\ by using a second modality which takes care of states. Regarding
\Cl, we show that we can  map \Cl-models into \Pe-models with an
additional modality. The very fact that one can  interpret \Cl \
into \Pe \ by means of an additional modality does not seem to be
previously known and might be of independent interest. In both
cases, \Pe \ and \Cl, we can define a decision procedure and
obtain also a complexity bound for these logics, namely that they
are both {\bf coNP}-complete. In case of \Cl \ this bound is new,
to the best of our knowledge.

We treat \Cu \ in a similar way: we can establish a  mapping
between Cumulative models and a kind of bi-modal models. However,
because of the lack of transitivity, the target modal logic is no
longer G. The reason is that the  {\it smoothness condition} (for
any formula $A$, if a state satisfies $A$, then either it is
minimal or it admits a smaller minimal state satisfying $A$)  can
no longer be identified with the finite-chain condition of G. As a
matter of fact, the smoothness condition for \Cu \ cannot be
identified with any  property of the accessibility relation, as it
involves unavoidably  the evaluation of formulas in worlds.  We
can still derive a tableau calculus based on our semantic mapping.
But we pay a price: as a difference with \Pe \ and \Cl \ the
calculus for \Cu \ requires  a sort of (analytic) cut rule to
account for the smoothness condition. This calculus gives
nonetheless a decision procedure for \Cu.

\begin{blu}
Finally, we consider the case of the strongest logic \Ra; as for
the other weaker systems, our approach is based on an
interpretation of \Ra \ into an extension of modal logic G,
including modularity of the preference relation (previous
approaches \cite{CroccoLamarre:92,Boutilier:94} take S4.3 as the
modal counterpart of \Ra). As a difference with the tableau
calculi introduced for \Pe, \Cl, and \Cu, here we develop a
\emph{labelled} tableau system, which seems to be the most natural
approach in order to capture the modularity of the preference
relation. The calculus defines a systematic procedure which allows
the satisfiability problem for \Ra \ to be decided in
nondeterministic polynomial time, in accordance with the known
complexity results for this logic.
\end{blu}

All the calculi presented in this paper have been implemented in
SICStus Prolog. To the best of our knowledge, our theorem prover,
called KLMLean, is the first one for KLM logics\footnote{The
theorem prover KLMLean is not presented here. A description can be
found in \cite{m4m4} and in \cite{tesidottorato}. KLMLean can be
downloaded at
\texttt{http://www.di.unito.it/$\thicksim$pozzato/klmlean2.0}.}.

The plan of the paper is as follows: in section 2, we recall KLM
logics (from the strongest to the weakest): \Ra, \Pe, \Cl, and
\Cu, and we show how their semantics can be represented by
standard Kripke models. In section 3 we give a tableau calculus
for \Pe. We then elaborate this calculus in order to obtain a
terminating procedure. We propose a further refinement that gives
a {\bf coNP} decision procedure. The latter is based on a tighter
semantics of \Pe \ in terms of {\em multi-linear models}. In
section 4 we propose similar calculi for \Cl. In section 5, we
give a tableau calculus for \Cu.  As mentioned above, the calculus
requires a form of cut-rule. We prove however that we can restrict
its application in an analytic way (namely, it is needed only for
formulas $A$ that are antecedents of conditionals $A \ent B$
contained in the initial set of formulas).
\begin{blu}In section 6 we describe a labelled tableau calculus
for \Ra, then we refine it in order to obtain a terminating
procedure and to describe a {\bf coNP} decision
procedure.\end{blu}

\section{KLM Logics}
We briefly recall the axiomatizations and semantics of the  KLM
systems. For the sake of exposition, we present the systems in the
order from the strongest to the weakest: \Ra, \Pe, \Cl, and \Cu.
For a complete picture of KLM systems, see
\cite{KrausLehmannMagidor:90,whatdoes}. The language of KLM logics
consists just of conditional assertions $A \cond B$. We consider a
richer language allowing boolean combinations of assertions and
propositional formulas. Our language $\elle$ is defined from a set
of propositional variables $\mathit{ATM}$, the boolean connectives
and the conditional operator $\cond$. We use $A, B, C,...$ to
denote propositional formulas (that do not contain conditional
formulas), whereas $F, G, ...$ are used to denote all formulas
(including conditionals); $\Gamma, \Delta,...$ represent sets of
formulas, whereas $X, Y, ...$ denote sets of sets of formulas. The
formulas of $\elle$ are defined as follows: if $A$ is a
propositional formula, $A \appartiene \elle$; if $A$ and $B$ are
propositional formulas, $A \cond B \appartiene \elle$; if $F$ is a
boolean combination of formulas of $\elle$, $F \appartiene \elle$.


\subsection{Rational Logic \Ra}\label{sezione R}
The axiomatization of \Ra \ consists of all axioms and rules of
propositional calculus together with the following axioms and
rules. We use $\prova_{PC}$ to denote provability in the
propositional calculus, whereas $\prova$ is used to denote
provability in \Ra:

\begin{itemize}
  \item REF. $A \cond A$ (reflexivity)
  \item LLE. If $\prova_{PC} A \sse B$, then $\prova (A \cond C) \rightarrow (B \cond C)$
(left logical equivalence)
  \item RW. If $\prova_{PC} A \imp B$, then $\prova (C \cond A) \rightarrow (C \cond B)$
(right weakening)
  \item CM. $((A \cond B) \andd (A \cond C)) \imp (A \andd B \cond  C)$
(cautious monotonicity)
  \item AND. $((A \cond B) \andd (A \cond C)) \imp (A \cond B
  \andd C)$
  \item OR. $((A \cond C) \andd (B \cond C)) \imp (A \orr B \cond C)$
  \item RM. $((A \cond B) \andd \nott (A \cond \nott C)) \imp ((A \andd C) \cond
  B)$ (rational monotonicity)
\end{itemize}

\noindent REF states that $A$ is always a default conclusion of
$A$. LLE states that the syntactic form of the antecedent of a
conditional formula is irrelevant. RW describes a similar property
of the consequent. This allows to combine default and logical
reasoning \cite{halpern-plausibility}. CM states that if $B$ and
$C$ are two default conclusions of $A$, then adding one of the two
conclusions to $A$ will not cause the retraction of the other
conclusion. AND states that it is possible to combine two default
conclusions. OR states that it is allowed to reason by cases: if
$C$ is the default conclusion of two premises $A$ and $B$, then it
is also the default conclusion of their disjunction. RM is the
rule of \emph{rational monotonicity}, which characterizes the
logic \Ra\footnote{As we will see in section \ref{sottosezione
semantica P}, the axiom system of the weaker logic \Pe \ can be
obtained from the axioms of \Ra \ without RM.}: if $A \cond B$ and
$\nott (A \cond \nott C)$ hold, then one can infer $A \andd C
\cond B$. This rule allows a conditional to be inferred from a set
of conditionals in absence of other information. More precisely,
``it says that an agent should not have to retract any previous
defeasible conclusion when learning about a new fact the negation
of which was not previously derivable'' \cite{whatdoes}.

The semantics of \Ra \ is defined by considering possible world
structures with a preference relation (a strict partial order,
i.e. an irreflexive and transitive relation) $w < w'$, whose
meaning is that $w$ is preferred to $w'$. The preference relation
is also supposed to be \emph{modular}: for all $w, w_1$ and $w_2$,
if $w_1 < w_2$ then either $w_1<w$ or $w < w_2$. We have that $A
\cond B$ holds in a model $\emme$ if $B$ holds in all
\emph{minimal worlds} (with respect to the relation $<$) where $A$
holds. This definition makes sense provided minimal worlds for $A$
exist whenever there are $A$-worlds. This is ensured by the
\emph{smoothness condition} in the next definition.

\begin{definition}[Semantics of \Ra, Definition 14 in \cite{whatdoes}]\label{semantica R}
  A rational model is a triple

  \[\emme= \sx \WW, <, V \dx\]

\noindent   where:

  \begin{itemize}

    \item$\WW$ is a non-empty set of items called worlds;

    \item $<$ is an irreflexive, transitive and modular relation
    on $\WW$;

    \item  $V$ is a function $V: \WW \longmapsto pow(\mathit{ATM})$, which
    assigns to every world $w$ the set of atoms holding in
    that world.

  \end{itemize}

\noindent   We define the truth conditions for a formula $F$ as
follows:

\begin{itemize}
\item If $F$ is a boolean combination of formulas, $\emme, w
\modello F$ is defined as for propositional logic;
  \item Let $A$ be a propositional formula; we define
$Min_{<}(A)=\{w\appartiene \WW
     \tc \emme, w \modello A $ and $\perogni w'$, $w' < w$ implies $\emme,
w' \not\modello
     A\}$;

\item $\emme, w \modello A \cond B$ if for all $w'$, if $w' \appartiene
  Min_{<}(A)$ then $\emme, w' \modello B$.
\end{itemize}

\noindent \emph{(Smoothness Condition).} The relation $<$
satisfies the following condition, called \emph{smoothness}: if
$\emme, w \modello A$, then $w \appartiene Min_{<}(A)$ or $\esiste
w' \appartiene
    Min_{<}(A)$ such that $w'<w$.

\noindent We say that a formula $F$ is \emph{valid in a model}
$\emme$, denoted with $\emme \modello F$, if $\emme, w \modello F$
for every $w \appartiene \WW$. A formula is \emph{valid} if it is
valid in every model $\emme$. A formula  $F$ is satisfiable if
there exists a model $\emme$ such that $\emme \modello F$.
\end{definition}

\noindent Observe that the above definition of rational model
extends the one given by KLM to boolean combinations of formulas.

Notice also that the truth conditions for conditional formulas are
given with respect to single possible worlds for uniformity sake.
Since the truth value of a conditional only depends on global
properties of $\emme$, we have that: $\emme, w \modello A \cond B$
iff $\emme \modello A \cond B$.

By the transitivity of $<$, the smoothness condition is equivalent
to the following {\em Strong Smoothness Condition}, namely that
for all $A$ and $w$, if there is a world $w'$ preferred to $w$
that satisfies $A$ (i.e. if $\exists w': w' < w$ and $\emme, w'
\models A$), then there is also a {\em minimal} such world (i.e.
$\exists w'': w'' \in Min_{<}(A)$ and $w'' < w$). This follows
immediately: by the smoothness condition, since $\emme, w' \models
A$, either $w' \in Min_{<}(A)$ (and the property immediately
follows) or $\exists w''$ s.t. $w'' < w'$ and $w'' \in
Min_{<}(A)$; in turn, by transitivity $w'' < w$. Observe that this
holds for all $A$, whether $\emme,w \models A$ or not. In turn,
this entails that $<$ does not have infinite descending chains.
\hide{ This property is also established by Corollary
\ref{finitezza delle labels} in section \ref{sezione calcolo R}.\\
}Observe also that by the modularity of $<$ it follows that
possible worlds of $\WW$ are \emph{clustered} into equivalence
classes, each class consisting of worlds that are incomparable to
one another; the classes are totally ordered\footnote{Notice that
the worlds themselves may be incomparable since the relation $<$
is not assumed to be (weakly) connected.}. In other words the
property of modularity determines a \emph{ranking} of worlds so
that the semantics of \Ra \ can be specified equivalently in terms
of \emph{ranked} models \cite{whatdoes}. By means of the
modularity condition on the preference relation, we can also prove
the following theorem, whose detailed proof can be found in the
Appendix. We write $A \cond B \appartiene_{+} \Gamma$ (resp. $A
\cond B \appartiene_{-} \Gamma$) if $A \cond B$ occurs positively
(resp. negatively) in $\Gamma$, where positive and negative
occurrences are defined in the standard way.

\begin{theorem}[Small Model Theorem]\label{small model} For any $\Gamma \subseteq
\elle$, if $\Gamma$ is satisfiable in a rational model, then it is
satisfiable in a rational model containing at most $n$ worlds,
where $n$ is the size of $\Gamma$, i.e. the length of the string
representing $\Gamma$.
\end{theorem}


\begin{sketch}
Let $\Gamma$ be satisfiable in a rational model $\emme=\sx \WW, <,
V \dx$, i.e. $\emme, x_0 \models \Gamma$ for some $x_0 \appartiene
\WW$. We build the model $\emme' = \langle \WW', <', V'\rangle$ as
follows.
\begin{itemize}
\item We build the set of worlds $\WW'$ by means of the following procedure:
\begin{enumerate}
    \item $\WW' \longleftarrow \{x_0\}$;

    \item {\bf for each} $A_i \cond B_i \in_- \Gamma$ {\bf do}

      \begin{itemize}
        \item choose  $x_i \in \WW$ s.t. $x_i \in Min_<(A_i)$ and $\emme, x_i \not\models B_i$;
        \item $\WW' \longleftarrow \WW' \cup \{x_i\}$;
      \end{itemize}

    \item {\bf for each} $A_i \cond B_i \in_+ \Gamma$ {\bf do}

     \quad {\bf if} $Min_<(A_i) \neq \emptyset$, and there is no $x_i$
    s.t. $x_i \in Min_<(A_i)$ and

    \qquad $x_i$ is already in $\WW'$ {\bf then}
     \begin{itemize}
     \item choose any  $x_i \in Min_<(A_i)$;
     \item $\WW' \longleftarrow \WW' \cup \{x_i\}$;
     \end{itemize}
\end{enumerate}

\item For all $x_i, x_j \in \WW'$, we let $x_i <' x_j$ if $x_i < x_j$;

\item For all $x_i \in \WW'$, we let $V'(x_i) = V(x_i)$.
\end{itemize}

We can easily show that $\WW'$ is a rational model satisfying
$\Gamma$.
\end{sketch}


\noindent In the calculus for \Ra, that we will introduce in
section \ref{sezione calcolo R}, we need a slightly extended
language $\elle_R$. $\elle_R$ extends $\elle$ by formulas of the
form $\bbox A$, where $A$ is propositional, whose intuitive
meaning is that $\bbox A$ holds in a world $w$ if $A$ holds in all
the worlds preferred to $w$ (i.e. in all $w'$ such that $w' < w$).
We extend the notion of rational model to provide an evaluation of
boxed formulas as follows:

\begin{definition}[Truth condition of modality $\bbox$]\label{verit di
box}
  We define the truth condition of a boxed formula as follows:

  \begin{center}
    $\emme, w \modello \bbox A$ if, for every $w' \appartiene \WW$, if
$w' <
    w$ then $\emme, w' \modello A$
  \end{center}
\end{definition}

\noindent From definition of $Min_{<}(A)$ in Definition
\ref{semantica R} above, and Definition \ref{verit di box}, it
follows that  for any formula $A$, $w \appartiene Min_{<}(A)$ iff
$\emme, w \modello A \andd \bbox \nott A$.

Notice that by the Strong Smoothness Condition, it holds that if
$\emme, w \not\models \bbox \neg A$, then $\exists w' < w$:
$\emme, w' \models A \land \bbox \neg A$. If we regard the
relation $<$ as the inverse of the accessibility relation $R$
(thus $x R y$ if $y<x$), it immediately follows that the Strong
Smoothness Condition is an instance of the property G (restricted
to $A$ propositional). Hence it turns out that the modality
$\bbox$ has the properties of modal system G, in which the
accessibility relation
 is transitive and does not have infinite ascending chains.

 \hide{The correspondence
 between rational models and models in which
$<$ does not have infinite-descending chains is also independently
stated by Corollary \ref{infinite-chainsR} in section \ref{sezione
calcolo R}.}

Since we have introduced boxed formulas for capturing a notion of
minimality among worlds, in the rest of the paper we will only use
this modality in front of negated formulas. Hence, to be precise,
the language $\elle_R$ of our tableau extends $\elle$ with modal
formulas of the form $\bbox \nott A$.


\subsection{Preferential Logic \Pe}\label{sottosezione semantica P}

The axiomatization of \Pe \ \begin{blu}can be obtained from the
axiomatization of \Ra \ by removing the axiom RM.
\end{blu} As for \Ra, the semantics of \Pe \ is defined by considering
possible world structures with a preference relation (an
irreflexive and transitive relation), which is no longer assumed
to be modular.

\begin{definition}[Semantics of P, Definition 16 in \cite{KrausLehmannMagidor:90}]\label{semantica P}
  A preferential model is a triple

  \[\emme= \sx \WW, <, V \dx\]

\noindent   where $\WW$ and $V$ are defined as for rational models
 in Definition \ref{semantica
  R}, and  $<$ is an irreflexive and transitive relation
    on $\WW$. The truth conditions for a formula $F$, the smoothness condition,
    and the notions of validity of a formula are defined as for rational models in Definition \ref{semantica
  R}.
\end{definition}

\noindent As for rational models, we have extended the definition
of preferential models given by KLM in order to deal with boolean
combinations of formulas.

Even in this case, we define the satisfiability of conditional
formulas with respect to worlds rather than with respect to models
for uniformity sake. As for \Ra, by the transitivity of $<$, the
smoothness condition is equivalent to the Strong Smoothness
Condition. In turn, this entails that $<$ does not have infinite
descending chains.

\hide{\begin{blu}This property is also established by Corollary
\ref{infinite-chainsP} in section \ref{sezione calcolo
P}.\end{blu}}

Here again, we consider the language $\elle_P$ of the calculus
introduced in section \ref{sezione calcolo P}; $\elle_P$
corresponds to the language $\elle_R$, i.e. it extends $\elle$ by
boxed formulas of the form $\bbox \nott A$. It follows that, even
in \Pe, we can prove that, for any formula $A$, $w \appartiene
Min_{<}(A)$ iff $\emme, w \modello A \andd \bbox \nott A$.

\hide{The correspondence
 between preferential models and models in which
$<$ does not have infinite-descending chains is also independently
stated by Corollary \ref{infinite-chainsP} in section \ref{sezione
calcolo P}.}

\subsubsection{Multi-linear models for \Pe}\label{sezione multilineare}
In the following of the paper we will need a special kind of
preferential models, that we call \emph{multi-linear}. As we will
see, these models will be useful in order to provide an optimal
calculus for \Pe. \begin{blu}Indeed, as we will see in section
\ref{sezione complessità P}, our calculus for \Pe \ based on
multi-linear models will allow us to define proof search
procedures for testing the satisfiability of a set of formulas in
\Pe \ in nondeterministic polynomial time. This result matches the
known complexity results for \Pe, according to which the problem
of validity for \Pe \ is in coNP.
\end{blu}

\begin{definition}\label{definizione modello multi lineare}
A finite preferential model $\emme = (\WW, <, V)$ is {\it
multi-linear} if the set of worlds $\WW$ can be partitioned into a
set of components $\WW_i$ for $i=1,\ldots, n$, that is $\WW =
\WW_1 \cup \ldots \cup \WW_n$ and
\begin{enumerate}
\item the relation $<$ is a total order on each $\WW_i$;
\item the elements in  two different components $\WW_i$ and $\WW_j$ are
incomparable with respect to $<$.
\end{enumerate}
\end{definition}

\noindent The following theorem shows that  we could restrict our
consideration to multi-linear models and generalizes Lemma 8 in
\cite{whatdoes}. The proof can be found in the Appendix.

\begin{theorem}\label{teorema nicola}
Let $\Gamma$ be any set of formulas, if $\Gamma$ is satisfiable
then it has a multi-linear model.
\end{theorem}



\subsection{Loop Cumulative Logic {\bf CL}}\label{sezione CL}

The next KLM logic we consider is {\bf CL}, weaker than \Pe. The
axiomatization of {\bf CL}  can be obtained from the
axiomatization of {\bf P} by removing the axiom OR and by adding
the following infinite set of LOOP axioms:
$$\mbox{LOOP.} \ (A_0 \cond A_1) \andd (A_1 \cond
A_2) ... (A_{n-1} \cond A_n) \andd (A_n \cond A_0) \imp (A_0 \cond
A_n)$$ and the following axiom CUT:
$$\mbox{CUT.} \ ((A \cond B) \andd (A
\andd B \cond  C)) \imp (A \cond C)$$

\noindent Notice that these axioms are derivable in \Pe \ (and
therefore in \Ra).

The following Definition is essentially the same as Definition 13
in \cite{KrausLehmannMagidor:90}, but it is extended to boolean
combinations of conditionals.

\begin{definition}[Semantics of \Cl]\label{loopcummodels}
 A loop-cumulative model is a tuple

 \[\emme=\langle S, \WW, l, <, V
\rangle\]

\noindent where:

 \begin{itemize}
   \item $S$ is a set, whose elements are called states;
   \item $\WW$ is a set of possible worlds;
   \item $l: S \mapsto \mathit{pow}(\WW)$ is a
function that labels every state with a nonempty set of worlds;
  \item $<$ is an irreflexive and transitive relation on $S$;
  \item $V$ is a
valuation function $V: {\WW} \longmapsto pow(\mathit{ATM})$, which
assigns to every world $w$ the atoms holding in that world.
\end{itemize}

\begin{blu}
\noindent For $s \appartiene S$ and $A$ propositional, we let
$\emme, s \mid\hspace{-1pt}\equiv A$ if $\perogni w \appartiene
l(s)$, $\emme, w \modello A$, where $\emme, w \modello A$ is
defined as for propositional logic. Let $Min_<(A)$ be the set of
minimal states $s$ such that $\emme, s \mid\hspace{-1pt}\equiv A$.
We define $\emme, s \mid\hspace{-1pt}\equiv A \cond B$ if
$\perogni s' \appartiene Min_<(A)$, $\emme, s'
\mid\hspace{-1pt}\equiv B$. The relation $\mid\hspace{-1pt}\equiv$
can be extended to boolean combinations of conditionals in the
standard way. We assume that $<$ satisfies the smoothness
condition.
\end{blu}
\end{definition}

\noindent The above notion of cumulative model extends the one
given by KLM to boolean combinations of conditionals. A further
extension to arbitrary boolean combinations will be provided by
the notion of CL-preferential model below.

Here again, we define satisfiability of conditionals with respect
to states rather than with respect to models for uniformity
reasons. Indeed, a conditional is satisfied by a state of a model
only if and only if it is satisfied by all the states of that
model, hence by the whole model.

As for \Pe \ and \Ra, by the transitivity of $<$, the smoothness
condition is equivalent to the Strong Smoothness Condition. In
turn, this entails that $<$ does not have infinite descending
chains.

\hide{This property is also established by Corollary
\ref{infinite-chainsCL} in section \ref{sezione calcolo CL}.}

We show that we can map loop-cumulative models into preferential
models extended with an additional accessibility relation $R$. We
call these preferential models {\em CL-preferential models}. The
idea is to represent states as sets of possible worlds related by
$R$ in such a way that a formula is satisfied in a state $s$ just
in case it is satisfied in all possible worlds $w'$ accessible
from a world $w$ corresponding to $s$. The syntactic counterpart
of the extra accessibility relation $R$ is a modality $L$. Given a
loop-cumulative model $\emme$ and the corresponding
CL-preferential model $\emme'$, $\emme, s \mid\hspace{-1pt}\equiv
A$ iff for a world $w \in \emme'$ corresponding to $s$, we have
that $\emme', w \models LA$. As we will see, this mapping enables
us to use a variant of the tableau calculus for \Pe \ to deal with
system {\bf CL}. As for \Pe, the tableau calculus for \Cl \ will
use boxed formulas. In addition, it will also use $L$-formulas.
Thus, the formulas that appear in the tableaux for \Cl \ belong to
the language $\lan_L$ obtained from $\lan$ as follows: $(i)$ if
$A$ is propositional, then $A \in \lan_L$; $LA \in \lan_L$; $\bbox
\neg LA \in \lan_L$; $(ii)$ if $A$, $B$ are propositional, then $A
\cond B \in \lan_L$; $(iii)$ if $F$ is a boolean combination of
formulas of $\elle_L$, then $F \appartiene \elle_L$. Observe that
the only allowed combination of $\bbox$ and $L$ is in formulas of
the form $\bbox \nott L A$ where $A$ is propositional.

We can map loop-cumulative models into preferential models with an
additional accessibility relation as defined below:

\begin{definition}[CL-preferential models]\label{semantica_CL}
  A CL-preferential model has the form

  \[\emme= \sx \WW, R, <, V \dx\]

\noindent   where:

  \begin{itemize}
    \item $\WW$ and $V$ are defined as for preferential models in Definition
  \ref{semantica P};

    \item $<$ is an irreflexive and transitive relation on $\WW$;

     \item $R$ is a serial accessibility relation;
   \end{itemize}
\noindent      We add to the truth conditions for preferential
models in Definition \ref{semantica P} the following clause:
\begin{center}
$\emme, w \models LA$ if, for all $w'$, $w R w'$ implies $\emme,
w' \models A$
\end{center}

\noindent The relation $<$ satisfies the following
\emph{Smoothness Condition}: if $\emme, w \modello LA$, then $w
\appartiene Min_{<}(LA)$ or $\esiste w' \appartiene
    Min_{<}(LA)$ such that $w'<w$.

\noindent Moreover, we need to change the truth condition for
conditional formulas as follows: $\emme, w \modello A \cond B$ if
for all $w' \appartiene Min_{<}(LA)$ we have $\emme, w' \modello
LB$.
\end{definition}

\begin{rosso}
\noindent We can prove that, given a loop-cumulative model
$\emme=\sx S, \WW, l, <, V \dx$ satisfying a boolean combination
of conditional formulas, one can build a CL-preferential model
$\emme'=\sx \WW', R, <', V' \dx$ satisfying the same combination
of conditionals. We build a CL-preferential model $\emme' =
\langle \WW', R, <', V' \rangle$ as follows:
\begin{itemize}
\item $\WW' = \{(s,w): s \in S$ and $w \in l(s)\}$;
\item $(s, w) R (s, w')$ for all $(s, w), (s, w') \in \WW'$;
\item $(s, w) <' (s', w')$ if $s < s'$;
\item $V'(s, w)$ = $V(w)$.
\end{itemize}

\noindent Viceversa, given a CL-preferential model $\emme=\sx \WW,
R, <, V\dx$ satisfying a boolean combination of conditional
formulas, one can build a loop-cumulative model $\emme' = \langle
S, \WW, l, <', V' \rangle$ satisfying the same combination of
conditional formulas. The model $\emme'$ is defined as follows (we
define $Rw=\{w' \in \WW \tc (w,w') \in R \}$):

\begin{itemize}
\item $S = \{(w, Rw) \tc w \in \WW\}$;
\item $l((w, Rw)) = Rw$;
\item $(w, Rw) <' (w', Rw')$ if $w < w'$;
\item $V'(w)$ = $V(w)$.
\end{itemize}

\noindent This is stated in a rigorous manner by the following
proposition, whose proof can be found in the Appendix:
\begin{proposition}\label{correspondence CL}
A boolean combination of conditional formulas is satisfiable in a
loop-cumulative model $\emme = \langle S, \WW , l, <, V \rangle$
iff it is satisfiable in a CL-preferential model $\emme'= \langle
\WW', R, <', V' \rangle$.
\end{proposition}


\begin{ultimissimi}
Similarly to what done for \Pe, we define multi-linear
CL-preferential models as follows:

\begin{definition}\label{definizione modello multi lineare in CL}
A finite CL-preferential model $\emme = (\WW, R, <, V)$ is {\it
multi-linear} if the set of worlds $\WW$ can be partitioned into a
set of components $\WW_i$ for $i=1,\ldots, n$ (that is $\WW =
\WW_1 \cup \ldots \cup \WW_n$), and in each $\WW_i$:
\begin{enumerate}
\item we can distinguish a chain of worlds $w_1,
w_2, \dots, w_h$ totally ordered w.r.t. $<$ (i.e. $w_1 < w_2 <
\dots < w_h$) such that all other worlds $w \in \WW_i$ are
$R$-accessible from some $w_l$ in the chain, i.e. $\perogni w \in
\WW_i$ such that $w \diverso w_1, w_2, \dots, w_h$, we have that
$w_1 R w \orr w_2 R w \orr \dots \orr w_h R w$;
\item for all $w_l, w_j, w_k$, if $w_l < w_j$ and $w_j R w_k$,
then $w_l < w_k$.
\end{enumerate}
Moreover, the elements of different $\WW_i$ are incomparable
w.r.t. $<$.
\end{definition}

\noindent We can easily prove the following Theorem:

\begin{theorem}\label{teorema nicola per CL}
Let $\Gamma$ be any set of formulas, if $\Gamma$ is satisfiable in
a CL-preferential model, then it has a multi-linear model.
\end{theorem}

\end{ultimissimi}

\hide{ \noindent Notice that, in a CL-preferential model, the
accessibility relation $R$ is defined to be serial. However, in
the above construction that, given a loop-cumulative model, builds
a ``corresponding'' (i.e. satisfying the same conditionals)
CL-preferential model, $R$ is an equivalence relation. We should
define a CL-preferential model in which $R$ is not only serial
(for instance, $R$ should be an euclidean or an equivalence
relation); we have chosen that $R$ is only serial since this is
sufficient to prove a correspondence between loop-cumulative
models and CL-preferential models. Moreover, as we will see in
section \ref{sezione calcolo CL}, this allows us to describe a
simpler tableau system, in which the rule taking care of the
modality $L$ corresponds to the rule of a modal logic whose
accessibility relation is serial: this is enough, since we do not
have to consider formulas containing nested $L$-formulas.

}

\end{rosso}


\subsection{Cumulative Logic \Cu}

The weakest logical system considered by KLM
\cite{KrausLehmannMagidor:90}  is Cumulative Logic {\bf C}. System
{\bf C} is weaker than {\bf CL} considered above since it does not
have the set of (LOOP) axioms. At a semantic level, the difference
between {\bf CL} models and {\bf C} models is that in {\bf CL}
models the relation $<$ is transitive, whereas in {\bf C} it is
not. Thus, cumulative {\bf C} models are defined as follows :
\begin{definition}[Semantics of \Cu, Definitions 5, 6, 7 in \cite{KrausLehmannMagidor:90}]
 A cumulative model is a tuple

\begin{center}
 $\emme=\langle S, \WW, l, <, V \rangle$
\end{center}

\noindent  where $S$, $\WW$, $l$, and $V$ are defined as for
loop-cumulative models in Definition \ref{loopcummodels}, whereas
  $<$ is an irreflexive
relation on $S$. The truth definitions of formulas are as for
loop-cumulative models in Definition \ref{loopcummodels}. We
assume that $<$ satisfies the smoothness condition.
\end{definition}

\noindent Since $<$ is no longer transitive, the smoothness
condition is no longer equivalent to the Strong Smoothness
Condition; hence, in this case, we cannot show that $<$ does  not
have infinite descending chains. Indeed, the relation $<$ might
have cycles (leading to infinite descending chains): it can be
easily seen that in \Cu \ we may have sequences of worlds such as:
a minimal $A$-world followed by a minimal $B$-world followed by a
minimal $A$-world and so on. This sequence respects the smoothness
condition. However, one can legitimately wonder what minimal means
in this case, the notion having lost its intuitive meaning.

In order to be convinced that (1) the Strong Smoothness Condition
and (2) the smoothness condition are not equivalent, consider the
following set of formulas: $\{\nott (C \cond B), C \cond A, A
\cond B, B \cond C\}$. This set of formulas is unsatisfiable in a
model satisfying (1) whereas it is satisfiable in a model only
satisfying (2), hence it is satisfiable in \Cu.

 Similarly to
what done for loop-cumulative models, we can establish a
correspondence between cumulative models and preferential models
augmented with an accessibility relation in which the preference
relation $<$ is an irreflexive relation satisfying the smoothness
condition. We call these models C-preferential models.

\begin{definition}[C-preferential models]\label{semantica_C}
  A C-preferential model has the form $\emme= \sx \WW, R, <, V \dx$
  where: $\WW$ is a non-empty set of items called worlds;
     $R$ is a serial accessibility relation;
    $<$ is an irreflexive relation
    on $\WW$ satisfying the smoothness condition for L-formulas;
    $V$ is a function $V: \WW \longmapsto pow(ATM)$, which
    assigns to every world $w$ the atomic formulas holding in
    that world. The truth conditions for the boolean cases are defined in
  the obvious way. Truth conditions for modal and conditional
  formulas are the same as in CL-preferential models in Definition \ref{semantica_CL}, thus:
\begin{itemize}
\item $M, w \models LA$ if for all $w'$,  $w R w'$ implies $\emme,
w' \modello A$
 \item $\emme, w \modello A
\cond B$ if for all $w' \appartiene Min_<(LA)$, we have $\emme, w'
\modello LB$.
\end{itemize}
\end{definition}

\noindent The correspondence between cumulative and preferential
models is established by the following proposition. Its proof is
the same as the proof of Proposition \ref{correspondence CL}
(except for transitivity) and is therefore omitted.

\begin{proposition}\label{corrispondenza-C} A boolean combination of
conditional formulas is satisfiable in a cumulative model $\emme =
\langle S, \WW, l, <, V \rangle$ iff it is satisfiable in a
C-preferential model $\emme'= \langle \WW', R, <', V' \rangle$.
\end{proposition}

In the following sections we present the tableaux calculi for the
logics introduced. We start by presenting the calculus for \Pe,
which is the simpler and more general one. The calculi for \Cl,
\Cu, and \Ra \ will become more understandable once the calculus
for \Pe \ is known.


\section{The Tableau Calculus for Preferential Logic \Pe}\label{sezione
calcolo P}
In this section we present a tableau calculus for \Pe \
called $\calcoloP$, then we analyze it in order to obtain a
decision procedure for this logic. We also give an explicit
complexity bound for \Pe.

As already mentioned in section \ref{sottosezione semantica P}, we
consider the language $\elle_P$, which extends $\elle$ by boxed
formulas of the form $\bbox \nott A$.

\begin{definition}[The calculus $\calcoloP$]\label{Definizioni Gamma box ecc.}
  The rules of the calculus manipulate sets of formulas $\Gamma$.
  We write $\Gamma, F$ as a shorthand for $\Gamma \unione \{F\}$. Moreover, given $\Gamma$ we
  define the following sets:
  \begin{itemize}
    \item $\Gamma^{\bbox}=\{ \bbox \nott A \tc \bbox \nott A \appartiene \Gamma
    \}$
    \item $\Gamma^{\bbox^{\freccia}}=\{ \nott A \tc \bbox \nott A \appartiene \Gamma \}$
    \item $\Gamma^{\cond^{+}}=\{A \cond B \tc A \cond B \appartiene
    \Gamma\}$
    \item $\Gamma^{\cond^{-}}=\{\nott(A \cond B) \tc\nott(A \cond B)\appartiene
    \Gamma\}$
    \item $\Gamma^{\cond\pm}=\Gamma^{\cond^{+}} \unione \Gamma^{\cond^{-}}$
  \end{itemize}
\noindent The tableau rules are given in Figure \ref{Figura
calcolo preferential}. A tableau is a tree whose nodes are sets of
formulas $\Gamma$. Therefore, a branch is a sequence of sets of
formulas $\Gamma_1, \Gamma_2, \dots, \Gamma_n, \dots$ Each node
$\Gamma_i$ is obtained by its immediate predecessor $\Gamma_{i-1}$
by applying a rule of $\calcoloP$, having $\Gamma_{i-1}$ as the
premise and $\Gamma_i$ as one of its conclusions. A branch is
closed if one of its nodes is an instance of $({\bf AX})$,
otherwise it is open. We say that a tableau is closed if all its
branches are closed.
\end{definition}

\begin{figure}[h]

\linea

\mbox{ \(
\begin{scriptsize}
\begin{array}{l@{\quad\quad\quad\quad\quad\quad\quad\quad\quad\quad\quad\quad}l}\\
{\bf (AX)} \ \Gamma, P, \nott P \quad \mbox{with $P \appartiene
\mathit{ATM}$} {} & {\bf (\nott)} \ \irule{\Gamma, \nott \nott F}
{\Gamma, F} {}
\\
\\
\\
{\bf (\andd^{+})} \ \irule{\Gamma, F \andd G}
  {\Gamma, F, G} {} & {\bf (\andd^{-})} \ \irule{\Gamma, \nott(F \andd G)}
  {\Gamma, \nott F \quad\quad\quad \Gamma, \nott G} {}
\\
\\
\\
{\bf (\orr^{+})} \ \irule{\Gamma, F \orr G}
  {\Gamma, F \quad\quad\quad \Gamma, G} {} &
  {\bf (\orr^{-})} \ \irule{\Gamma, \nott(F \orr G)}
  {\Gamma, \nott F, \nott G} {}
\\
\\
\\
{\bf (\imp^{+})} \ \irule{\Gamma, F \imp G}
  {\Gamma, \nott F \quad\quad\quad \Gamma, G} {} &
  {\bf (\imp^{-})} \ \irule{\Gamma, \nott(F \imp G)}
  {\Gamma, F, \nott G} {}
\\
\\
\\
\end{array}
\end{scriptsize}
\) }

\mbox{ \(
\begin{scriptsize}
\begin{array}{ll}
\quad\quad\quad\quad\quad {\bf (\cond^{+})} \ \irule{\Gamma, A
\cond B}
  {\Gamma, \nott A, A \cond B
    \quad\quad\quad \Gamma, \nott \bbox \nott A, A \cond B
    \quad\quad\quad \Gamma, B, A \cond B
  } {}
\\
\\
\\
\end{array}
\end{scriptsize}
\) }

\mbox{ \(
\begin{scriptsize}
\begin{array}{l@{\quad\quad\quad\quad\quad\quad\quad\quad\quad\quad\quad\quad\quad\quad\quad}l}
{\bf (\cond^{-})} \ \irule{\Gamma, \nott(A \cond B)}%
{A, \bbox \nott A, \nott B, \Gamma^{\cond\pm}} {}%
 & {\bf (\bbox^{-})} \
\irule{\Gamma, \nott \bbox \nott A}%
{\Gamma^{\bbox}, \Gamma^{\bbox^{\freccia}}, \Gamma^{\cond\pm}, A, \bbox \nott A} {}%
\\
\\
\end{array}
\end{scriptsize}
\) }

\linea

\caption{Tableau system $\calcoloP$.} \label{Figura calcolo
preferential}
\end{figure}

\noindent \begin{rosso}The rues for the boolean propositions are
the usual ones. According to the rule $(\cond^{-})$, if a negated
conditional $\nott (A \ent B)$ holds in a world, then there is a
minimal $A$-world (i.e. in which $A$ and $\bbox \nott A$ hold)
which falsifies $B$. According to the rule $(\cond^{+})$, if a
positive conditional $A \cond B$ holds in a world, then either the
world falsifies $A$ or it is not minimal for $A$ (i.e. $\nott
\bbox \nott A$ holds) or it is a $B$-world. According to the rule
$(\bbox^{-})$, if a world satisfies $\nott \bbox \nott A$, by the
strong smoothness condition there must be a preferred minimal
$A$-world, i.e. a world in which $A$ and $\bbox \nott A$ hold. In
our calculus $\calcoloP$, axioms are restricted to atomic formulas
only. It is easy to extend axioms to a generic formula $F$, as
stated by the following Proposition:\end{rosso}

\begin{proposition}\label{assiomi estesi a formule qualsiasi}
  Given a formula $F$ and a set of formulas $\Gamma$, then
  $\Gamma, F, \nott F$ has a closed tableau.
\end{proposition}
\begin{provaallettore}
  By an easy inductive argument on the structure of the formula
  $F$.
\end{provaallettore}

\begin{definition}\label{consistenza}
Given a set of formulas $\Gamma$, $\Gamma$ is consistent if no
tableau for $\Gamma$ is closed.
\end{definition}

\begin{figure}
{\centerline{\includegraphics[angle=0,width=4.8in]{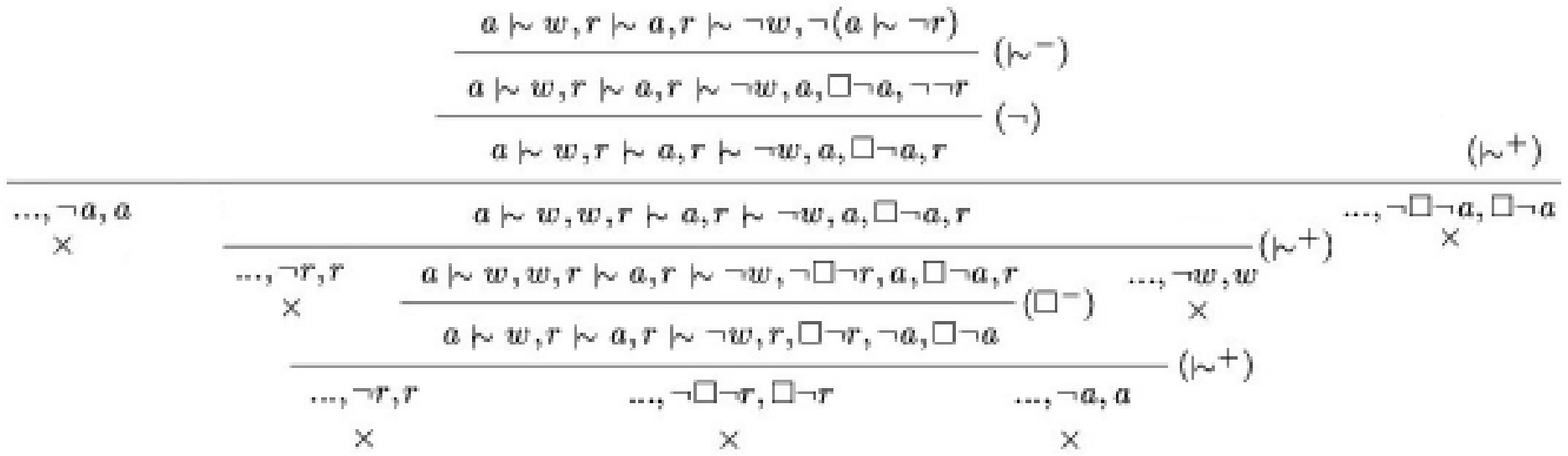}}}
 \caption{A derivation of $adult \cond
worker, retired \cond adult, retired \cond \nott worker,$
 $\nott (adult \cond \nott retired)$. For readability, we use $a$ to denote $adult$, $r$ for $retired$,
 and $w$ for $worker$.} \label{Esmepio di derivazione}
\end{figure}

\begin{blu}
\noindent As an example, we show that $adult \ent \nott retired$
can be inferred from a knowledge base containing the following
assertions: $adult \cond worker, retired \cond adult, retired
\cond \nott worker$. Figure \ref{Esmepio di derivazione} shows a
derivation for the initial set of formulas $adult \cond worker,
retired \cond adult, retired \cond \nott worker,
 \nott (adult \cond \nott retired)$.
\end{blu}

\noindent Our tableau calculus $\calcoloP$ is based on a runtime
translation of conditional assertions into modal logic G. As we
have seen in section \ref{sottosezione semantica P}, this allows a
characterization of the minimal worlds satisfying a formula $A$
(i.e., the worlds in $Min_{<}(A)$) as the worlds $w$ satisfying
the formula $A \andd \bbox \nott A$. It is tempting to provide a
full translation of the conditionals in the logic G, and then to
use the standard tableau calculus for G. To this purpose, we can
exploit the transitivity properties of G frames in order to
capture the fact that conditionals are global to all worlds by the
formula $\bbox(A \andd \bbox \nott A\rightarrow  B)$. Hence, the
overall translation of a conditional formula $A \cond B$ could be
the following one: $(A \andd \bbox \nott A \rightarrow  B) \andd
\bbox(A \andd \bbox \nott A \rightarrow  B)$. However, there are
significant differences between the calculus resulting from the
translation and our calculus.

Using the standard tableau rules for G on the translation, we get
the rule $(\cond^{+})$ as a derived rule. Instead, the rule for
dealing with negated conditionals (which would be translated in G
as a disjunction of two formulas, namely $(A \andd \bbox \nott A
\andd \neg B) \vee \Diamond(A \andd \bbox \nott A \andd  \neg
B)$), is rather different.

Let us first observe that the rule $(\cond^{-})$ we have
introduced precisely captures the intuition that conditionals are
global, hence (1) all conditionals are kept in the conclusion of
the rule and (2) when moving to a new minimal world, all the boxed
formulas (positive and negated) are removed. Conversely, when the
tableau rules for G are applied to the translation of the negated
conditionals, we get two branches (due to the disjunction). None
of the branches can be eliminated. In both branches all the boxed
formulas are kept, while negated conditionals are erased. This is
quite different from our rule $(\cond^{-})$, and it is not that
obvious that the calculus obtained by the translation of \Pe \
conditionals in G is equivalent to $\calcoloP$. \begin{rosso}
Roughly speaking, point (2) can be explained as follows: when a
negated conditional $\nott (A \ent B)$ is evaluated in a world
$w$, this corresponds to finding a minimal $A$-world $w'$
satisfying $\nott B$ (a world satisfying $A, \bbox \nott A, \nott
B$). $w'$ does not depend from $w$ (since conditionals are
global), hence boxed formulas, keeping information about $w$, can
be removed.
\end{rosso}

Also observe that, from the semantic point of view, the model
extracted from an open tableau has the structure of a forest,
while the model constructed by applying the tableau for G to the
translation of conditionals has the structure of a tree. This
difference is due to the fact that the above translation of \Pe \
in G uses the same modality $\bbox$ both for capturing the
minimality condition and for modelling the fact that conditionals
are global. For this reason, a translation to G as the one
proposed above for \Pe, would not be applicable to the cumulative
logic \Cu, as the relation $<$ is not transitive in \Cu. Moreover,
the treatment of both the logics \Cu \ and \Cl \ would anyhow
require the addition to the language of a new modality to deal
with states. The advantage of the runtime translation we have
adopted is that of providing a uniform approach to deal with the
different logics.

The system $\calcoloP$ is sound and complete with respect to the
semantics.
\begin{theorem}[Soundness of $\calcoloP$]\label{correttezza}
  The system $\calcoloP$ is sound with respect to preferential models,
  i.e. if there is a closed tableau for a set $\Gamma$, then $\Gamma$
  is unsatisfiable.
\end{theorem}

\begin{provaposu} As usual, we proceed by induction on the
structure of the closed tableau having the set $\Gamma$ as a root.
The base case is when the tableau consists of a single node; in
this case, both $P$ and $\nott P$ occur in $\Gamma$, therefore
$\Gamma$ is obviously unsatisfiable. For the inductive step, we
have to show that, for each rule $r$, if all the conclusions of
$r$ are unsatisfiable, then the premise is unsatisfiable too. We
show the contrapositive, i.e. we prove that if the premise of $r$
is satisfiable, then at least one of the conclusions is
satisfiable. \hide{Consider the rule $(\andd^{-})$ as an example:
to say that $\Gamma, \nott (A \andd B)$ is satisfiable means that
there is a model $\emme=\sx W, <, V\dx$ with some $w \appartiene
W$ such that $\emme, w \modello \Gamma, \nott (A \andd B)$, i.e.
$\emme, w \modello \Gamma$ and $\emme, w \not\modello A \andd B$;
this means that $\emme, w \not\modello A$ or $\emme, w
\not\modello B$: in the first case the left conclusion of
$(\andd^{-})$ is satisfiable, in the second case the right
conclusion is satisfiable.} Boolean cases are easy and left to the
reader. We present the cases for conditional and box rules:

\begin{itemize}
  \item $(\cond^{+})$: if $\Gamma, A \cond B$ is satisfiable, then there exists a model $\emme = \sx \WW, <, V\dx$ with
  some world $w \appartiene \WW$ such that $\emme, w \modello \Gamma, A
  \cond B$. We distinguish the two following cases:
    \begin{itemize}
      \item $\emme, w \not\modello
      A$, thus $\emme, w \modello \nott A$: in this case, the left
     conclusion of the $(\cond^{+})$ rule is satisfied ($\emme, w \modello \Gamma, A \cond B, \nott A$);
     \item $\emme, w \modello A$: we consider two subcases:
       \begin{itemize}
         \item $w \appartiene Min_{<}(A)$: by the definition
  of $\emme, w \modello A \cond B$, we have that for all $w' \appartiene
  Min_{<}(A)$, $\emme, w' \modello B$. Therefore, we have
  that $\emme, w \modello B$ and the right
  conclusion of $(\cond^{+})$ is satisfiable;
         \item $w \not\appartiene Min_{<}(A)$: by the smoothness condition,
         there exists a world $w' < w$ such that $w' \appartiene
         Min_{<}(A)$; therefore, $\emme, w \modello \nott \bbox
         \nott A$ by the definition of $\bbox$. The central
         conclusion of the $(\cond^{+})$ rule is then
         satisfiable.
       \end{itemize}
    \end{itemize}

   \item $(\cond^{-})$: if $\Gamma, \nott (A \cond B)$ is satisfiable,
   then $\emme, w \modello \Gamma$ and $(*) \emme, w \not\modello A \cond
   B$ for some world $w$. By $(*)$, there is a world $w'$ in
   the model $\emme$ such that $w' \appartiene Min_{<}(A)$ (i.e. $(1)\emme, w' \modello A$
   and $(2)\emme, w' \modello \bbox \nott A$) and
   $(3)\emme, w' \not\modello B$. By $(1), (2)$ and $(3)$,
   we have that $\emme, w' \modello A, \bbox \nott A, \nott B$\footnote{We
   use $\emme, w' \modello F_1, F_2, \dots, F_n$ to denote that $\emme, w' \modello F_1$,
   $\emme, w' \modello F_2$, \dots, and $\emme, w' \modello
   F_n$.}.
   We conclude $\emme, w' \modello A, \bbox \nott A, \nott B,
   \Gamma^{\cond\pm}$, since conditionals are ``global'' in a
   model.

   \item $(\bbox^{-})$: if $\Gamma, \nott \bbox \nott A$ is satisfiable, then
   there is a model $\emme$ and some world $w$ such that $\emme, w
   \modello \Gamma, \nott \bbox \nott A$, then $\emme, w \not\modello \bbox
   \nott A$. By the truth definition of $\bbox$, there exists a
   world $w'$ such that $w' < w$ and $\emme, w'
   \modello A$. By the Strong Smoothness Condition, we can assume that $w'$ is a \emph{minimal} $A$-world.
   Therefore, $\emme, w' \modello \bbox \nott A$
   by the truth definition of $\bbox$. It is easy to conclude
   that $\emme, w' \modello A, \bbox \nott A, \Gamma^{\cond\pm},
   \Gamma^{\bbox^{\freccia}}, \Gamma^{\bbox}$, since 1. conditionals are global in a
   model, then $\emme, w' \modello \Gamma^{\cond\pm}$, 2. formulas in
   $\Gamma^{\bbox^{\freccia}}$ are true in $w'$ since $w' < w$ and 3. the
   $<$ relation is transitive, thus boxed formulas holding in $w$
   (i.e. $\Gamma^{\bbox}$) also hold in $w'$.

\end{itemize}

\end{provaposu}

\noindent To prove the completeness of $\calcoloP$ we have to show
that if $F$ is unsatisfiable, then there is a closed tableau
starting with $F$. We prove the contrapositive, that is: if there
is no closed tableau for $F$, then there is a model satisfying
$F$. This proof is inspired by \cite{gorè}. First of all, we
distinguish between \emph{static} and \emph{dynamic} rules. The
rules $(\cond^{-})$ and $(\bbox^{-})$ are called \emph{dynamic},
since their conclusion represents another world with respect to
the premise; the other rules are called \emph{static}, since the
world represented by premise and conclusion(s) is the same.
Moreover, we have to introduce the \emph{saturation} of a set of
formulas $\Gamma$. Given a set of formulas $\Gamma$, we say that
it is saturated if all the  rules have been applied.

\begin{definition}[Saturated sets]\label{Insiemi saturati}
  A set of formulas $\Gamma$ is saturated with respect to the static
  rules if the following conditions hold:

  \begin{itemize}
    \item if $F \andd G \appartiene \Gamma$ then $F \appartiene \Gamma$ and $G \appartiene \Gamma$;
    \item if $\nott (F \andd G) \appartiene \Gamma$
then $\nott F
    \appartiene \Gamma$ or $\nott G \appartiene \Gamma$;
    \item if $F \orr G \appartiene \Gamma$ then $F \appartiene
    \Gamma$ or $G \appartiene \Gamma$;
    \item if $\nott(F \orr G) \appartiene \Gamma$ then $\nott F \appartiene
    \Gamma$ and $\nott G \appartiene \Gamma$;
    \item if $F \imp G \appartiene \Gamma$ then $\nott F
    \appartiene \Gamma$ or $G \appartiene \Gamma$;
    \item if $\nott (F \imp G) \appartiene \Gamma$ then $F \appartiene \Gamma$ and $\nott
    G \appartiene \Gamma$;
    \item if $\nott \nott F \appartiene \Gamma$ then $F \appartiene \Gamma$;
    \item if $A \cond B \appartiene \Gamma$ then $\nott A \appartiene
    \Gamma$ or $\nott \bbox \nott A \appartiene \Gamma$ or $B \appartiene
    \Gamma$.
  \end{itemize}

\end{definition}

\noindent It is easy to observe that the following Lemma holds
(the proof can be found in the Appendix):

\begin{lemma}\label{lemma saturazione}
  Given a consistent finite set of formulas $\Gamma$, there is a consistent, finite, and
  saturated set $\Gamma' \supseteq \Gamma$.
\end{lemma}


\noindent By Lemma \ref{lemma saturazione}, we can think of having
a function which, given a consistent set $\Gamma$, returns one
fixed consistent saturated set, denoted by \texttt{SAT}$(\Gamma)$.
Moreover, we denote by \texttt{APPLY}$(\Gamma,F)$ the result of
applying to $\Gamma$ the rule for the principal connective in $F$.
In case the rule for $F$ has several conclusions (the case of a
branching), we suppose that the function \texttt{APPLY} chooses
one consistent conclusion in an arbitrary but fixed manner.

\begin{theorem}[Completeness of $\calcoloP$]\label{completezza}
  $\calcoloP$ is complete w.r.t. preferential models, i.e. if a set
  of formulas $\Gamma$ is unsatisfiable, then it has a
  closed tableau in $\calcoloP$.
\end{theorem}

\begin{provaposu}
We assume that no tableau for $\Gamma_0$ is closed, then we
construct a model for $\Gamma_0$. We build $X$, the set of worlds
of the model, as follows:

\vspace{0.2cm}

\begin{small}

  \noindent 1. initialize $X=\{$\texttt{SAT}$(\Gamma_0)\}$; mark
  \texttt{SAT}$(\Gamma_0)$ as unresolved;

  \noindent {\bf while} $X$ contains unresolved nodes do

  \indent 2. choose an unresolved $\Gamma$ from $X$;

  \indent 3. {\bf for} each formula $\nott (A \cond B) \appartiene \Gamma$

  \indent \indent 3a. let $\Gamma_{\nott(A \cond B)}=$\texttt{SAT}(\texttt{APPLY}$(\Gamma,\nott(A \cond
  B)))$;

  \indent\indent 3b. {\bf if} $\Gamma_{\nott(A \cond B)} \not\appartiene X$ {\bf then} $X=X
  \unione \{\Gamma_{\nott(A \cond B)}\}$;

  \indent 4. {\bf for} each formula $\nott \bbox \nott A \appartiene \Gamma$, let
  $\Gamma_{\nott \bbox \nott A}=$\texttt{SAT}(\texttt{APPLY}$(\Gamma,\nott
  \bbox \nott A))$;

  \indent\indent 4a. add the relation $\Gamma_{\nott \bbox \nott A} < \Gamma$;

  \indent\indent 4b. {\bf if} $\Gamma_{\nott \bbox \nott A} \not\appartiene X$ {\bf then} $X=X
  \unione \{\Gamma_{\nott \bbox \nott A}\}$.

  \indent 5. mark $\Gamma$ as resolved;

 \noindent {\bf endWhile};

\end{small}

 \vspace{0.2cm}

\noindent This procedure terminates, since the number of possible
sets of formulas that can be obtained by applying $\calcoloP$'s
rules to an initial finite set $\Gamma$ is finite. We construct
the model $\emme=\sx X, <_X, V\dx$ for $\Gamma$ as follows:

\begin{quote}
  $\bullet$ $<_X$ is the transitive closure of the relation $<$;\\
  $\bullet$ $V(\Gamma)=\{P \tc P \appartiene \Gamma \intersezione \mathit{ATM}\}$
\end{quote}

\noindent In order to show that $\emme$ is a preferential model
for $\Gamma$, we prove the following facts:

\begin{fact}\label{Fatto 1}
The relation $<_X$ is acyclic.
\end{fact}

\begin{pf*}{Proof of Fact \ref{Fatto 1}.}
If there were a loop, there would be $\Gamma_1$ and $\Gamma_3$ in
$X$, s.t. $\Gamma_3 <_X \Gamma_1$, and $\Gamma_1$ is obtained
again from $\Gamma_3$ by applying step 4 (i.e. $\Gamma_1 <_X
\Gamma_3$). However, this situation, presented in Figure
\ref{Immagine con il loop scongiurato}, will never happen. Indeed,
since $\Gamma_3 <_X \Gamma_1$, $\Gamma_3$ has been generated by a
sequence of applications of $(\bbox^{-})$, starting from an
initial application of $(\bbox^{-})$ to some formula $\nott \bbox
\nott A$ in $\Gamma_1$. By the $(\bbox^{-})$ rule, $\bbox \nott A
\appartiene \Gamma_3$. If $\Gamma_1$ were to be generated again
from $\Gamma_3$ by an application of $(\bbox^{-})$, then $\bbox
\nott A \appartiene \Gamma_1$, which contradicts the fact that
$\Gamma_1$ is consistent. We can reason in the same way for loops
of any length.

\begin{figure}
{\centerline{\includegraphics[angle=0,width=4.8in]{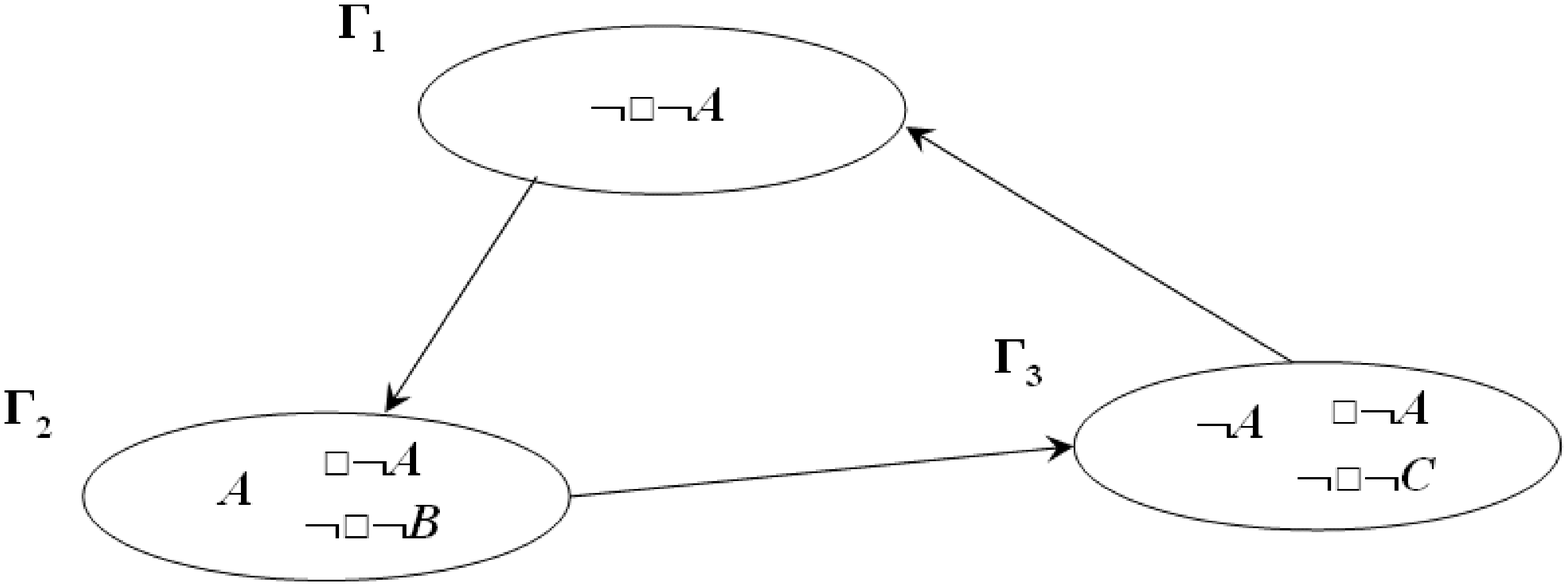}}}
 \caption{}
\label{Immagine con il loop scongiurato}
\end{figure}
\provafatto{\ref{Fatto 1}}
\end{pf*}

\begin{fact}\label{fatto aggiunto in smoothness}
The relation $<_X$ is irreflexive, transitive, and satisfies the
smoothness condition.
\end{fact}

\begin{pf*}{Proof of Fact \ref{fatto aggiunto in smoothness}.}
Transitivity follows by construction. Irreflexivity follows the
acyclicity. As there are finitely many worlds, and the relation
$<_X$ is acyclic, it follows that there cannot be infinitely
descending chains. This fact, together with the transitivity of
$<_X$, entails that $<_X$ satisfies the smoothness condition.
\provafatto{\ref{fatto aggiunto in smoothness}}
\end{pf*}

\noindent The only rules introducing a new world in $X$ in the
procedure above are $(\cond^{-})$ and $(\bbox^{-})$. Since these
two rules keep positive conditionals in their conclusions, it
follows that any positive conditional $A \cond B$ belonging to
\texttt{SAT($\Gamma_0$)}, where $\Gamma_0$ is the initial set of
formulas, also belongs to each world introduced in $X$.
Furthermore, it can be easily shown that \emph{only} the
conditionals in \texttt{SAT($\Gamma_0$)} belong to possible worlds
in $X$. Indeed, all worlds in $X$ are generated by the application
of a dynamic rule, followed by the application of static rules for
saturation. It can be shown that this combination of rules does
never introduce a new conditional. This gives the following Fact:

\begin{fact}\label{Fatto aggiunto sui condizionali}
  Given a world $\Delta \appartiene X$ and any positive conditional
  $A \cond B$, we have that $A \cond B \appartiene
  \Delta$ iff $A \cond B \appartiene$
  \emph{\texttt{SAT($\Gamma_0$)}}.
\end{fact}

\noindent We conclude by proving the following Fact:
\begin{fact}\label{Fatto 2}
 For all formulas $F$ and for all sets $\Gamma \appartiene X$ we have
 that:\\
 (i) if $F \appartiene \Gamma$ then $\emme, \Gamma \modello F$;
 (ii) if $\nott F \appartiene \Gamma$ then $\emme, \Gamma \not\modello
   F$.
 \end{fact}

\begin{pf*}{Proof of Fact \ref{Fatto 2}.}
By induction on the structure of $F$. If $F$ is an atom $P$, then
$P \appartiene \Gamma$ implies $\emme, \Gamma \modello P$ by
definition of $V$. Moreover, $\nott P \appartiene \Gamma$ implies
that $P \not\appartiene \Gamma$ as $\Gamma$ is consistent; thus,
$\emme, \Gamma \not\modello P$ (by definition of $V$). For the
inductive step we only consider the case of $(\nott) \cond$ and
$(\nott) \bbox$:

\begin{itemize}
  \item $\bbox \nott A \appartiene \Gamma$. Then, for all $\Gamma_i <_X \Gamma$ we have
   $\nott A \in \Gamma_i$ by definition of $(\bbox^{-})$, since $\Gamma_i$ has been generated by a sequence of
applications of $(\bbox^{-})$. By
  inductive hypothesis $\emme, \Gamma_i \not\modello A$ for all $\Gamma_i <_X \Gamma$, whence $\emme, \Gamma \modello \bbox
  \nott A$.

  \item $\neg \bbox \nott A \appartiene \Gamma$. By construction there is a
  $\Gamma'$ s.t. $\Gamma' <_X \Gamma$ and $A \appartiene \Gamma'$. By inductive
  hypothesis $\emme, \Gamma'
  \models A$. Thus, $\emme, \Gamma \not\models \bbox \nott  A$.

  \item $A \cond B \appartiene \Gamma$.
  Let $\Delta \appartiene Min_{<_X}(A)$; \begin{rosso}one can observe that $(1)\nott A \appartiene
  \Delta$ or $(2)\nott \bbox \nott A \appartiene \Delta$ or $(3)B \appartiene \Delta$, since $A \cond B \appartiene \Delta$
  by Fact \ref{Fatto aggiunto sui condizionali}, and since $\Delta$ is
  saturated\end{rosso}. $(1)$ cannot be the case, since otherwise by inductive hypothesis $\emme, \Delta \not\modello A$,
  which contradicts the definition of $Min_{<_X}(A)$. If $(2)$, by
  construction of $\emme$ there exists a set $\Delta' <_X \Delta$ such
  that $A \appartiene \Delta'$. By inductive hypothesis $\emme, \Delta' \modello A$, which contradicts
  $\Delta \appartiene Min_{<_X}(A)$. Thus it must be that
  $(3) B \appartiene \Delta$, and by inductive hypothesis
  $\emme, \Delta \modello B$. Hence, we can conclude $\emme, \Gamma \modello A \cond
  B$.

  \item $\nott (A \cond B) \appartiene \Gamma$: by construction of $X$,
  there exists $\Gamma' \appartiene X$ such that $A, \bbox \nott A, \nott B \appartiene \Gamma'$.
  By inductive hypothesis  we have that $\emme, \Gamma'
  \modello A$ and $\emme, \Gamma' \modello \bbox \nott A$. It
  follows that $\Gamma' \appartiene Min_{<_X}(A)$. Furthermore,
  always by induction, $\emme, \Gamma' \not\modello B$. Hence,
  $\emme, \Gamma \not\modello A \cond B$.
\end{itemize}
\provafatto{\ref{Fatto 2}}
\end{pf*}

\noindent  By the above Facts the proof of the completeness of
$\calcoloP$ is over, since $\emme$ is a model for the initial set
$\Gamma_0$.

\end{provaposu}

\noindent By Theorem \ref{correttezza} above and by the
construction of the model done in the proof of Theorem
\ref{completezza} just above, we can show the following Corollary.

\begin{corollary}[Finite model property]\label{corollario-finitezzaP} \Pe \ has the finite model
property.
\end{corollary}
\begin{provaposu}
By Theorem \ref{correttezza}, if $\Gamma$ is satisfiable, then
there is no closed tableau for $\Gamma$. By the construction in
the proof of Theorem \ref{completezza},  if there is no closed
tableau for $\Gamma$, then $\Gamma$ is satisfiable in a finite
model.
\end{provaposu}

\hide{
\noindent By reasoning in the same way (from Fact
\ref{Fatto 1} in the proof above) we can show that:
\begin{corollary}\label{corollario-aciclicitàP} If $\Gamma$ is
satisfiable in \Pe, then it is satisfiable in an acyclic model.
\end{corollary}

\noindent From the two Corollaries above it immediately follows
that:
\begin{corollary}\label{infinite-chainsP}
If $\Gamma$ is satisfiable in \Pe , then it is satisfiable in a
model in which $<$ is transitive and does not have infinite
descending chains.
\end{corollary}
}

\noindent A relevant property of the calculus that will be useful
to estimate the complexity of logic \Pe \ is the so-called
\emph{disjunction property} of conditional formulas:

\begin{rosso}
\begin{proposition}[Disjunction property]\label{Proprietà disgiuntiva}
  If there is a closed tableau for $\Gamma, \nott (A \cond B), \nott (C \cond
  D)$, then there is a closed tableau either  for $\Gamma, \nott (A \cond B)$ or
  for $\Gamma, \nott (C \cond D)$.
\end{proposition}

\hide{
\begin{provaposu}
Consider two applications of $(\cond^{-})$ first on $\nott (A
\cond B)$ and then on $\nott (C \cond D)$, as follows:
\[
  \begin{prooftree}
    \[
       \Gamma, \nott (A \cond B), \nott (C \cond D)
      \justifies \Gamma^{\cond\pm}, A, \bbox \nott A, \nott B, \nott (C
      \cond D) \using (\cond^{-})
    \]
    \justifies \Gamma^{\cond\pm}, C, \bbox \nott C, \nott D \using (\cond^{-})
  \end{prooftree}
\]

\noindent In this case, the set $\Gamma, \nott (C \cond D)$ is
also derivable, as follows:

\[
  \begin{prooftree}
    \Gamma, \nott (C \cond D)
    \justifies \Gamma^{\cond\pm}, C, \bbox \nott C, \nott D \using (\cond^{-})
  \end{prooftree}
\]

\noindent Obviously, if $(\cond^{-})$ is applied first to $\nott
(C \cond D)$ then on $\nott (A \cond B)$, then the set $\Gamma,
\nott (A \cond B)$ is also derivable (the proof is symmetric).

\end{provaposu}
}

\begin{provaposu}
  Consider a closed tableau for $\Gamma, \nott (A \cond B), \nott (C \cond
  D)$. If the tableau does not contain any application of
  $(\cond^{-})$, then the property immediately follows. The same
  holds if either $\nott (A \cond B)$ or $\nott (C \cond D)$ are
  not used in the tableau. Consider the case in which there is an
  application of $(\cond^{-})$ first to $\nott (C \cond D)$, and
  then to $\nott (A \cond B)$. We show that in this case there is also
  a closed tableau for $\Gamma,
  \nott (A \cond B)$. We can build a tableau of the
  form:

  \[
    \begin{prooftree}
      \[
      \shortstack{$\Gamma, \nott (A \cond B), \nott (C \cond D)$
      \\ $\Pi_1$ \\ $\Gamma', \nott (A \cond B), \nott (C \cond D)$}
      \justifies \shortstack{$\Gamma'^{\cond^\pm}, \nott (A \cond B), C, \bbox \nott C, \nott D$ \\
      $\Pi_2$ \\ $\Gamma'', \nott (A \cond B)$} \using (\cond^{-})
      \]
      \justifies
       \Gamma''^{\cond^\pm}, A, \bbox \nott A, \nott B
      \using (\cond^{-})
    \end{prooftree}
  \]

\noindent  Since $C$ and $D$ are propositional formulas, $C, \bbox
\nott C,
  \nott D$ and, eventually, their subformulas introduced by the
  application of some boolean rules in $\Pi_2$, will be
  removed by the application of $(\cond^{-})$ on $\nott (A \cond
  B)$. Therefore, one can obtain a closed tableau of $\Gamma, \nott (A
  \cond B)$ as follows:
  \[
    \begin{prooftree}
      \shortstack{$\Gamma, \nott (A \cond B)$
      \\ $\Pi_1'$ \\ $\Gamma', \nott (A \cond B)$ \\ $\Pi_2'$ \\
      $\Gamma^{*}, \nott (A \cond B)$}
      \justifies
       \Gamma''^{\cond^\pm}, A, \bbox \nott A, \nott B
      \using (\cond^{-})
    \end{prooftree}
  \]

\noindent  $\Pi_1'$ is obtained by removing $\nott (C \cond D)$
from all
  the nodes of $\Pi_1$;$\Pi_2'$ is obtained by removing from $\Pi_2$
  the application of rules on $C, \nott D$ and their subformulas.
  The symmetric case, corresponding to the case in which $(\cond^{-})$
  is applied first on $\nott (A \cond B)$ and then on $\nott (C \cond
  D)$, can be proved in the same manner, thus we can conclude that $\Gamma, \nott (C
  \cond D)$ has a closed tableau.
\end{provaposu}

\end{rosso}

\noindent The reason why this property holds is that the
$(\cond^{-})$ rule discards all the other formulas that could have
been introduced by its previous application.

\subsection{Decidability and Complexity of \Pe}\label{sezione complessità P}

\subsubsection{Terminating procedure for \Pe}
In general, non-termination in tableau calculi can be caused by
two different reasons: 1. some rules copy their principal formula
in the conclusion, and can thus be reapplied over the same formula
without any control; 2. dynamic rules may generate infinitely-many
worlds, creating infinite branches.

Concerning the second source of non-termination (point 2.), notice
that infinitely-many worlds cannot be generated on a branch by
$(\cond^{-})$ rule, since this rule can be applied only once to a
given negated conditional on a branch. Another possible source of
infinite branches could be determined by the interplay between
rules $(\cond^{+})$ and $(\bbox^{-})$. We show that this cannot
occur, once we introduce the following standard restriction on the
order of application of the rules.

\begin{definition}[Restriction on the calculus]\label{restrizione sull'ordine di applicazione delle regole}
Building a tableau for a set of formulas $\Gamma$, the application
of the $(\bbox^{-})$ rule must be postponed to the application of
the propositional rules and to the verification that $\Gamma$ is
an instance of $(\bf AX)$.
\end{definition}

\noindent It is easy to observe that, without the restriction
above, point 2. could occur; for instance, consider the following
trivial example, showing a branch of a tableau starting with $P
\cond Q$, with $P, Q \in \mathit{ATM}$:

\[
  \begin{prooftree}
  \[
  \[
    \[
      \[
        \[
          \[
            P \cond Q
            \justifies \nott \bbox \nott P, P \cond Q  \quad\quad \dots \using
            (\cond^{+})
          \]
          \justifies P, \bbox \nott P, P \cond Q \using
          (\bbox^{-})
        \]
        \justifies  P, \bbox \nott P, \nott \bbox \nott P, P \cond
        Q \quad\quad \dots \using (\cond^{+})
      \]
      \justifies (*) \nott P, P, \bbox \nott P, P \cond Q \using
      (\bbox^{-})
    \]
    \justifies \nott P, P, \bbox \nott P, \nott \bbox \nott P, P
    \cond Q \quad\quad \dots \using (\cond^{+})
  \]
    \justifies \nott P, P, \bbox \nott P, P \cond Q \using
    (\bbox^{-})
  \]
    \justifies \shortstack{$\nott P, P, \bbox \nott P, \nott \bbox \nott P, P
    \cond Q \quad\quad \dots$ \\\\ $\dots$} \using (\cond^{+})
  \end{prooftree}
\]

\noindent In the above example, the $(\bbox^{-})$ rule is applied
systematically before the other rules, thus generating an infinite
branch. However, if the restriction in Definition \ref{restrizione
sull'ordine di applicazione delle regole} is adopted, as it is
easy to observe, the procedure terminates at the step marked as
$(*)$. Indeed, the test that $\nott P, P, \bbox \nott P, P \cond
Q$ is an instance of the axiom $(\bf AX)$ succeeds before applying
$(\bbox^{-})$ again, and the branch is considered to be closed.

\noindent As already mentioned, with the above restriction at
hand, we can show (Lemma \ref{lemma diminuzione misura} and
Theorem \ref{terminazione del caclolo preferential}) that the
interplay between $(\cond^{+})$ and $(\bbox^{-})$ does not
generate branches containing infinitely-many worlds. Intuitively,
the application of $(\bbox^{-})$ to a formula $\nott \bbox \nott
A$ (introduced by $(\cond^{+})$) adds the formula $\bbox \nott A$
to the conclusion, so that $(\cond^{+})$ can no longer
consistently introduce $\nott \bbox \nott A$. This is due to the
properties of $\bbox$ in G, and would not hold if $\bbox$ had
weaker properties (e.g. K4 properties).

Concerning point 1. the above calculus $\calcoloP$ does not ensure
a terminating proof search due to $(\cond^{+})$, which can be
applied without any control. We ensure the termination by putting
some constraints on $\calcoloP$. The intuition is as follows: one
does not need to apply $(\cond^{+})$ on the same conditional
formula $A \cond B$ \emph{more than once in the same world},
therefore we keep track of positive conditionals already used by
moving them in an additional set $\Sigma$ in the conclusions of
$(\cond^{+})$, and restrict the application of this rule to unused
conditionals only. The dynamic rules re-introduce formulas from
$\Sigma$ in order to allow further applications of $(\cond^{+})$
in the other worlds. This machinery is standard.

Theorem \ref{terminazione del caclolo preferential} below shows
that no additional machinery is needed to ensure termination.
Notice that this would not work in other systems (for instance, in
K4 one needs a more sophisticated loop-checking as described in
\cite{zimmermann}).

The terminating calculus $\calcoloPterminante$ is presented in
Figure \ref{Figura calcolo preferential TERMINANTE}. Observe that
the tableau nodes are now pairs $\Gamma;\Sigma$. The calculus
$\calcoloPterminante$ is sound and complete with respect to the
semantics:

\begin{theorem}[Soundness and completeness of
$\calcoloPterminante$]\label{correttezza e completeza calcolo
terminante}
  Given a set of formulas $\Gamma$, it is unsatisfiable iff it has a closed tableau in $\calcoloPterminante$.
\end{theorem}

\begin{provaposu}
The soundness is immediate and left to the reader. The
completeness easily follows from the fact that two applications of
$(\cond^{+})$ to the same conditional in the same world are
useless. Indeed, given a proof in $\calcoloP$, if $(\cond^{+})$ is
applied \emph{twice} to $\Gamma, A \cond B$ in the same world,
then we can assume, without loss of generality, that the two
applications are consecutive. Therefore, the second application of
$(\cond^{+})$ is useless, since each of the conclusions has
already been obtained after the first application, and can be
removed.

\end{provaposu}

  \begin{figure}

\linea

\mbox{ \(
\begin{scriptsize}
\begin{array}{ll}\\
\quad\quad\quad\quad {\bf (\cond^{+})} \ \irule{\Gamma, A \cond B;
\Sigma}
  {\Gamma, \nott A; \Sigma, A \cond B
    \quad\quad\quad\quad \Gamma, \nott \bbox \nott A; \Sigma, A \cond B
    \quad\quad\quad\quad \Gamma, B; \Sigma, A \cond B
  } {}
  \\
  \\
  \\
\end{array}
\end{scriptsize}
\) }

\mbox{ \(
\begin{scriptsize}
\begin{array}{l@{\quad\quad\quad\quad\quad\quad\quad\quad\quad\quad}l}
{\bf (\cond^{-})} \ \irule{\Gamma, \nott(A \cond B); \Sigma}%
{\Sigma, A, \bbox \nott A, \nott B, \Gamma^{\cond\pm};\vuoto}%
{} & {\bf (\bbox^{-})} \
\irule{\Gamma, \nott \bbox \nott A; \Sigma}%
{\Sigma, \Gamma^{\bbox}, \Gamma^{\bbox^{\freccia}}, \Gamma^{\cond\pm}, A, \bbox \nott A; \vuoto}%
{}
\\
\\
\end{array}
\end{scriptsize} \)  }

\linea

\caption{The calculus $\calcoloPterminante$. Propositional rules
are as in Figure \ref{Figura calcolo preferential}  adding
$\Sigma$.} \label{Figura calcolo preferential TERMINANTE}
\end{figure}

\begin{rosso}
Let us introduce a property of the tableau which will be crucial
in many of the following proofs. Let us first define the notion of
regular node.

\begin{definition}\label{nodo regolare}
A node $\Gamma;\Sigma$ is called regular if the following
condition holds:
\begin{center}
  if $\nott \bbox \nott A \appartiene \Gamma$, then there is
  $A \cond B \appartiene \Gamma \unione \Sigma$
\end{center}
\end{definition}

\noindent It is easy to see that all nodes in a tableau starting
from a pair $\Gamma_0;\vuoto$ are regular, when $\Gamma_0$ is a
set of formulas of $\elle$. This is stated by the following
Proposition, whose proof can be found in the Appendix:

\begin{proposition}\label{proposizione nodi regolari}
Given a pair $\Gamma_0;\vuoto$, where $\Gamma_0$ is a set of
formulas of $\elle$, all the tableaux obtained by applying
$\calcoloPterminante$'s rules only contain regular nodes.
\end{proposition}

\noindent From now on, we can assume without loss of generality
that only regular nodes may occur in a tableau.
\end{rosso}

\noindent In order to prove that $\calcoloPterminante$ ensures a
terminating proof search, we define a complexity measure on a set
of formulas $\Gamma$ and the corresponding set of positive
conditionals already used $\Sigma$, denoted by $m(\Gamma;
\Sigma)$, which consists of four measures $c_1, c_2, c_3$ and
$c_4$ in a lexicographic order. We denote by $cp(F)$ the
complexity of a formula $F$, defined as follows:

\begin{ultimissimi}
\begin{definition}[Complexity of a formula]\label{complessità formule}
  $\quad$

  \begin{itemize}
    \item $cp(P)=1$, where $P \in \mathit{ATM}$
    \item $cp(\nott F)=1 + cp(F)$
    \item $cp(F \bigotimes G)=1 + cp(F) + cp(G)$, where
    $\bigotimes$ is any binary boolean operator
    \item $cp(\bbox \nott A)=1+cp(\nott A)$
    \item $cp(A \ent B)=3+cp(A)+cp(B)$.
  \end{itemize}
\end{definition}
\end{ultimissimi}

\begin{definition}\label{definizione misura di complessità}
  We define $m(\Gamma; \Sigma)=\sx c_1, c_2, c_3, c_4 \dx$ where:
  \begin{itemize}
  \item $c_1=\tc\{A \cond B \appartiene_{-}
  \Gamma\}\tc$
  \item $c_2=\tc \{ A \cond B \appartiene_{+} \Gamma \unione \Sigma \tc \bbox \nott A \not\appartiene \Gamma
    \}\tc$
  \item $c_3= \tc \{A \cond B \appartiene_{+} \Gamma \}
    \tc$
  \item $c_4=\sum_{F \appartiene \Gamma} cp(F)$
  \end{itemize}
  \begin{blu}
  We consider the \emph{lexicographic order} given by $m(\Gamma;
  \Sigma)$, that is to say: given
  $m(\Gamma;\Sigma)=\sx c_1,c_2,c_3,c_4\dx$ and $m(\Gamma';\Sigma')=\sx c_1',c_2',c_3',c_4'
  \dx$, we say that $m(\Gamma;\Sigma) < m(\Gamma';\Sigma')$ iff
  there exists $i$, $i=1,2,3,4$, such that the following conditions hold:
  \begin{itemize}
    \item $c_i < c_i'$
    \item for all $j$, $0<j<i$, we have that $c_j=c_j'$
  \end{itemize}
  \end{blu}

\end{definition}

\noindent Intuitively, $c_1$ is the number of negated conditionals
to which the $(\cond^{-})$ rule can still be applied. An
application of $(\cond^{-})$ reduces $c_1$. $c_2$ represents the
number of positive conditionals \emph{which can still create a new
world}. The application of $(\bbox^{-})$ reduces $c_2$: indeed, if
$(\cond^{+})$ is applied to $A \cond B$, this application
introduces a branch containing $\nott \bbox \nott A$; when a new
world is generated by an application of $(\bbox^{-})$ on $\nott
\bbox \nott A$, it contains $A$ and $\bbox \nott A$. If
$(\cond^{+})$ is applied to $A \cond B$ once again, then the
conclusion where $\nott \bbox \nott A$ is introduced leads to a
closed branch, by the presence of $\bbox \nott A$ in that branch.
$c_3$ is the number of positive conditionals not yet considered in
that branch. $c_4$ is the sum of the complexities of the formulas
in $\Gamma$; an application of a boolean rule reduces $c_4$.

To prove that $\calcoloPterminante$ ensures a terminating proof
search, we show that the tableau cannot contain an open branch of
infinite length. To this purpose we need the following Lemma:

\begin{lemma}\label{lemma diminuzione misura}
  Let $\Gamma'; \Sigma'$ be obtained by an application of a rule of $\calcoloPterminante$ to a
  premise $\Gamma; \Sigma$. Then, \begin{rosso}we have that either $m(\Gamma'; \Sigma')<m(\Gamma;
  \Sigma)$ or $\calcoloPterminante$ leads to the construction of a closed tableau for $\Gamma'; \Sigma'$.\end{rosso}
\end{lemma}

\begin{provaposu}
 We consider each rule of the calculus
$\calcoloPterminante$:
\begin{itemize}
  \item $(\cond^{-})$: one can easily observe that the conditional
  formula $\nott (A \cond B)$ to which this rule is applied
  does not belong to the only conclusion. Hence the measure $c_1$ in $m(\Gamma'; \Sigma')$, say $c_{1'}$,
  is smaller than $c_1$ in $m(\Gamma, \Delta)$, say $c_1$;

  \item $(\bbox^{-})$: no negated conditional is added nor
  deleted in the conclusions, thus $c_1=c_1'$. Suppose we are considering an
  application of $(\bbox^{-})$
  on a formula $\nott \bbox \nott A$. We can observe the following facts:
    \begin{itemize}
      \item the formula $\nott \bbox \nott A$ has been introduced
      by an application of $(\cond^{+})$, being this one the only rule introducing a boxed formula in the
      conclusion; more precisely, it derives from an application of $(\cond^{+})$ on a
      conditional formula $A \cond B$;
      \item $A \cond B$ belongs to both $\Gamma; \Sigma$ and $\Gamma'; \Sigma'$, since no rule of $\calcoloPterminante$
      removes positive conditionals (at most, the $(\cond^{+})$ rule \emph{moves} conditionals from $\Gamma$ to
      $\Sigma$);
      \item $A \cond B$ does not ``contribute'' to $c_{2'}$, since
      the application of $(\bbox^{-})$ introduces
      $\bbox \nott A$ in the conclusion $\Gamma'$ (remember that $c_{2'}=\tc
      \{A \cond B \appartiene_{+} \Gamma' \unione \Sigma' \tc \bbox \nott A \not\appartiene \Gamma' \}\tc$).
    \end{itemize}

  \begin{rosso}
  We distinguish two cases:
    \begin{enumerate}
      \item $\bbox \nott A$ does \emph{not} belong to the premise
      of $(\bbox^{-})$: in this case, by the above facts, we can
      easily conclude that $c_{2'}<c_2$, since $\bbox \nott A$
      belongs only to the conclusion;

      \item $\bbox \nott A$ belongs to the premise of
      $(\bbox^{-})$: we are considering a derivation of the
      following type:

      \[
        \begin{prooftree}
          \Gamma, \bbox \nott A, \nott \bbox \nott A
          \justifies \Gamma^{\cond\pm}, \Gamma^{\bbox},
          \Gamma^{\bbox^{\freccia}}, \nott A, A, \bbox \nott A
          \using (\bbox^{-})
        \end{prooftree}
      \]

      In this case, $c_{2'}=c_2$; however, we can conclude that
      the tableau built for $\Gamma^{\cond\pm}, \Gamma^{\bbox},
          \Gamma^{\bbox^{\freccia}}, \nott A, A, \bbox \nott A$ is closed,
          since:
          \begin{itemize}
            \item $A$ is a propositional formula
            \item the restriction in Definition \ref{restrizione sull'ordine di applicazione delle
          regole} leads to a proof in which the propositional
          rules and $(\bf AX)$ are applied to $A$ and $\nott A$ \emph{before}
          $(\bbox^{-})$ is further applied. The resulting
          tableau is closed;
          \end{itemize}
    \end{enumerate}
  \end{rosso}

  \item $(\cond^{+})$: we have that $c_1=c_1'$, since we have the same negated conditionals
  in the premise as in all the conclusions.
  The same for $c_2$, since the
  formula $A \cond B$ to which the rule is applied is also maintained
  in the conclusions (it moves from unused to already used conditionals).
  We conclude that $m(\Gamma'; \Sigma')<m(\Gamma; \Sigma)$, since
  $c_{3'} < c_3$. Indeed, the $(\cond^{+})$ rule moves $A \cond B$
  from $\Gamma$ to the set $\Sigma$ of already considered conditionals;

  \item rules for the boolean connectives: it is easy to observe
  that $c_1, c_2$ and $c_3$ are the same in the premise and in any
  conclusion, since conditional formulas are side formulas
  in the application of these rules. We conclude that
  $m(\Gamma'; \Sigma')<m(\Gamma; \Sigma)$ since $c_{4'}<c_4$.
  Indeed, the complexity of the formula to which the rule is applied is
  greater than (the sum of) the complexity of its subformula(s)
  introduced in the conclusion(s).

\end{itemize}

\end{provaposu}

\noindent Now we have all the elements to prove that
$\calcoloPterminante$ ensures termination in a proof search:

 \begin{theorem}[Termination of $\calcoloPterminante$]\label{terminazione del caclolo preferential}
   $\calcoloPterminante$ ensures a terminating proof search.
 \end{theorem}

 \begin{provaposu}
By Lemma \ref{lemma diminuzione misura}
 we know that, starting from $\Gamma_0; \vuoto$, the value of $m(\Gamma;\Sigma)$ decreases each time
 a tableau rule is applied or leads to a closed tableau.
 Therefore, a finite number of applications of the rules leads
 either to build a closed tableau or to nodes $\Gamma;\Sigma$ such
 that $m(\Gamma;\Sigma)$ is \emph{minimal}. In particular, we observe that, when
 the branch does not close, $m(\Gamma;\Sigma)=\sx 0,0,0, c_{4_{\mathit{min}}}\dx$, and the following facts hold:
 \begin{itemize}
   \item no negated conditional belongs to $\Gamma$, since $c_1=0$;
   \item for each $A \cond B \appartiene \Gamma \unione \Sigma$, we have that $\bbox
   \nott A \appartiene \Gamma$, since $c_2=0$;
   \item all positive conditionals $A \cond B$ have been moved in
   $\Sigma$ since $c_3=0$;
   \item $\Gamma$ is saturated with respect to the propositional rules, since $c_4$ assumes its minimal value $c_{4_{\mathit{min}}}$.
 \end{itemize}

\noindent   By the above facts it is easy to see that, in this
case, either $\Gamma; \Sigma$ is closed
  or no rule, with the exception of $(\bbox^{-})$, is applicable to $\Gamma; \Sigma$.
  Indeed, $(\cond^{-})$ rule is not applicable, since no negated conditional belongs to
  $\Gamma$. If $(\bbox^{-})$ is applicable, then there is
  $\nott \bbox \nott A \appartiene \Gamma$, to which the rule is
  applied. However, since $c_2=0$, we have also that $\bbox \nott
  A \appartiene \Gamma$. Therefore, the conclusion of an application of $(\bbox^{-})$ contains both $A$ and $\nott A$
  and, by the restriction in Definition
  \ref{restrizione sull'ordine di applicazione delle regole} and since $A$ is propositional, the
  procedure terminates building a closed tableau.
  $(\cond^{+})$ is not applicable, since no positive conditionals
  belong to $\Gamma$ (all positive conditionals $A \cond B$ have been moved in
   $\Sigma$). Last, no rule for the boolean connectives is
   applicable. For a contradiction, suppose one boolean rule is still
   applicable: by Lemma \ref{lemma diminuzione misura}, the sum of
   the complexity of the formulas in the conclusion(s) decreases,
   i.e. $c_4$ in the conclusion(s) is smaller than in the premise $\Gamma; \Sigma$,
   against the minimality of this measure in $\Gamma; \Sigma$.
\end{provaposu}

\subsubsection{Optimal Proof Search Procedure for
\Pe}\label{optimal proof search}
We conclude this section with a
complexity analysis of $\calcoloPterminante$, in order to prove
that validity in \Pe \ is {\bf coNP}-complete. Intuitively, we can
obtain a {\bf coNP} decision procedure, by taking advantage of the
following facts:
\begin{enumerate}
\item Negated conditionals do not interact with the current world, nor they interact among themselves
(by the disjunction property). Thus they can be handled separately
and eliminated always as a first step.
\item We can replace the $(\bbox^{-})$  which is responsible of backtracking in the tableau construction
by a stronger rule that does not need backtracking.
\end{enumerate}

\noindent Regarding (1), by the disjunction property we can
reformulate the $(\cond^{-})$ rule as follows:

\[
  \begin{prooftree}
    \Gamma, \nott (A \cond B); \Sigma
    \justifies \Sigma, A, \bbox \nott A, \nott B, \Gamma^{\cond^{+}}; \vuoto \using (\cond^{-})
  \end{prooftree}
\]

\noindent This rule reduces the length of a branch at the price of
making the proof search more non-deterministic.

Regarding (2), we can adopt the following strengthened version of
$(\bbox^{-})$. We use $\Gamma_{-i}^{\bbox^{-}}$ to denote $\{\nott
\bbox \nott A_j \orr A_j \tc \nott \bbox \nott A_j \appartiene
\Gamma \andd j \diverso i\}$:

\[
  \begin{prooftree}
    \Gamma, \nott \bbox \nott A_1, \nott \bbox \nott A_2, ...,
    \nott \bbox \nott A_n; \Sigma
    \justifies \Sigma, \Gamma^{\cond\pm}, \Gamma^{\bbox},
    \Gamma^{\bbox^{\freccia}}, A_1, \bbox \nott
    A_1,\Gamma_{-1}^{\bbox^{-}}; \vuoto \tc \dots \tc \Sigma, \Gamma^{\cond\pm}, \Gamma^{\bbox},
    \Gamma^{\bbox^{\freccia}}, A_n, \bbox \nott
    A_n,\Gamma_{-n}^{\bbox^{-}}; \vuoto
    \using (\bbox^{-}_s)
  \end{prooftree}
\]

\noindent The advantage of  this rule  over the original
$(\bbox^{-})$ rule is that no backtracking on the choice of the
formula $\nott \bbox \nott A_i$ is needed. The reason is that  all
alternatives are kept in the conclusion. As we will see below, by
using this rule we can provide a tableau construction algorithm
with no backtracking.

We call $L\calcoloPterminante$ the calculus obtained by replacing
in $\calcoloPterminante$ the initial rules $(\cond^{-})$ and
$(\bbox^{-})$ with the ones reformulated above. We can prove that
$L\calcoloPterminante$ is sound and complete w.r.t. the
preferential models. To prove soundness, we consider the
multi-linear models introduced in section \ref{sezione
multilineare}.

\begin{theorem}
The rule $(\bbox^{-}_s)$ is sound.
\end{theorem}

\begin{provaposu}
Let $\Gamma =  \Gamma', \nott \bbox \nott A_1, \nott \bbox \nott
A_2, \dots, \nott \bbox \nott A_n$. We omit $\Sigma$ for
readability reasons. We prove that if $\Gamma$ is satisfiable then
also one conclusion of the rule
$$\Gamma^{\cond\pm}, \Gamma^{\bbox}, \Gamma^{\bbox^{\freccia}},
A_i, \bbox \nott A_i,\Gamma_{-i}^{\bbox^{-}}$$ is satisfiable. By
Theorem \ref{teorema nicola}, we can assume that $\Gamma$ is
satisfiable in a multi-linear model $\emme = \sx \WW, <, V \dx$,
let $\emme, x \models \Gamma$. Then there are $z_1 < x, \ldots ,
z_n < x$, such that $z_i \in Min_{<}(A_i)$; thus $\emme,z_i
\models A_i \land \bbox\neg A_i$; we easily have also that $\emme,
z_i \models \Gamma^{\cond\pm}, \Gamma^{\bbox},
\Gamma^{\bbox^{\freccia}}$. Being $\emme$ a multi-linear model,
\begin{blu} the $z_i$, $i=1, 2, \dots, n$, whenever distinct, are totally ordered: we
have that $z_i < x$,
\end{blu} so that they must belong to the same component. Let
$z_k$ be the maximum of $z_i$, for a certain $1\leq k \leq n$. We
have that for each $z_i$ ($i\not=k$) either (i) $z_i = z_k$, so
that $\emme, z_k \models A_i$, or (ii) $z_i < z_k$, so that
$\emme, z_k \models \neg \bbox \neg A_i$. We have shown that for
each $i\not=k$, $\emme,z_k \models A_i \lor \neg \bbox \neg A_i$.
We can conclude that $\emme, z \models \Gamma_{-k}^{\bbox^{-}}$.
Thus
$$\emme,
z_k \models \Gamma^{\cond\pm}, \Gamma^{\bbox},
\Gamma^{\bbox^{\freccia}}, A_k, \bbox \nott
A_k,\Gamma_{-k}^{\bbox^{-}}$$ which is one of the conclusions of
the rule.

\end{provaposu}

\noindent We can prove that the calculus obtained by replacing the
$(\bbox^{-})$ rule with its stronger version $(\bbox^{-}_s)$ is
complete  w.r.t. the semantics:

\begin{theorem}
The calculus $L\calcoloPterminante$ is complete.
\end{theorem}

\begin{provaposu}
  We repeat the same construction as in the proof of Theorem
  \ref{completezza}, in order to build a preferential model, more
  precisely a multi-linear model, of a set of formulas $\Gamma_0$ for
  which there is no closed tableau. We
denote by \texttt{APPLY}$(\Gamma,\mathit{RuleName})$ the result of
applying the rule corresponding to $\mathit{RuleName}$ to
$\Gamma$. As a difference with the
  construction in Theorem \ref{completezza}, we replace point 4.
  by the points $4_{strong}$, $4a_{strong}$, $4b_{strong}$, and
  $4c_{strong}$, obtaining the following procedure ($X$ is the set of
worlds of the model):

\hide{ \vspace{0.5cm}
  \noindent $4_{strong}$. {\bf if} there is $\nott \bbox \nott A \appartiene
  \Gamma$ {\bf then}

  \indent $4a_{strong}$. let
  $\Gamma'=$\texttt{SAT(APPLY(}$\Gamma,\bbox^{-}_s$\texttt{))};

  \indent $4b_{strong}$. add the relation $\Gamma' < \Gamma$;

  \indent $4c_{strong}$. {\bf if} $\Gamma' \not\appartiene X$ {\bf
  then}
  $X=X \unione  \{\Gamma'\}$;
\vspace{0.5cm}

\noindent resulting in the following procedure ($X$ is the set of
worlds of the model): }

\begin{rosso}

\vspace{0.3cm}

\begin{small}

  \noindent 1. initialize $X=\{$\texttt{SAT}$(\Gamma_0)\}$; mark
  \texttt{SAT}$(\Gamma_0)$ as unresolved;

  \noindent {\bf while} $X$ contains unresolved nodes do

  \indent 2. choose an unresolved $\Gamma$ from $X$;

  \indent 3. {\bf for} each formula $\nott (A \cond B) \appartiene \Gamma$

  \indent \indent 3a. let $\Gamma_{\nott(A \cond B)}=$\texttt{SAT}(\texttt{APPLY}$(\Gamma,\nott(A \cond
  B)))$;

  \indent\indent 3b. {\bf if} $\Gamma_{\nott(A \cond B)} \not\appartiene X$ {\bf then} $X=X
  \unione \{\Gamma_{\nott(A \cond B)}\}$;

  \indent $4_{strong}$. {\bf if} there is $\nott \bbox \nott A \appartiene
  \Gamma$ {\bf then}

  \indent\indent $4a_{strong}$. let
  $\Gamma'=$\texttt{SAT(APPLY(}$\Gamma,\bbox^{-}_s$\texttt{))};

  \indent\indent $4b_{strong}$. add the relation $\Gamma' < \Gamma$;

  \indent\indent $4c_{strong}$. {\bf if} $\Gamma' \not\appartiene X$ {\bf
  then}
  $X=X \unione  \{\Gamma'\}$;

  \indent 5. mark $\Gamma$ as resolved;

 \noindent {\bf endWhile};

\end{small}

 \vspace{0.5cm}
\end{rosso}

\noindent Facts \ref{Fatto 1} and \ref{fatto aggiunto in
smoothness} can be proved as in Theorem \ref{completezza}. This
holds also for Fact \ref{Fatto 2} with one difference, for what
concerns the case in which $\nott \bbox \nott A \appartiene
\Gamma$. In this case, by construction there is a $\Gamma'$ such
that $\Gamma' <_X \Gamma$. We can prove by induction on the length
$n$ of the chain $<_X$ starting from $\Gamma$ that if $\nott \bbox
\nott A \appartiene \Gamma$, then $\emme, \Gamma \not\modello
\bbox \nott A$. If $n=1$, it must be the case that $A \appartiene
\Gamma'$; hence, by inductive hypothesis on the structure of the
formula, $\emme, \Gamma' \modello A$, thus $\emme, \Gamma
\not\modello \bbox \nott A$. If $n>1$, by the  $(\bbox^{-}_s)$
rule, either $A \appartiene \Gamma'$ and we conclude as in the
previous case, or $\nott \bbox \nott A \appartiene \Gamma'$ and
the Fact holds by inductive hypothesis on the length.

\end{provaposu}

\noindent We give a non-deterministic algorithm for testing
satisfiability in
  \Pe \ that: $(i)$ takes a set of formulas $\Gamma$ as input;
  $(ii)$ returns \texttt{SAT} iff $\Gamma$ is satisfiable.
By using the new version of $(\cond^{-})$ rule, we can consider a
negated conditional at a time. Indeed, for $\Gamma, \nott (A \cond
B), \nott (C \cond
  D)$ to be satisfiable, it is sufficient that both $\Gamma, \nott (A \cond B)$ and
 $\Gamma, \nott (C \cond D)$, separately considered, are satisfiable. For each negated conditional,
the algorithm \texttt{GENERAL-CHECK} applies the rule
$(\cond^{-})$ to it, and calls the algorithm \texttt{CHECK} on the
resulting set of formulas. \texttt{CHECK} is  a non-deterministic
algorithm that tests satisfiability in \Pe \ of a set of formulas
not containing negated conditionals.

Let \texttt{EXPAND}($\Gamma$) be a procedure that returns one
saturated expansion of $\Gamma$ w.r.t. all static rules. In case
of a branching rule, \texttt{EXPAND} nondeterministically selects
(guesses) one conclusion of the rule. The algorithm below allows
the satisfiability of a set of formulas (not containing negated
conditionals) to be decided. In brackets we give the complexity of
each operation, considering that $n=\tc \Gamma \tc$.

\vspace{0.2cm}

\linea

\begin{small}
\texttt{CHECK}($\Gamma$)

1. $\Gamma \longleftarrow$ \texttt{EXPAND}$(\Gamma)$; ($O(n)$)

2. {\bf if} $\Gamma$ contains an axiom {\bf then return}
\texttt{UNSAT}; ($O(n^2)$)

3. {\bf if} $\{\nott \bbox \nott A \tc \nott \bbox \nott A
\appartiene \Gamma\}=\emptyset$
{\bf then return} \texttt{SAT};

\hide{4. {\bf else} {\bf if} ($\{\nott \bbox \nott A \tc \nott
\bbox \nott A \appartiene \Gamma\} \not=\emptyset$) {\bf then}}

\hide{\indent\indent  4a. let $\{\neg \bbox \neg A_1, \ldots,\neg
\bbox \neg A_k\}$ be all the negated boxed formulas in $\Gamma$;}

4. {\bf else return} \texttt{CHECK(APPLY($\Gamma,\bbox^{-}_s$))};

\end{small}

\linea

\vspace{0.2cm}

\noindent Notice that the execution of
\texttt{APPLY($\Gamma,\bbox^{-}_s$)} chooses the branch generated
by the application of $(\bbox^{-}_s)$ to $\Gamma$.

To see that \texttt{CHECK} is a nondeterministic polynomial
procedure to decide the satisfiability of a set of formulas (not
containing negated conditionals), observe that: (1) the complexity
of each call to the procedure \texttt{EXPAND} is polynomial.
Indeed, as the number of different subformulas is at most $O(n)$,
\texttt{EXPAND} makes at most $O(n)$ applications of the static
rules. (2) The test that a set $\Gamma$ (of size $O(n)$) of
formulas contains an axiom has at most complexity $O(n^2)$. (3)
The number of recursive calls to the procedure \texttt{CHECK} is
at most $O(n)$, since in a branch the rule $(\bbox^{-}_s)$ can be
applied only once for each formula $\neg \bbox \neg A_i$, and the
number of different negated box formulas is at most $O(n)$.

Let us now define a procedure to decide whether an arbitrary set
of formulas $\Gamma$ (possibly containing negated conditionals) is
satisfiable:

\vspace{0.2cm}

\linea

\begin{small}
\texttt{GENERAL-CHECK}($\Gamma$)

1. $\Gamma \longleftarrow$ \texttt{EXPAND}$(\Gamma);$ ($O(n)$)

2. let $\neg (A_1 \cond B_1),\ldots, \neg (A_k \cond B_k)$ be all
negated conditionals in $\Gamma$;

\indent \indent 2.1.  {\bf for all} $i=1,\ldots,k$
 result[i] $\longleftarrow$ \texttt{CHECK}(\texttt{APPLY}($\Gamma, \neg(A_i\cond B_i$))) ;

3. {\bf if for all} $i=1,\ldots,k$ result[i]\texttt{==SAT}  {\bf
then return} \texttt{SAT};

\indent \indent {\bf else} {\bf return} \texttt{UNSAT};

\end{small}

\linea

\vspace{0.2cm}

\noindent By the subformula property, the number of negated
conditionals which can occur in $\Gamma$ is at most $O(n)$. Hence,
the procedure \texttt{GENERAL-CHECK} calls to the algorithm
\texttt{CHECK} at most $O(n)$ times.

\begin{theorem}[Complexity of {\bf P}]\label{complessità di P}
  The problem of deciding validity for preferential logic {\bf P} is
  {\bf coNP}-complete.
\end{theorem}

\begin{provaposu}
The procedure \texttt{GENERAL-CHECK} allows the
satisfiability of a set of formulas of logic \Pe \ to be decided
in nondeterministic polynomial time.
 The validity problem for \Pe \ is therefore in {\bf coNP}.
As {\bf coNP}-hardness is immediate (this logic includes classical
propositional logic), we conclude that the validity problem for
logic \Pe \ is {\bf coNP}-complete.

\end{provaposu}

\noindent This result matches the known complexity results for
logic \Pe \ \cite{whatdoes}. Due to the {\bf coNP} lower bound,
the above method provides a computationally optimal reasoning
procedure for logic \Pe.


\section{The Tableau Calculus for Loop Cumulative Logic \Cl}\label{sezione calcolo CL}

In this section we develop a tableau calculus $\calcoloCL$ for
{\bf CL}, and we show that it can be turned into a terminating
calculus. This provides a decision procedure for {\bf CL} and a
{\bf coNP}-membership upper bound for validity in {\bf CL}.

The calculus $\calcoloCL$ can be obtained from the calculus
$\calcoloP$ for preferential logics, by adding a suitable rule
$(L^{-})$ for dealing with the modality $L$ introduced in section
\ref{sezione CL}. As already mentioned in section \ref{sezione
CL}, the formulas that appear in the tableaux for \Cl \ belong to
the language $\lan_L$ obtained from $\lan$ as follows: $(i)$ if
$A$ is propositional, then $A \in \lan_L$; $LA \in \lan_L$; $\bbox
\neg LA \in \lan_L$; $(ii)$ if $A$, $B$ are propositional, then $A
\cond B \in \lan_L$; $(iii)$ if $F$ is a boolean combination of
formulas of $\elle_L$, then $F \appartiene \elle_L$. Observe that
the only allowed combination of $\bbox$ and $L$ is in formulas of
the form $\bbox \nott L A$ where $A$ is propositional.

We define:

\begin{center}
$\Gamma^{L^{\freccia}}=\{ A \tc L A \appartiene \Gamma \}$
\end{center}

\noindent Our tableau system $\calcoloCL$ is shown in Figure
\ref{Figura calcolo CL}. Observe that rules $(\cond^{+})$ and
$(\cond^{-})$ have been changed as they introduce the modality $L$
in front of the propositional formulas $A$ and $B$ in their
conclusions. This straightforwardly corresponds to the semantics
of conditionals in CL preferential models (see Definition
\ref{semantica_CL}). The new rule $(L^{-})$ is a dynamic rule.

\begin{figure}

\linea

\mbox{ \(
\begin{scriptsize}
\begin{array}{ll}\\
\quad\quad\quad\quad\quad\quad{\bf (\cond^{+})} \ \irule{\Gamma, A
\cond B}
  {\Gamma, \nott L A, A \cond B
    \quad\quad\quad \Gamma, \nott \bbox \nott L A, A \cond B
    \quad\quad\quad \Gamma, L B, A \cond B
  } {} &
\\
\\
\\
\end{array}
\end{scriptsize}
\) }

\mbox{ \(
\begin{scriptsize}
\begin{array}{l@{\quad\quad\quad\quad\quad\quad\quad\quad\quad\quad}l}
{\bf (\cond^{-})} \ \irule{\Gamma, \nott(A \cond B)}%
{L A, \bbox \nott L A, \nott L B, \Gamma^{\cond\pm}}%
{} & {\bf (\bbox^{-})} \
\irule{\Gamma, \nott \bbox \nott L A}%
{\Gamma^{\bbox}, \Gamma^{\bbox^{\freccia}}, \Gamma^{\cond\pm}, LA, \bbox \nott L A}%
{}
\\
\\
\\
\end{array}
\end{scriptsize}
\) }

\mbox{ \(
\begin{scriptsize}
\begin{array}{ll}
\quad\quad\quad\quad\quad\quad\quad\quad\quad\quad (L^{-}) \
\irule{\Gamma, \nott L A}%
{ \Gamma^{L^{\freccia}}, \nott A }
;\ \irule{\Gamma}%
{\Gamma^{L^{\freccia}}}%
{\mbox{if $\Gamma$ does not contain negated $L$-formulas}} {}
\\
\\
\end{array}
\end{scriptsize} \)
 }

\linea

 \caption{Tableau system $\calcoloCL$. If there are no negated $L$-formulas $\nott LA$
in the premise of $(L^{-})$, then the rule allows to step from
$\Gamma$ to $\Gamma^{L^{\freccia}}$. To save space, the boolean
rules are omitted.} \label{Figura calcolo CL}
\end{figure}

\begin{theorem}[Soundness of
$\calcoloCL$]\label{correttezzaCL}
  The system $\calcoloCL$ is sound with respect to CL-preferential
  models, i.e. if there is a closed tableau for a set of formulas $\Gamma$, then
  $\Gamma$ is unsatisfiable.
\end{theorem}

\begin{provaposu}
 We show that for all the rules in $\calcoloCL$, if
the premise is satisfiable by a CL-preferential model then also
one of the conclusions is. As far as the rules already present in
$\calcoloP$ are concerned, the proof is very similar, with the
only exception that we have to substitute $A$ in the proof by
$LA$.

We consider now the new rule $(L^-)$. Let $\emme, w \models
\Gamma, \nott L A$ where $\emme = \langle \WW, R, <, V \rangle$ is
a CL-preferential model. Then there is $w': wRw'$  and $\emme, w'
\models \neg A$. Furthermore, $\emme, w' \models
\Gamma^{L^{\freccia}}$. It follows that the conclusion of the rule
is satisfiable. If $\Gamma$ does not contain negated $L$-formulas,
since $R$ is serial, we still have that  $\exists w': wRw'$, and
$\emme, w' \models \Gamma^{L^{\freccia}}$, hence the conclusion is
still satisfiable.

\end{provaposu}

\noindent Soundness with respect to loop-cumulative models in
Definition \ref{loopcummodels} follows from the correspondence
established by Proposition \ref{correspondence CL}.

The proof of the completeness of the calculus can be done as for
the preferential case, provided we suitably modify the procedure
for constructing a model for a finite consistent set of formulas
$\Gamma$ of $\lan_L$. First of all, we modify the definition of
saturated sets as follows:

\begin{quote}
  \begin{itemize}
    \item if $A \cond B \appartiene \Gamma$ then $\nott L A \appartiene
    \Gamma$ or $\nott \bbox \nott L A \appartiene \Gamma$ or $L B \appartiene
    \Gamma$
  \end{itemize}
\end{quote}

\noindent For this notion of saturated set of formulas we can
still prove Lemma \ref{lemma saturazione} for language $\lan_L$.

\begin{theorem}[Completeness of $\calcoloCL$]\label{completezza calcolo CL}
  $\calcoloCL$ is complete with respect to CL-preferential models, i.e. if a set
  of formulas $\Gamma$ is unsatisfiable, then it has a closed tableau
  in $\calcoloCL$.
\end{theorem}

\begin{provaposu}
We define a procedure for constructing a model satisfying a
consistent set of formulas $\Gamma_0 \appartiene \lan_L$ by
modifying the procedure for the preferential logic {\bf P}.
\begin{rosso}We add to the procedure used in the proof of Theorem
\ref{completezza} the new steps 4' and 4'' between step 4 and step
5, obtaining the following procedure:

\hide{
\vspace{0.2cm}
\begin{small}
  4'. {\bf if} $\{\nott LA \tc \nott LA \appartiene \Gamma\} \diverso
  \vuoto$ {\bf then}

  \indent \indent\indent {\bf for} each $\nott L A \appartiene \Gamma$, let
  $\Gamma_{\nott L A}=$\texttt{SAT}(\texttt{APPLY}$(\Gamma,\nott L
  A))$;

    \indent\indent\indent \indent  4'\ a. add the relation $\Gamma \ R \ \Gamma_{\nott L A}$;

   \indent \indent\indent\indent   4'\ b. {\bf if} $\Gamma_{\nott L A} \not\appartiene X$ {\bf then} $X=X
  \unione \{\Gamma_{\nott L A}\}$;

  4''. {\bf else if} $\Gamma^{L^{\freccia}} \diverso \vuoto$ {\bf then} let
  $\Gamma'=$\texttt{SAT}(\texttt{APPLY}$(\Gamma, L^{-}))$;

    \indent \indent\indent  4''\ a. add the relation $\Gamma \ R \ \Gamma'$;

   \indent \indent\indent   4''\ b. {\bf if} $\Gamma' \not\appartiene X$ {\bf then} $X=X
  \unione \{\Gamma'\}$;

\end{small}
\vspace{0.2cm}
}

\vspace{0.2cm}

\begin{small}

  \noindent 1. initialize $X=\{$\texttt{SAT}$(\Gamma_0)\}$; mark
  \texttt{SAT}$(\Gamma_0)$ as unresolved;

  \noindent {\bf while} $X$ contains unresolved nodes do

  \indent 2. choose an unresolved $\Gamma$ from $X$;

  \indent 3. {\bf for} each formula $\nott (A \cond B) \appartiene \Gamma$

  \indent \indent 3a. let $\Gamma_{\nott(A \cond B)}=$\texttt{SAT}(\texttt{APPLY}$(\Gamma,\nott(A \cond
  B)))$;

  \indent\indent 3b. {\bf if} $\Gamma_{\nott(A \cond B)} \not\appartiene X$ {\bf then} $X=X
  \unione \{\Gamma_{\nott(A \cond B)}\}$;

  \indent 4. {\bf for} each formula $\nott \bbox \nott L A \appartiene \Gamma$, let
  $\Gamma_{\nott \bbox \nott L A}=$\texttt{SAT}(\texttt{APPLY}$(\Gamma,\nott
  \bbox \nott L A))$;

  \indent\indent 4a. add the relation $\Gamma_{\nott \bbox \nott L A} < \Gamma$;

  \indent\indent 4b. {\bf if} $\Gamma_{\nott \bbox \nott L A} \not\appartiene X$ {\bf then} $X=X
  \unione \{\Gamma_{\nott \bbox \nott L A}\}$.

  4'. {\bf if} $\{\nott LA \tc \nott LA \appartiene \Gamma\} \diverso
  \vuoto$ {\bf then}

  \indent \indent\indent {\bf for} each $\nott L A \appartiene \Gamma$, let
  $\Gamma_{\nott L A}=$\texttt{SAT}(\texttt{APPLY}$(\Gamma,\nott L
  A))$;

    \indent\indent\indent \indent  4'\ a. add the relation $\Gamma \ R \ \Gamma_{\nott L A}$;

   \indent \indent\indent\indent   4'\ b.\hide{{\bf if} $\Gamma_{\nott L A} \not\appartiene X$ {\bf then}} $X=X
  \unione \{\Gamma_{\nott L A}\}$;

  4''. {\bf else if} $\Gamma^{L^{\freccia}} \diverso \vuoto$ {\bf then} let
  $\Gamma'=$\texttt{SAT}(\texttt{APPLY}$(\Gamma, L^{-}))$;

    \indent \indent\indent  4''\ a. add the relation $\Gamma \ R \ \Gamma'$;

   \indent \indent\indent   4''\ b.\hide{{\bf if} $\Gamma' \not\appartiene X$ {\bf then}} $X=X
  \unione \{\Gamma'\}$;

  \indent 5. mark $\Gamma$ as resolved;

 \noindent {\bf endWhile};

\end{small}

 \vspace{0.2cm}

\noindent This procedure terminates. \begin{ultimissimi}Observe
that, although an application of $(L^{-})$ may introduce in $X$
several copies of the same world (set) of propositional formulas,
each of these worlds cannot lead to generate any further world by
means of a dynamic rule.\end{ultimissimi}

We construct the model $\emme=\sx X,R_X, <_X, V\dx$ by defining
$X$ and $V$ as in the case of \Pe. We then define $R_X$ as the
relation obtained from $R$ augmented with the following
conditions:
\begin{itemize}

\item[(i)] all the pairs $(\Gamma, \Gamma)$ such that $\Gamma \appartiene X$
and $\Gamma$ has no $R$-successor.

\item[(ii)] all the pairs $(\Gamma, \Gamma')$ such that $(\Gamma^{''}, \Gamma) \in R$ and $(\Gamma^{''}, \Gamma') \in R$ for some
$\Gamma^{''};$

\end{itemize}

\noindent  Last, we define $<_X$ as follows:
 \begin{itemize}
\item[(iii)] if $\Gamma' < \Gamma$, then $\Gamma' <_X \Gamma;$
\item[(iv)] if $\Gamma' < \Gamma$, and $\Gamma R_X \Gamma^{''}$, then $\Gamma' <_X \Gamma^{''};$
\item[(v)] if $\Gamma' <_X \Gamma$ and $\Gamma <_X \Gamma^{''}$, then $\Gamma' <_X
\Gamma^{''}$, i.e. $<_X$ is transitive.
\end{itemize}

\noindent Notice that the above conditions on $R_X$ and $<_X$ are
needed since the procedure builds two different kinds of worlds:
\begin{itemize}
  \item \emph{bad} worlds, obtained by an application of
  $(L^{-})$;
  \item \emph{good} worlds: the other ones.
\end{itemize}

\noindent \emph{Bad} worlds are those obtained by an application
of $(L^{-})$. These worlds ``forget'' the positive conditionals in
the initial set of formulas; for instance, if $\Gamma=\{\nott C,
D\}$ is a bad world obtained from $\Gamma'=\{A \ent B, \nott LC,
LD\}$, then it is ``incomplete'' by the absence of $A \ent B$. The
above supplementary conditions on $R_X$ and $<_X$ are needed in
order to prove that, even in presence of bad worlds, we can build
a CL-preferential model satisfying the initial set of formulas, as
shown below by Fact \ref{propr(iii)}. \hide{This is not the only
way to proceed: as an alternative, one should define $R_X$ as
obtained from $R$ only by adding $(\Gamma,\Gamma)$ for each
$\Gamma$ having no $R$-successors (in order to build a serial
accessibility relation), then describe a saturation process for
the bad worlds.}

It is easy to show that the following properties hold for $\emme$:

\begin{fact}\label{propr(iii)}
For all $\Gamma, \Gamma^{'} \appartiene X$, if $(\Gamma,
\Gamma^{'}) \appartiene R_X$ and $L A \appartiene \Gamma$ then $A
\appartiene \Gamma^{'}$.
\end{fact}
\begin{pf*}{Proof of Fact \ref{propr(iii)}.}
\begin{rosso}In case $\Gamma \not = \Gamma'$, then
$(\Gamma,\Gamma')\appartiene R$ and it has been added to $R$ by
step 4' or step 4'' of the procedure above. Indeed, for all
$(\Gamma, \Gamma')$ that have been introduced because
$(\Gamma^{''}, \Gamma) \in R$ and $(\Gamma^{''}, \Gamma') \in R$,
both $\Gamma$ and $\Gamma^{'}$ derive from the application of
$(L^-)$ to $\Gamma^{''}$, hence  they only contain propositional
formulas and do not contain any $LA$.

Hence, we have two different cases:

\begin{itemize}
  \item the relation $(\Gamma, \Gamma^{'})$ has been
  added to $R$ by step 4': in this case, we have that $\neg L B \appartiene \Gamma$.
  We can conclude that
  $A \appartiene \Gamma'$ by construction, since for each $LA
  \appartiene \Gamma$ we have that $A \appartiene \Gamma^{'}$ as a
  result of the application of \texttt{SAT(APPLY($\Gamma, \nott
  LB$))};
  \item the relation $(\Gamma, \Gamma^{'})$ has been
  added to $R$ by step 4'': similarly to previous case, for each
  $LA \appartiene \Gamma$, we have that $A$ is added to
  $\Gamma^{'}$ by construction.
\end{itemize}

\noindent In the case $\Gamma =\Gamma'$, then it must be that
$(\Gamma,\Gamma)$ has been added to $R_X$, as $\Gamma$ has no
$R$-successors. This means that $\Gamma$ does not contain formulas
of the form $LA, \nott LA$, otherwise it would have an
$R$-successor.\end{rosso} \provafatto{\ref{propr(iii)}}
\end{pf*}

\begin{fact}\label{Fatto 3}
 For all formulas $F$ and for all sets $\Gamma \appartiene X$ we
 have that:\\
 (i) if $F \appartiene \Gamma$ then $\emme, \Gamma \modello F$;
 (ii) if $\nott F \appartiene \Gamma$ then $\emme, \Gamma \not\modello
   F$.
\end{fact}
\begin{pf*}{Proof of Fact \ref{Fatto 3}.}
The proof is similar to the one for the preferential case. If $F$
is an atom or a boolean combination of formulas,  the proof is the
same as the proof for Fact \ref{Fatto 2} of the preferential case.
We consider the following cases:

\begin{itemize}\item $L A \appartiene \Gamma$: we have to show that $\emme, \Gamma
  \modello L A$, that is, we must show that, for all $\Delta
  \appartiene X$, if $(\Gamma, \Delta)\appartiene R_X$ then $\emme, \Delta  \modello
  A$.
  Let $\Delta$ be such that $(\Gamma, \Delta)\appartiene R_X$. Then, by
  Fact~\ref{propr(iii)},
  as $L A \appartiene \Gamma$, we can conclude $A \appartiene \Delta$.
  By inductive hypothesis, then $\emme, \Delta  \modello A$.

  \item $\neg L A \appartiene \Gamma$: we have to show that $\emme, \Gamma
  \not \modello L A$, that is, we must show that there exists $\Delta
  \appartiene X$ such that $(\Gamma, \Delta)\appartiene R_X$ and $\emme, \Delta  \not \modello
  A$. As $\neg L A \appartiene \Gamma$, by construction (step $4'$ in the
  procedure) there must be a $\Delta \appartiene X$ such that $\neg A
  \appartiene \Delta$. By inductive hypothesis, $\emme, \Delta  \not \modello
  A$, which concludes the proof that $\emme, \Gamma
  \not \modello L A$.

  \begin{rosso}
  \item $\bbox \nott LA \appartiene \Gamma$. Then, for all $\Gamma_i <_X \Gamma$ we have
   $\nott LA \in \Gamma_i$ by the definition of $(\bbox^{-})$, since $\Gamma_i$ has been generated by a sequence of
applications of $(\bbox^{-})$ (notice that point (iv) in the
definition of  $<_X$ above does not play any role here, since this
point only concerns
 sets of formulas $\Gamma$ that are propositional and do not contain boxed or negated box formulas). By
  inductive hypothesis $\emme, \Gamma_i \not\modello LA$ for all $\Gamma_i <_X \Gamma$, whence
  $\emme, \Gamma \modello \bbox
  \nott LA$.

  \item $\neg \bbox \nott LA \appartiene \Gamma$. By construction there is a
  $\Gamma^{'}$ s.t. $\Gamma^{'} <_X \Gamma$ and $LA \appartiene \Gamma^{'}$. By inductive
  hypothesis $\emme, \Gamma^{'}
  \models LA$. Thus, $\emme, \Gamma \not\models \bbox \nott  LA$.

  \item $A \cond B \appartiene \Gamma$. Let $\Delta \appartiene Min_{<_X}(LA)$. We distinguish two
  cases:

  \begin{itemize}
    \item  $A \cond B \in \Delta$,
   one can observe that $(1)\nott LA \appartiene
  \Delta$ or $(2)\nott \bbox \nott LA \appartiene \Delta$ or $(3)LB \appartiene \Delta$, since $\Delta$ is
  saturated. $(1)$ cannot be the case, since by inductive hypothesis $\emme, \Delta \not\modello LA$,
  which contradicts the definition of $Min_{<_X}(LA)$. If $(2)$, by
  construction of $\emme$ there exists a set $\Delta^{'} <_X \Delta$ such
  that $LA \appartiene \Delta^{'}$. By inductive hypothesis $\emme, \Delta^{'} \modello LA$, which contradicts
  $\Delta \appartiene Min_{<_X}(LA)$. Therefore, it must be that
  $(3) LB \appartiene \Delta$, and by inductive hypothesis
  $\emme, \Delta \modello LB$.

  \item $A \cond B \not\in \Delta$. Since all the rules apart from $(L^-)$
preserve the conditionals,  $\Delta$ must have been generated by
applying $(L^-)$ to $\Delta'$, i.e. $\Delta$ is a \emph{bad
world}. Hence, $\Delta' R_X \Delta$.
   In turn, it can be easily shown that $\Delta'$ itself cannot have been
generated by $(L^-)$,  hence $A \cond B \in \Delta'$, and, since
$\Delta'$ is
  saturated, either
$(1) \nott LA \appartiene
  \Delta'$ or $(2)\nott \bbox \nott LA \appartiene \Delta'$ or $(3)LB \appartiene \Delta'$. (1) is not possible,
  since by inductive hypothesis, it
  would entail that
   $\emme, \Delta' \not \modello LA$, i.e. there is
  $\Delta^{''}$ such that $\Delta^{'} R_X \Delta^{''}$ and $\emme, \Delta^{''}
  \not\modello A$. By point (ii) in the definition of $R_X$ above,
  also $\Delta R_X \Delta^{''}$, hence also $\emme, \Delta \not\modello LA$, which contradicts $\Delta \in
Min_{<_X}(LA)$.
  If (2),  by
  construction of $\emme$ there exists a set $\Delta^{''} <_X \Delta'$ such
  that $LA \appartiene \Delta^{''}$. By point (iv) in the definition
  of $<_X$ above, $\Delta^{''} <_X \Delta$, which contradicts
  $\Delta \in Min_{<_X}(LA)$, since by inductive hypothesis $\emme, \Delta^{''} \modello LA$.
  It follows that $LB \in \Delta'$. By inductive hypothesis $\emme, \Delta' \modello LB$, hence also
$\emme, \Delta \modello
  LB$ (indeed, since $\Delta$ does not contain any $L$-formula, by construction of the model and by point
  (ii) in the definition of $R_X$ above,
  $\Delta R_X \Delta^{''}$ just in case $\Delta' R_X \Delta^{''}$,
  from which the result follows).

Hence, we can conclude $\emme, \Gamma \modello A \cond
  B$.

\end{itemize}

\end{rosso}

  \item $\nott (A \cond B) \appartiene \Gamma$: by construction of $X$,
  there exists $\Gamma^{'} \appartiene X$ such that $LA, \bbox \nott LA, \nott LB \appartiene \Gamma^{'}$.
  By inductive hypothesis  we have that $\emme, \Gamma^{'}
  \modello LA$ and $\emme, \Gamma^{'} \modello \bbox \nott LA$. It
  follows that $\Gamma^{'} \appartiene Min_{<_X}(LA)$. Furthermore,
  always by induction, $\emme, \Gamma^{'} \not\modello LB$. Hence,
  $\emme, \Gamma \not\modello A \cond B$.
\end{itemize}

\provafatto{\ref{Fatto 3}}
\end{pf*}

\noindent Similarly to the case of \Pe, it is easy to prove the
following Fact:

\begin{fact}\label{fatto aggiunto in smoothness CL}
The relation $<_X$ is irreflexive, transitive, and satisfies the
smoothness condition.
\end{fact}

\noindent Moreover:

\begin{fact}\label{CL: fatto proprietà R}
The relation $R_X$ is serial.
\end{fact}

\noindent From the above Facts, we can conclude that $\emme=\sx
X,R_X,<_X,V \dx$ is a CL-preferential model satisfying $\Gamma_0$,
which concludes the proof of completeness.

\end{rosso}

\end{provaposu}

\noindent From the above Theorem \ref{completezza calcolo CL},
together with Proposition \ref{correspondence CL}, it follows that
for any boolean combination of conditionals $\Gamma_0$, if it does
not have any closed tableau, then it is satisfiable in a
loop-cumulative model.

 Similarly to what
done for \Pe, we can prove the following Corollary.

\begin{corollary}[Finite model property]\label{corollario-finitezzaCL} {\bf CL} has the finite model
property.
\end{corollary}

\hide{
\begin{corollary}\label{corollario-aciclicitàCL} If $\Gamma$ is
satisfiable in {\bf CL}, then it is satisfiable in an acyclic model.
\end{corollary}

\begin{corollary}\label{infinite-chainsCL}
If $\Gamma$ is satisfiable in {\bf CL} then it is satisfiable in a
model in which $<$ is irreflexive, transitive and does not have
infinite descending chains.
\end{corollary}
}

\subsection{Decision Procedure for \Cl}

Let us now  analyze the calculus $\calcoloCL$ in order to obtain a
decision procedure for \Cl \ logic. First of all, we reformulate
the calculus as we have done for \Pe, obtaining a system called
$\calcoloCLterminante$: we reformulate the $(\cond^{+})$ rule so
that it applies only once to each conditional in each world, by
adding  an extra set $\Sigma$. We reformulate the other rules
accordingly. \begin{rosso}Moreover, we adopt the same restriction
on the order of application of the rules in Definition
\ref{restrizione sull'ordine di applicazione delle regole}.

Notice that the rule $(L^{-})$ can only be applied a finite number
of times. Indeed, if it is applied to a premise $\Gamma, \nott L
A$, then the conclusion only contains propositional formulas
$\Gamma^{L^{\freccia}}, \nott A$, and the rule $(L^{-})$ is no
further applicable. The same in the case $(L^{-})$ is applied to a
premise $\Gamma, L A$. Notice that $(L^{-})$ can also be applied
to a set of formulas \emph{not containing} any formula $LA$ or
$\nott LA$: in this case, the conclusion of the rule corresponds
to an empty set of formulas. Therefore, we reformulate $(L^{-})$
only by adding the extra set $\Sigma$ of conditionals; the
reformulated rule is shown in Figure \ref{riformulazione regola
L}.

\begin{figure}

\linea

 \mbox{ \(
\begin{scriptsize}
\begin{array}{ll}\\
\quad\quad\quad\quad\quad (L^{-}) \
\irule{\Gamma, \nott L A; \Sigma}%
{ \Gamma^{L^{\freccia}}, \nott A; \vuoto }
;\ \irule{\Gamma; \Sigma}%
{\Gamma^{L^{\freccia}}; \vuoto}%
{\mbox{if $\Gamma$ does not contain negated $L$-formulas}} {}
\\
\\
\end{array}
\end{scriptsize} \)
 }

\linea

 \caption{The rule $(L^{-})$ reformulated in $\calcoloCLterminante$.} \label{riformulazione regola L}
\end{figure}
\end{rosso}

\noindent Exactly as we made for \Pe, we consider a lexicographic
order given by $m(\Gamma; \Sigma)=\sx c_1, c_2, c_3, c_4 \dx$, and
easily prove that each application of the rules of
$\calcoloCLterminante$ reduces this measure, as stated by the
following Lemma:

\begin{lemma}\label{prova diminuzione misura in cumulative}
  Consider an application of any rule of $\calcoloCLterminante$ to
  a premise $\Gamma; \Sigma$ and be $\Gamma'; \Sigma'$ any conclusion obtained; we have
  that either $m(\Gamma'; \Sigma')<m(\Gamma; \Sigma)$ or $\calcoloCLterminante$ leads
  to the construction of a closed tableau for $\Gamma'; \Sigma'$.
\end{lemma}

\begin{provaposu}
Identical to the proof of Lemma \ref{lemma diminuzione misura}.
Just observe that if $(L^{-})$ is applied, then $c_1$, $c_2$, and
$c_3$ become $0$, since conditional formulas are not kept in the
conclusion. If the premise contains only $L$-formulas, then $c_1$,
$c_2$ and $c_3$ are already equal to $0$ in both the premise and
the conclusion, but $c_4$ decreases, since (at least) one formula
$(\nott)LA$ in the premise is removed, and a formula with a lower
complexity ($\nott A$ or $A$) is introduced in the conclusion.

\end{provaposu}

\noindent Thus, $\calcoloCLterminante$ ensures termination.
Furthermore, the decision algorithm for \Pe \ described in section
\ref{optimal proof search} can be adapted to \Cl. To this aim, we
observe that the disjunction property holds for \Cl, and this
allows us to change the rule for negated conditionals in order to
treat them independently as we have done for \Pe. Moreover, we can
replace the $(\bbox^{-})$ rule by a stronger rule that does not
require backtracking in the tableau construction. The rule is the
following ($\Gamma_{-i}^{\bbox^{-}}$ is used to denote $\{\nott
\bbox \nott LA_j \orr LA_j \tc \nott \bbox \nott LA_j \appartiene
\Gamma \andd j \diverso i\}$):

\[
  \begin{prooftree}
    \Gamma, \nott \bbox \nott LA_1, \nott \bbox \nott LA_2, ...,
    \nott \bbox \nott LA_n; \Sigma
    \justifies \Sigma, \Gamma^{\cond\pm}, \Gamma^{\bbox},
    \Gamma^{\bbox^{\freccia}}, LA_1, \bbox \nott
    LA_1,\Gamma_{-1}^{\bbox^{-}}; \vuoto \tc \dots \tc \Sigma, \Gamma^{\cond\pm}, \Gamma^{\bbox},
    \Gamma^{\bbox^{\freccia}}, LA_n, \bbox \nott
    LA_n,\Gamma_{-n}^{\bbox^{-}}; \vuoto
    \using (\bbox^{-}_s)
  \end{prooftree}
\]

\noindent By reasoning similarly to what done for \Pe, we can show
that the calculus in which $(\bbox^{-})$ is replaced by
$(\bbox^{-}_s)$ is sound and complete w.r.t. multi-linear
CL-preferential models introduced in Definition \ref{definizione
modello multi lineare in CL}. We get a decision procedure as for
\Pe, structured in a top-level \texttt{GENERAL-CHECK} procedure,
taking care of multiple negated conditionals, and in a
\texttt{CHECK} procedure of tableau expansion. The procedure
\texttt{CHECK} has to be modified by introducing steps 2' and 2''
between steps 2 and 3 in the procedure for \Pe \ as follows:

\hide{ \vspace{0.2cm}

\begin{small}

2'. {\bf else if} $\{\nott LA \tc \nott LA \appartiene \Gamma\}
\diverso \vuoto$ {\bf then}

\indent \indent 2'a.  {\bf for all} $\nott LA_i \appartiene
\Gamma$ {\bf do} result[i] $\longleftarrow$
\texttt{CHECK(APPLY($\Gamma,\nott LA_i$))};

\indent \indent 2'b. {\bf if for all} $i$ result[i]==\texttt{SAT}
{\bf then return} \texttt{SAT};

\indent \indent 2'c.  {\bf else} {\bf return} \texttt{UNSAT};

2''. {\bf else if} $\{LA \tc LA \appartiene \Gamma\} \diverso
\vuoto$ {\bf then}

\indent\indent  2''a. {\bf return} \texttt{CHECK(APPLY(} $\Gamma,
L^{-}$\texttt{))};

\end{small}

\vspace{0.2cm}

\noindent The resulting procedure \texttt{CHECK} is as follows:

}

\begin{rosso}

 \vspace{0.2cm}

\linea

\begin{small}
\texttt{CHECK}($\Gamma$)

1. $\Gamma \longleftarrow$ \texttt{EXPAND}$(\Gamma)$; ($O(n)$)

2. {\bf if} $\Gamma$ contains an axiom {\bf then return}
\texttt{UNSAT}; ($O(n^2)$)

2'. {\bf if} $\{\nott LA \tc \nott LA \appartiene \Gamma\}
\diverso \vuoto$ {\bf then}

\indent \indent 2'a.  {\bf for all} $\nott LA_i \appartiene
\Gamma$ {\bf do} result[i] $\longleftarrow$
\texttt{CHECK(APPLY($\Gamma,\nott LA_i$))};

\indent \indent 2'b. {\bf if for some} $i$
result[i]\texttt{==UNSAT} {\bf then return} \texttt{UNSAT};

2''. {\bf else if} $\{LA \tc LA \appartiene \Gamma\} \diverso
\vuoto$ {\bf then}

\indent\indent  2''a. {\bf if} \texttt{CHECK(APPLY(} $\Gamma,
L^{-}$\texttt{))==UNSAT} {\bf then return} \texttt{UNSAT};

3. {\bf if} $\{\nott \bbox \nott L A \tc \nott \bbox \nott L A
\appartiene \Gamma\}=\emptyset$
{\bf then return} \texttt{SAT};

\hide{ \indent\indent  4a. let $\{\neg \bbox \neg L A_1,
\ldots,\neg \bbox \neg L A_k\}$ be all the negated boxed formulas
in $\Gamma$;}

4. {\bf else return} \texttt{CHECK(APPLY($\Gamma,\bbox^{-}_s$))};

\end{small}

\linea

 \vspace{0.2cm}

\noindent The top-level procedure \texttt{GENERAL-CHECK} is the
same as the one in section \ref{optimal proof search}. For a
better readability, we rewrite this procedure here below:

\vspace{0.2cm}

\linea

\begin{small}
\texttt{GENERAL-CHECK}($\Gamma$)

1. $\Gamma \longleftarrow$ \texttt{EXPAND}$(\Gamma);$ ($O(n)$)

2. let $\neg (A_1 \cond B_1),\ldots, \neg (A_k \cond B_k)$ be all
negated conditionals in $\Gamma$;

\indent \indent 2.1.  {\bf for all} $i=1,\ldots,k$
 result[i] $\longleftarrow$ \texttt{CHECK}(\texttt{APPLY}($\Gamma, \neg(A_i\cond B_i$))) ;

3. {\bf if for all} $i=1,\ldots,k$ result[i]\texttt{==SAT}  {\bf
then return} \texttt{SAT};

\indent \indent {\bf else} {\bf return} \texttt{UNSAT};

\end{small}

\linea

 \vspace{0.2cm}

\end{rosso}

\noindent Observe that the two recursive calls of \texttt{CHECK}
in 2'a and 2''a do not generate further recursive calls. By this
reason, we can argue similarly to what done for \Pe, then we
obtain the following result:

\begin{theorem}[Complexity of \Cl]\label{complessità cumulative}
  The problem of deciding validity for \Cl \ is {\bf coNP}-complete.
\end{theorem}


\section{The Tableau Calculus for Cumulative Logic
\Cu}\label{sezione calcolo C} In order to provide a calculus for
the weaker logic \Cu, we have to replace the rule $(\bbox^{-})$
with the weaker $(\bbox^{\bf C-})$:

$$ {\bf (\bbox^{C-})}
 \irule{\Gamma, \nott \bbox \nott LA}
{\Gamma^{\bbox^{\freccia}}, \Gamma^{\cond\pm},  LA}{}
$$

Observe that, if we ignore conditionals, this rule is nothing else
than the standard rule of modal logic K. This rule is weaker than
the corresponding rule of the two other systems in two respects:
(i) transitivity is not assumed thus we no longer have
$\Gamma^{\bbox}$ in the conclusion; (ii) the smoothness condition
does no longer ensure that if $\nott \bbox \nott LA$ is true in
one world, then there is a smaller {\em minimal} world satisfying
$LA$, this only happens if the world satisfies $LA$; thus  $\bbox
\nott LA$ is dropped from the conclusion as well.

Moreover, we add the following form of cut:
$${\bf {\rm \mbox{(Weak-Cut)}}}  \irule{\Gamma}
  {\Gamma, \nott LA
    \quad\quad \Gamma^{\cond\pm}, \Gamma^{\bbox^{\freccia}}, LA , \bbox \nott LA
    \quad\quad \Gamma, \bbox \nott LA} { }$$

\noindent Intuitively, this rule takes care of enforcing the
smoothness condition, and it can be applied to all $L$-formulas.

\hide{
\begin{figure}

\linea

\mbox{ \(
\begin{scriptsize}
\begin{array}{ll}\\
\quad\quad\quad\quad\quad\quad{\bf (\cond^{+})} \ \irule{\Gamma, A
\cond B}
  {\Gamma, \nott L A, A \cond B
    \quad \Gamma, \nott \bbox \nott L A, A \cond B
    \quad \Gamma, L B, A \cond B
  } {}
\\
\\
\\
\end{array}
\end{scriptsize}
\) }

\mbox{ \(
\begin{scriptsize}
\begin{array}{l@{\quad\quad\quad\quad\quad\quad\quad\quad\quad\quad\quad\quad\quad\quad\quad\quad}l}
{\bf (\cond^{-})} \ \irule{\Gamma, \nott(A \cond B)}%
{L A, \bbox \nott L A, \nott L B, \Gamma^{\cond\pm}}%
{} &
 {\bf (\bbox^{C-})} \
\irule{\Gamma, \nott \bbox \nott L A}%
{\Gamma^{\bbox^{\freccia}}, \Gamma^{\cond\pm}, LA}%
{}
\\
\\
\\
\end{array}
\end{scriptsize}
\) }

\mbox{ \(
\begin{scriptsize}
\begin{array}{ll}
\quad\quad\quad\quad {(\mbox{Weak-Cut})} \
\irule{\Gamma}%
{\Gamma, \nott LA \quad\quad \Gammas, LA, \bbox \nott LA \quad\quad \Gamma, \bbox \nott LA} %
{}
\\
\\
\\
\end{array}
\end{scriptsize}
\) }

\mbox{ \(
\begin{scriptsize}
\begin{array}{ll}
 \quad\quad\quad\quad (L^{-}) \
\irule{\Gamma, \nott L A}%
{ \Gamma^{L^{\freccia}}, \nott A }

;\ \irule{\Gamma}%
{\Gamma^{L^{\freccia}}}%
{\mbox{if $\Gamma$ does not contain negated $L$-formulas}} {}
\\
\\
\end{array}
\end{scriptsize} \)
 }

\linea

 \caption{Tableau system $\calcoloCtent$. The boolean rules are
omitted. } \label{Figura calcolo C tent}
\end{figure}

}

\begin{figure}
{\centerline{\includegraphics[angle=0,width=4.8in]{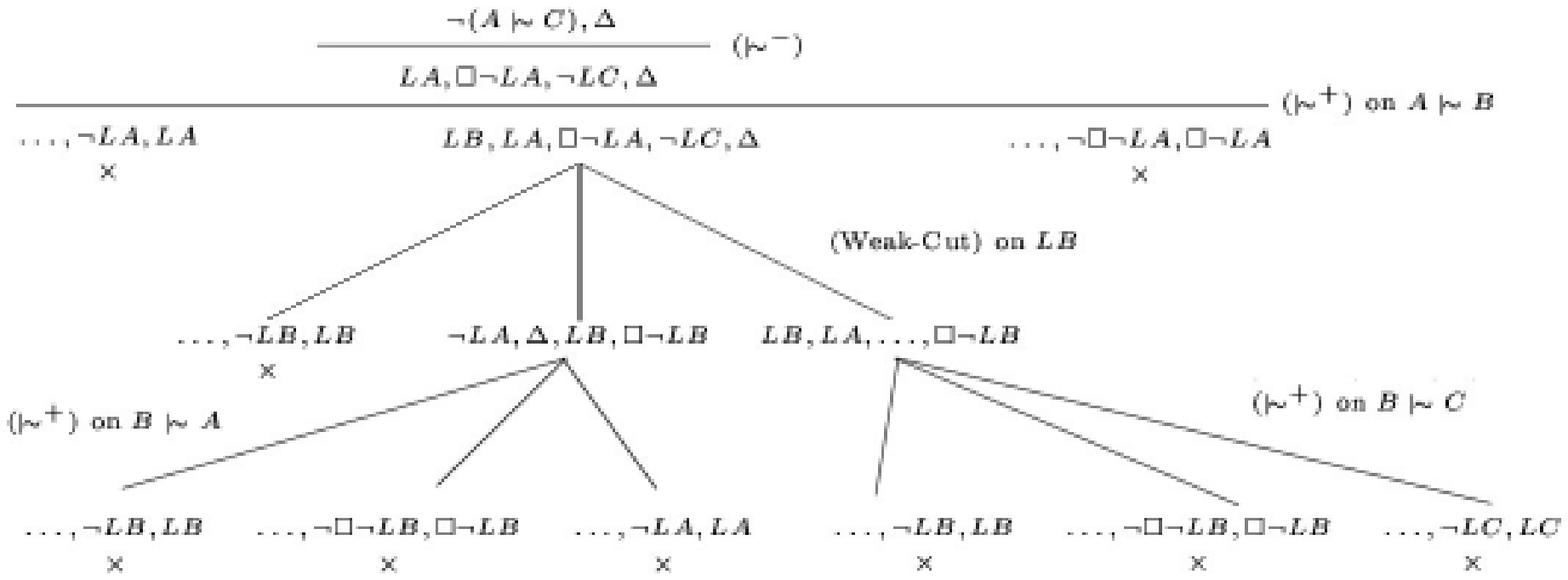}}}
\caption{A derivation of $\nott (A \cond C), A \cond B, B \cond A,
B \cond C$ in $\calcoloC$.
  To save space, we use $\Delta$ to denote the set of positive conditionals $A \cond B, B \cond A, B \cond C$.}
  \label{controesempio cut}
\end{figure}

\noindent The (Weak-Cut) rule is not completely eliminable, as
shown by the following example. Let $\Gamma = \{\neg (A \cond C),
A \cond B, B \cond A, B \cond C\}$: the set $\Gamma$ is
unsatisfiable in \Cu. $\Gamma$ has a closed tableau only if we use
(\mbox{Weak-Cut}) in the calculus, as shown by the derivation in
Figure \ref{controesempio cut}. Without (\mbox{Weak-Cut}), the
above set of formulas does not have any closed tableau.

\noindent The (\mbox{Weak-Cut}) rule makes the resulting calculus
not analytic. However, the rule can be restricted so that it only
applies to formulas $LA$ such that $A$ is the antecedent of a
positive conditional formula in $\Gamma$, thus making the
resulting calculus analytic. In order to prove this, we simplify
the calculus by incorporating the application of $(\bbox^{C-})$
and the restricted form of (Weak-Cut) in the $(\cond^{+})$ rule.
The resulting calculus, called $\calcoloC$ and given in Figure
\ref{Figura calcolo C}, is equivalent to the calculus that would
be obtained from the calculus $\calcoloCL$ by replacing
$(\bbox^{-})$ with $(\bbox^{C-})$ and by introducing (Weak-Cut)
restricted to antecedents of positive conditionals. The advantage
of the adopted formulation is that, being more compact, it allows
a simpler proof of the admissibility of the non-restricted
(Weak-Cut) (see Theorem \ref{eliminazione cut C} below).

\begin{figure}

\linea

\mbox{ \(
\begin{scriptsize}
\begin{array}{ll}\\
\quad{\bf (\cond^{+})} \ \irule{\Gamma, A \cond B}
  {\Gamma, A \cond B, \nott L A
    \quad\quad\quad \Gammas, A \cond B, LA, \bbox \nott LA
    \quad\quad\quad \Gamma, A \cond B, LA, \bbox \nott LA,  L B
  } {}
\\
\\
\\
\end{array}
\end{scriptsize}
\) }

\mbox{ \(
\begin{scriptsize}
\begin{array}{l@{\quad\quad\quad}l}
{\bf (\cond^{-})} \ \irule{\Gamma, \nott(A \cond B)}%
{L A, \bbox \nott L A, \nott L B, \Gamma^{\cond\pm}}%
{} & {(L^{-})} \
 \irule {\Gamma, \nott LA}{ \Gamma^{L^{\freccia}}, \nott A }

;\ \irule{\Gamma}%
{\Gamma^{L^{\freccia}}}%
{\mbox{if $\Gamma$ does not contain negated $L$-formulas}} {}
\\
\\
\\
\end{array}
\end{scriptsize}
\) }

\linea

 \caption{Tableau system $\calcoloC$. The boolean rules are
omitted. } \label{Figura calcolo C}
\end{figure}

\noindent Observe that the calculus does not contain any rule for
negated box formulas, as the modified $(\cond^{+})$ rule does no
longer introduce formulas of the form $\nott \bbox \nott LA$. The
resulting language $\elle_{L'}$ of the formulas appearing in a
tableau for \Cu \ extends $\elle$ by formulas $LA$ and $\bbox
\nott LA$ where $A$ is propositional, whereas negated box formulas
of the form $\nott \bbox \nott LA$ are not allowed. Therefore,
$\elle_{L'}$ is a restriction of $\elle_{L}$ used in $\calcoloCL$.

Notice also that, as a difference with $\calcoloCL$, the
$(\cond^{+})$ rule is neither a \emph{static} nor a \emph{dynamic}
rule. Indeed, the leftmost and rightmost conclusions of the rule
represent the same world as the world represented by the premise,
whereas the inner conclusion represents a world which is different
from the one represented by the premise.

We prove that $\calcoloC$ is sound and complete w.r.t. the
semantics.

\begin{theorem}[Soundness of $\calcoloC$]\label{correttezzaC}
  The system $\calcoloC$ is sound with respect to C-preferential
  models, i.e. if there is a closed tableau for a set of formulas
  $\Gamma$, then $\Gamma$ is unsatisfiable.
\end{theorem}

\begin{provaposu}
Given a set of formulas $\Gamma$, if there is a closed tableau for
$\Gamma$, then $\Gamma$
  is unsatisfiable in C-preferential models.
For the rules already present in $\calcoloCL$, the proof is the
same as the proof for Theorem \ref{correttezzaCL} (notice that in
the proof the transitivity of $<$ does not play any role). We only
consider here the rule $ (\cond^+)$. We show that if the premise
is satisfiable by a C-preferential model, then also one of the
conclusions is.

Let $\emme,w \models \Gamma, A \cond B$.  We distinguish the two
following cases:
    \begin{itemize}
      \item $\emme, w \not\modello
      LA$, thus $\emme, w \modello \nott LA$: in this case, the left
     conclusion of the $(\cond^{+})$ rule is satisfied ($\emme, w \modello \Gamma, A \cond B, \nott LA$);
     \item $\emme, w \modello LA$: we consider two subcases:
       \begin{itemize}
         \item $w \appartiene Min_{<}(LA)$, hence $\emme, w \modello LA, \bbox \neg LA$ . By the definition
  of $\emme, w \modello A \cond B$, we have that $\emme, w \modello LB$. Therefore, the right
  conclusion of $(\cond^{+})$ is satisfiable;
         \item $w \not\appartiene Min_{<}(LA)$: by the smoothness condition,
         there exists a world $w^{'} < w$ such that $w^{'} \appartiene
         Min_{<}(LA)$. It follows
that $(\emme, w') \models LA, \bbox \neg LA$. Furthermore, by the
semantics of $\cond$, $(\emme, w') \models \Gamma^{\cond\pm}$ and
since $w' < w$, $(\emme, w') \models \Gamma^{\bbox^{\freccia}}$.
       \end{itemize}
\end{itemize}

\end{provaposu}

\noindent Soundness with respect to cumulative models follows from
the correspondence established by Proposition
\ref{corrispondenza-C}.

\begin{rosso}

We can prove that the \mbox{(Weak-Cut)} rule is admissible in
$\calcoloC$; this is stated by the following Theorem
\ref{eliminazione cut C}. The non-trivial proof of this Theorem is
moved to the Appendix for space reasons.

\begin{theorem}\label{eliminazione cut C}
Given a set of formulas $\Gamma$ and a propositional formula $A$,
if  there is a closed tableau for each of the following sets of
  formulas:
  \begin{itemize}
    \item[$(1)$] $\Gamma, \nott LA$
    \item[$(2)$] $\Gammas, LA, \bbox \nott LA$
    \item[$(3)$] $\Gamma, \bbox \nott LA$
  \end{itemize}
\noindent   then  there is also a closed tableau for $\Gamma$,
i.e. the \emph{(Weak-Cut)} rule is admissible.\end{theorem}

\end{rosso}


We prove the completeness of our calculus by modifying the
procedure described in the proof of Theorem \ref{completezza}
above. The completeness is a consequence of the admissibility of
the (Weak-Cut) rule. Hence, in the following completeness proof,
we will make use of the (Weak-Cut) rule. We prove that given any
finite $\Gamma_0 \subseteq \lan_{L'}$, if it does not have any
closed tableau, then it is satisfiable in a C-preferential model.

In order to build  a {\em finite} model for $\Gamma_0$, we
consider a restricted version of $\lan_{L'}$, only containing  the
formulas in $\lan_{L'}$ made out of propositional formulas
appearing in $\Gamma_0$. We call this restricted language
$\lan_{\Gamma_0}$.

\begin{theorem}[Completeness of
$\calcoloC$]\label{completezzaC}
  $\calcoloC$ is complete with respect to C-preferential models, i.e. if
  a set of formulas $\Gamma$ is unsatisfiable, then it has a closed
  tableau in $\calcoloC$.
\end{theorem}

\begin{provaposu} We assume that no tableau for $\Gamma_0$ is closed, and we
construct a model for $\Gamma_0$.

Notice that the language $\elle_{\Gamma_0}$ may contain infinitely
many propositional formulas (obtained by boolean combinations of
the atomic propositions in $\Gamma_0$). In order to keep the
construction of the model finite, we define a notion of
equivalence between formulas w.r.t. their propositional part (or
{\em p-equivalence}) so to identify those formulas having the same
structure and containing equivalent propositional components. Let
$\equiv_{PC}$ be logical equivalence in the classical
Propositional Calculus. We define the notion of {\em
p-equivalence} between two formulas $F$ and $G$ (written
$F\equiv_p G$) as an equivalence relation satisfying the following
conditions:
\begin{itemize}
\item
if $F$ and $G$ are propositional formulas, then $F\equiv_p G$ iff
$F\equiv_{PC} G$;

\item
$LA\equiv_p LB$ iff $A\equiv_{PC} B$;

\item
$\neg LA\equiv_p \neg LB$ iff $A\equiv_{PC} B$;

\item
$\Box \neg LA \equiv_p \Box \neg LB$ iff $A\equiv_{PC} B$.

\end{itemize}

\noindent For instance, $LA \equiv_p L(A \wedge A)$, $\Box \neg LA
\equiv_p \Box \neg L(A \wedge A)$.

We say that two sets of formulas $\Gamma$ and $\Gamma'$ are
p-equivalent if for every formula in $\Gamma$ there is a
p-equivalent formula in $\Gamma'$, and viceversa.

Observe that this notion of p-equivalence is very weak and, for
instance, we do not recognize that the set $\{LA, LB\}$ is
equivalent to the set $\{L(A \wedge B)\}$. The notion of
p-equivalence has been introduced with the purpose of limiting the
application of the rule (\mbox{Weak-Cut}) so that it does not
generate infinitely many equivalent formulas. Moreover, as we will
see in the construction of the model below, this notion of
equivalence will be used to control the addition of a new set of
formulas to the current set of worlds.

In our construction of the model below we will identify those
p-equivalent sets of formulas $\Gamma$. Before adding a new set of
formulas $\Gamma$ to the current set of worlds $X$, we check that
$X$ does not already contain a set $\Gamma'$ which is p-equivalent
to $\Gamma$; for short, we will write $\Gamma \not\appartiene_P X$
in the case $X$ does not already contain such a $\Gamma'$.


We define the procedure $\texttt{SAT'}$ that for any $\Gamma
\subseteq \lan_{\Gamma_0}$ extends $\Gamma$ by applying the
transformations described below and, at the same time, builds a
set of formulas $\Gamma^{S}$ initially set to $\vuoto$. The
transformations below are performed in sequence:

\begin{itemize}
\item applies to $\Gamma$ the propositional
rules, once to each formula, as far as possible. In case of
branching, makes the choice that
 leads to an open tableau (this step saturates $\Gamma$ with respect to the static rules in {\bf
C});
\item for each $A \cond B \in \Gamma$, applies $(\cond^+)$ to it. If
the leftmost branch is open, then adds $\neg LA$ to $\Gamma$;
otherwise, if the rightmost branch is open, then adds $\bbox \neg
LA, LA, LB$ to $\Gamma$; if the only open branch is the inner one,
then adds the set $\{\Gammas, LA, \bbox \neg LA \}$ to a support
set $\Gamma^S$ associated with $\Gamma$, that is initially set to
$\emptyset$;
\item  for all $LA \in \lan_{\Gamma_0}$, if there is no $A \cond B \in \Gamma$, and there is no
$A'$ propositionally equivalent to $A$ on
  which (Weak-Cut) has been previously applied (in $\Gamma$),
  applies (Weak-Cut) to it. If the leftmost branch is open, then  adds $\neg LA$
to $\Gamma$; otherwise, if the rightmost branch is open, adds
$\bbox \neg LA$ to $\Gamma$; if the only open branch is the inner
one, then adds  $\{\Gammas, LA, \bbox \neg LA \}$ to the support
set $\Gamma^S$.
\end{itemize}

\noindent Observe that \texttt{SAT'} terminates, extends $\Gamma$
to a finite set of formulas, and produces a set $\Gamma^{S}$ which
a finite set of finite sets of formulas, since: (a) there are only
a finite number of conditionals  that can lead to create sets of
formulas in $\Gamma^{S}$; (b) there are only a finite number of
$p$-equivalent formulas that can lead to create a set in
$\Gamma^{S}$ by the third transformation above.

\noindent We build $X$, the set of worlds of the model, and $<$,
as follows:

\vspace{0.5cm}

\begin{small}

\noindent  1. initialize $X=\{\Gamma_0\}$; mark
  $\Gamma_0$ as unresolved;

\noindent  2. {\bf while} $X$ contains unresolved nodes {\bf do}

           3. choose an unresolved $\Gamma$ from $X$;

  4. {\bf for} each $\{\Gammas, LA, \bbox \neg LA \} \in \Gamma^S$, associated to $\Gamma$,

  \indent\indent\indent let
  $\Gamma_{LA} = \texttt{SAT'}(\{\Gamma^{\cond\pm}, \Gamma^{\bbox^{\freccia}},
 LA, \bbox \neg LA
  \})$;

  \indent\indent 4a. {\bf for} all $\Gamma' \in X$ p-equivalent to $\Gamma_{LA}$

  \indent\indent\indent add the relation $\Gamma' <
\Gamma$;

  \indent\indent 4b. {\bf if} $\Gamma_{LA} \not\appartiene_P X$
  {\bf then} let $X = X \cup \{\Gamma_{LA}\}$

  5. {\bf for} each formula $\nott L A \appartiene \Gamma$, let
  $\Gamma_{\nott L A}=$\texttt{SAT'}(\texttt{APPLY}$(\Gamma,\nott L
  A))$;

  \indent\indent 5a. {\bf for} all $\Gamma' \in X$ equivalent to $\Gamma_{\nott LA}$

  \indent\indent\indent add the relation $\Gamma' R \Gamma$;

  \indent\indent 5b. {\bf if} $\Gamma_{\nott LA} \not\appartiene_P X$
  {\bf then} let $X = X \cup \{\Gamma_{\nott LA}\}$

  6. {\bf for} each formula $\nott (A \cond B) \appartiene \Gamma$

  \indent\indent 6a. let $\Gamma_{\nott(A \cond B)}= $\texttt{SAT'}(\texttt{APPLY}$(\Gamma,\nott(A \cond
  B)))$;

  \indent\indent 6b. {\bf if} $\Gamma_{\nott(A \cond B)} \not\appartiene_P X$ {\bf then} $X=X
  \unione \{\Gamma_{\nott(A \cond B)}\}$;

  7. mark $\Gamma$ as resolved;

 \noindent {\bf endWhile};

\end{small}

\vspace{0.5cm}

\noindent If $\Gamma_0$ is finite, the procedure terminates.
Indeed, it can be easily seen that $\texttt{SAT'}$ terminates, as
there is only a finite number of propositional evaluations.
Furthermore, the whole procedure terminates, since the number of
possible different sets of formulas that can be added to $X$
starting from a finite set $\Gamma_0$ is finite. Indeed, the
number of non-p-equivalent sets of formulas that can be introduced
in $X$ is finite, as the number of p-equivalent classes is finite.


We construct the model $\emme=\sx X, R_X, <_X, V\dx$ as follows.

 \begin{itemize}
 \item $X$,$V$ and $R_X$ are defined as in  the completeness proof for $\calcoloCL$
 (Theorem \ref{completezza calcolo CL});
 \item  $<_X$ is defined as follows:
 \begin{itemize}
\item[(i)] if $\Gamma' < \Gamma$, then $\Gamma' <_X \Gamma;$
\item[(ii)] if $\Gamma' < \Gamma$, and $\Gamma R_X \Gamma^{''}$, then $\Gamma' <_X \Gamma^{''};$
\end{itemize}

\noindent As a difference from $<_X$ used in the completeness
proof for $\calcoloCL$ (Theorem \ref{completezza calcolo CL}),
$<_X$ is not transitive.
\end{itemize}

\noindent In order to show that $\emme$ is a \Cu-preferential
model for $\Gamma$, we prove the following:

\begin{fact}\label{Fact1-C}
The relation $<_X$ is irreflexive.
\end{fact}

\begin{pf*}{Proof of Fact \ref{Fact1-C}.}
By the procedure above, $\Gamma' <_X \Gamma$ only in two cases 1)
$\Gamma' = \Gamma_{LA}$. In this case, it can be easily seen that
$\Gamma \neq \Gamma'$. 2) $\Gamma' < \Gamma^{''}$, i.e. $\Gamma' =
\Gamma^{''}_{LA}$, and $\Gamma^{''} R_X \Gamma$. Also in this
case, it can be easily seen that $\Gamma \neq \Gamma'$, since
$\Gamma$ does not contain L-formulas, whereas $\Gamma'$ does.
\provafatto{\ref{Fact1-C}}
\end{pf*}

\noindent By reasoning analogously to what done for Facts
\ref{Fatto 2} and \ref{Fatto 3} above, we show that:
\begin{fact}\label{Fatto 2-C}
 For all formulas $F$ and for all sets $\Gamma \appartiene X$ we
 have that:\\
 (i) if $F \appartiene \Gamma$ then $\emme, \Gamma \modello F$;
 (ii) if $\nott F \appartiene \Gamma$ then $\emme, \Gamma \not\modello
   F$.
\end{fact}

\begin{pf*}{Proof of Fact \ref{Fatto 2-C}.}
The proof is very similar to the one of  Facts \ref{Fatto 2} and
\ref{Fatto 3} above. Obviously, the case of negated boxed formulas
disappears. Here we only consider the case of positive conditional
formulas since the rule $(\cond^+)$ in $\calcoloC$ is slightly
different than before. Let $\Delta \appartiene Min_{<_X}(LA)$. We
distinguish two cases:

  \begin{itemize}

    \item $A \cond B \in \Delta$. By definition of
  $\texttt{SAT'}$ and construction of the model above, either
   $(1)\nott LA \appartiene
  \Delta$ or $(2)$ there is $\Delta_{LA} \in X$ such that $LA \in \Delta_{LA}$ and $\Delta_{LA}< \Delta$,
  hence $\Delta_{LA}<_X \Delta$  or
(3)$LB \appartiene \Delta$. Similarly to what done in the proof
for Fact \ref{Fatto 3}, $(1)$ and $(2)$ cannot be the case, since
they both contradict the fact that
  $\Delta \appartiene Min_{<_X}(LA)$. Thus it must be that
  $(3) LB \appartiene \Delta$, and by inductive hypothesis
  $\emme, \Delta \modello LB$.

  \item $A \cond B \not\in \Delta$. We can reason in the same way than in the analogous case in the proof
  of Fact \ref{Fatto 3} above: $\Delta$ must have been generated by applying $(L^-)$ to $\Delta'$
  with  $A \cond B \in \Delta'$ , hence $\Delta' R_X
  \Delta$. By definition of $\texttt{SAT'}$ above, either
  $(1) \nott LA \appartiene
  \Delta'$ or  $(2)$ there is $\Delta'_{LA} \in X$ such that $LA \in \Delta'_{LA}$ and $\Delta'_{LA}< \Delta'$,
 or
(3) $LB \appartiene \Delta'$. (1) is not possible: by inductive
hypothesis, it would
  be $\emme, \Delta' \not \modello LA$, i.e. there is
  $\Delta^{''}$ such that $\Delta R_X \Delta^{''}$ and $\emme, \Delta^{''}
  \not\modello A$. By  definition of $R_X$ (see point (ii)
  in the definition of $R_X$, proof of Theorem \ref{completezza calcolo CL}),
  also $\Delta R_X \Delta^{''}$, hence also $\emme, \Delta \not\modello LA$, which contradicts $\Delta \in
Min_{<_X}(LA)$.
  If (2), by  definition of $<_X$, $\Delta'_{LA}<_X \Delta$, which contradicts
$\Delta \in Min_{<_X}(LA)$. It follows that $LB \in \Delta'$,
hence
  by inductive hypothesis $\emme, \Delta' \modello LB$, and also $\emme, \Delta \modello
  LB$ (indeed, since $\Delta$ does not contain any $L-$ formula, by construction of the model and by
  definition of $R_X$ - see point (ii) in the definition of $R_X$, proof of Theorem \ref{completezza calcolo
  CL} -  $\Delta R_X \Delta^{''}$ just in case $\Delta' R_X \Delta^{''}$,
  from which the result follows).

  \end{itemize}

\provafatto{\ref{Fatto 2-C}}

\end{pf*}

\noindent  Furthermore, we prove that:

 \begin{fact}\label{fatto smoothness su L-formulas}
  The relation $<$ satisfies the smoothness condition on  L-formulas.
 \end{fact}

\begin{pf*}{Proof of Fact \ref{fatto smoothness su L-formulas}.}
Let $\emme, \Gamma \models LA$. Then by Fact \ref{Fatto 2-C}
above, $\neg LA \not\in \Gamma$. By definition of \texttt{SAT'}
and point $4$ in the procedure above, either $\bbox \neg LA \in
\Gamma$ or there is $\Gamma' \in X$ s.t. $\Gamma' = \Gamma_{LA}$,
and $\Gamma' <_X \Gamma$. In the first case, by Fact \ref{Fatto
2-C}, $\emme, \Gamma \models \bbox \neg LA$, and it is minimal
w.r.t. the set of $LA$-worlds. In the second case $\emme,
\Gamma^{'} \models \bbox \neg LA$, it is minimal w.r.t. the set of
$LA$-worlds, and $\Gamma^{'} <_X \Gamma$. \provafatto{\ref{fatto
smoothness su L-formulas}}
\end{pf*}

\end{provaposu}

\noindent From the above Theorem \ref{completezzaC}, together with
Proposition \ref{corrispondenza-C}, it follows that for any
boolean combination of conditionals $\Gamma_0$, if it does not
have any closed tableau, then it is satisfiable in a cumulative
model.

Similarly to what done for \Pe \ and \Cl, we can show the
following Corollary.
\begin{corollary}[Finite model property]\label{corollario-finitezzaC} {\bf C} has the finite model
property.
\end{corollary}

\hide{ \noindent As a difference from \Pe \ and \Cl \ we cannot
prove a result analogous to Corollaries
\ref{corollario-aciclicitàP} and \ref{corollario-aciclicitàCL}.
This is due to the fact that in the construction above we cannot
prove that $<$ is acyclic. Hence we cannot show that $<$ does not
have infinite descending chains, as shown for \Pe \ and \Cl \ by
Corollaries \ref{infinite-chainsP} and \ref{infinite-chainsCL}. }

\noindent As a difference from \Pe \ and \Cl \ we cannot prove
that $<$ does not have infinite descending chains. This is due to
the fact that in the construction above we cannot prove that $<$
is acyclic.

\noindent In the case of logic \Cu, non-termination can be caused
by the generation of infinitely many worlds, producing infinite
branches. By Theorem \ref{eliminazione cut C}, only the formulas
occurring in the initial set $\Gamma_0$ of formulas can occur on a
branch. Hence, the number of possibly different sets of formulas
$\Gamma$ on the branch is finite (and they are exponentially many
in the size of $\Gamma_0$). A loop checking procedure can be used
in order to avoid that a given set of formulas is expanded again
on a branch, so to ensure the termination of the procedure.

The satisfiability problem for a set of formulas $\Gamma_0$  can
be solved by showing that all the tableaux for $\Gamma_0$ have an
open branch. As there are exponentially many tableaux that have to
be taken in consideration, each one of exponential size with
respect to the size of the initial set of formulas, our tableau
method provides an hyper exponential procedure to check the
satisfiability.

In further investigations it might be considered if this bound can
be improved. For this, a more accurate analysis of derivation
structures (and, in particular, an analysis of permutability of
the rules) might be required.


\section{The Tableau Calculus for Rational Logic \Ra}\label{sezione
calcolo R}

\begin{rosso}
In this section we present $\calcoloR$, a tableau calculus for
rational logic \Ra. We have already mentioned that, as a
difference with the calculi presented for the other weaker logics,
here we adopt a \emph{labelled} tableau calculus, which seems to
be a more natural approach. Indeed, in order to capture the
modularity condition, intuitively we must keep all worlds
generated by $(\bbox^{-})$ and we need to propagate formulas among
them according to all possible modular orderings. In an unlabelled
calculus, this might be achieved, for instance, by introducing ad
hoc modal operators (that act as a marker) or by adding additional
structures to tableau nodes similarly to hypersequents calculi
(see for instance \cite{avronipersequenti}). However, the
resulting calculus would be unavoidably rather cumbersome. In
contrast, by using world labels, we can easily keep track of
multiple worlds and their relations. This provides a much simpler,
intuitive and natural tableau calculus. On the other hand, even if
we use labels, we do not run into problems with complexity and
termination, and we are able to define an optimal decision
procedure for \Ra.
\end{rosso}

The calculus makes use of labels to represent possible worlds. We
consider a language $\elle_R$ and a denumerable alphabet of labels
$\mathcal{A}$, whose elements are denoted by \emph{x, y, z, ...}.
$\elle_R$ extends $\elle$ by formulas of the form $\bbox \nott A$
as for the other logics.

\hide{
\begin{definition}[Truth condition of $\bbox$]\label{verità box R}
$\emme, w \modello \bbox A$ if for every $w' \appartiene \WW$ if
$w' <  w$ then $\emme, w' \modello A$.
\end{definition}

\noindent From definition of $Min_{<}(A)$ in Definition
\ref{semantica R}, it follows that for any formula $A$, $w
\appartiene Min_{<}(A)$ iff $\emme, w \modello A \andd \bbox \nott
A$. $\calcoloR$ will only make use of boxed formulas with a
negated argument, i.e. with the form $x: \bbox \nott A$. }

Our tableau calculus includes two kinds of labelled formulas:
\begin{itemize}
  \item \emph{world formulas} $x: F$, whose meaning is that $F$ holds in
the possible world represented by $x$;
  \item \emph{relation formulas} of the form $x < y$, where $x, y \appartiene
  \mathcal{A}$, used to represent the relation
  $<$.
\end{itemize}
 \noindent  We denote by $\alpha, \beta\ldots$ a world or a relation
  formula.

\noindent We define:

$$\Gammam{x}{y}=\{y: \nott A, y: \bbox \nott A \tc x: \bbox \nott A
\appartiene \Gamma\}$$

\noindent The calculus $\calcoloR$ is presented in Figure
\ref{figura calcolo rational}. As for \Pe, the rules $(\cond^{-})$
and $(\bbox^{-})$ that introduce new labels in their conclusion
are called \emph{dynamic rules}; all the other rules are called
\emph{static rules}.

 \begin{figure}[h]

\linea

 \mbox{ \(
\begin{scriptsize}
\begin{array}{l@{\quad\quad\quad\quad}r}\\
\quad\quad\quad{\bf (AX)} \ \Gamma, x: P, x: \nott P \quad
\mbox{with $P \appartiene \mathit{ATM}$}

&  {\bf (AX)} \ \Gamma, x < y, y < x
\\
\\
\\
\quad\quad\quad{\bf (\andd^{+})} \ \irule{\Gamma, x: F \andd G}
  {\Gamma, x: F, x: G} {}

&  {\bf (\andd^{-})} \ \irule{\Gamma, x: \nott(F \andd G)}
  {\Gamma, x: \nott F \quad\quad\quad \Gamma, x: \nott G} {}
\\
\\
\\
\end{array}
\end{scriptsize}
 \)  }

\mbox{ \(
\begin{scriptsize}
\begin{array}{ll}
\quad\quad\quad{\bf (\nott)} \ \irule{\Gamma, x: \nott \nott F}
{\Gamma, x: F}
{}\\
\\
\\
\end{array}
\end{scriptsize}
 \)  }

\mbox{ \(
\begin{scriptsize}
\begin{array}{ll}
\quad\quad\quad{\bf (\cond^{+})} \ \irule{\Gamma, u: A \cond B}
  {\Gamma, x: \nott A, u: A \cond B
    \quad\quad\quad\quad \Gamma, x: \nott \bbox \nott A, u: A \cond B
    \quad\quad\quad\quad \Gamma, x: B, u: A \cond B
  } {}
\\
\\
\\
\end{array}
\end{scriptsize}
 \)  }

\mbox{ \(
\begin{scriptsize}
\begin{array}{l@{\quad\quad\quad}l}
\quad\quad\quad{\bf (\cond^{-})} \ \irule{\Gamma, u: \nott(A \cond B)}%
{\Gamma, x: A, x: \bbox \nott A, x: \nott B}%
{\shortstack{$x \ \mbox{new}$ \\ \mbox{label}}}

&  {\bf (\bbox^{-})} \
\irule{\Gamma, x: \nott \bbox \nott A}%
{\Gamma, y < x, \Gammam{x}{y}, y: A, y: \bbox \nott A}%
{\shortstack{$y \ \mbox{new}$ \\ \mbox{label}}}
\\
\\
\\
\end{array}
\end{scriptsize}
 \)  }

\mbox{ \(
\begin{scriptsize}
\begin{array}{ll}
\quad\quad\quad{\bf (<)} \ \irule{\Gamma, x < y}
  {\Gamma, x < y, z < y, \Gammam{y}{z} \quad\quad\quad\quad \Gamma, x < y, x < z,
  \Gammam{z}{x}}
 {}

& \shortstack{${\mbox{$z$ occurs in $\Gamma$ and}}$ \\ {\mbox{$\{x
< z,z < y\} \intersezione \Gamma = \vuoto$}} }
\\
\\
\\
\end{array}
\end{scriptsize}
 \)  }

\linea

 \caption{The calculus $\calcoloR$. To save space, rules for
 $\imp$ and $\orr$ are omitted.}
\label{figura calcolo rational}
\end{figure}

\begin{definition}[Truth conditions of formulas of $\calcoloR$]\label{verità labelled formulas}
Given a model $\emme=\sx \WW, <, V \dx$ and a labelled alphabet
$\mathcal{A}$, we consider a mapping $I: \mathcal{A} \mapsto \WW$.
Given a formula $\alpha$ of the calculus $\calcoloR$, we define
$\emme \modello_I \alpha$ as follows:
\begin{itemize}
  \item $\emme \modello_I x: F$ iff $\emme, I(x) \modello F$
  \item $\emme
\modello_I x < y$ iff $I(x) < I(y)$.
\end{itemize}

\noindent We say that a set of formulas $\Gamma$   is satisfiable
if, for all formulas $\alpha \appartiene \Gamma$, we have that
$\emme \modello_I \alpha$, for some model $\emme$ and some mapping
$I$.
\end{definition}

\begin{figure}
{\centerline{\includegraphics[angle=0,width=4.8in]{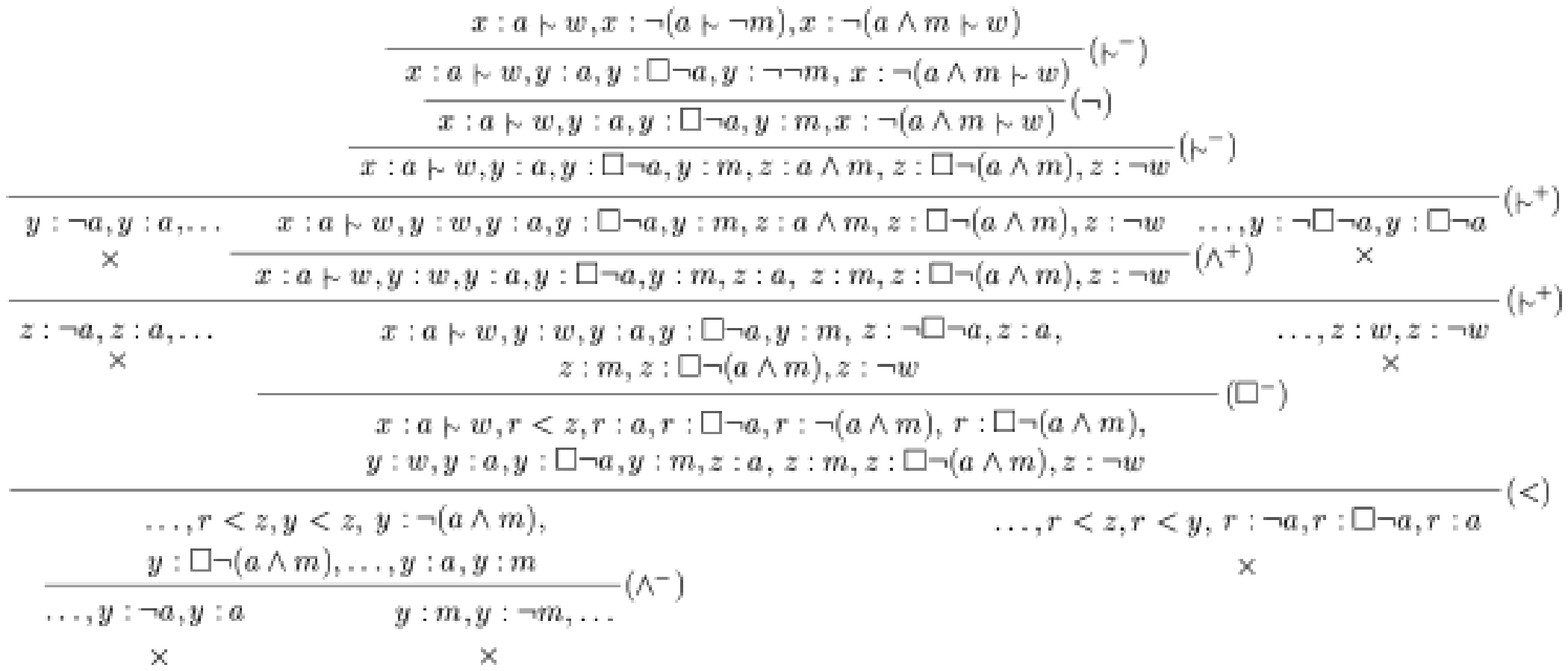}}}
 \caption{A derivation in $\calcoloR$ of $\{adult \ent worker$, $\nott (adult \cond \nott married)$,
$\nott (adult \andd married \cond worker)\}$. To save space, we
use $a$ for $adult$, $m$ for $married$, and $w$ for $worker$.}
\label{esempioDerivazione}
\end{figure}

In order to verify that a set of formulas $\Gamma$ is
unsatisfiable, we label all the formulas in $\Gamma$ with a new
label $x_0$, and verify that the resulting set of labelled
formulas has a closed tableau. For instance, in order to verify
that the set $\{adult \ent worker$, $\nott (adult \cond \nott
married)$, $\nott (adult \andd married \cond worker)\}$ is
unsatisfiable (and thus $adult \andd married \cond worker$ is
entailed by $\{adult \ent worker$, $\nott (adult \cond \nott
married)\}$), we can build the closed tableau in Figure
\ref{esempioDerivazione}.

\subsection{Soundness, Termination, and Completeness of $\calcoloR$}
In this section we prove that the calculus $\calcoloR$ is sound
and complete w.r.t. the semantics. Moreover, we prove that, with a
restriction on $(\cond^{+})$ based on the same idea of the one
adopted for the other calculi, the calculus guarantees
termination.

First of all, we reformulate the calculus, obtaining a terminating
calculus $\calcoloRterminante$. We will show in Theorem
\ref{terminazione del calcolo R} below that, as for the calculi
for \Pe, \Cl, and \Cu, non-termination of the procedure due to the
generation of infinitely-many worlds (thus creating infinite
branches) cannot occur. As a consequence, we will observe that
only a finite number of labels are introduced in a tableau
(Corollary \ref{finitezza delle labels}).

Similarly to the other cases, $\calcoloR$ does not ensure a
terminating proof search due to $(\cond^{+})$, which can be
applied without any control. We ensure the termination by putting
on $(\cond^{+})$ in $\calcoloR$ the same constraint used in the
other calculi. More precisely, it is easy to observe that it is
useless to apply the rule on the same conditional formula more
than once by using the same label $x$. By the invertibility of the
rules (Theorem \ref{invertibilità regole R}) we can assume,
without loss of generality, that two applications of $(\cond^{+})$
on $x$ are consecutive. We observe that the second application is
useless, since each of the conclusions has already been obtained
after the first application, and can be removed. We prevent
redundant applications of $(\cond^{+})$ by keeping track of labels
(worlds) in which a conditional $u: A \cond B$ has already been
applied in the current branch. To this purpose, we add to each
positive conditional a list of \emph{used} labels; we then
restrict the application of $(\cond^{+})$ only to labels not
occurring in the corresponding list.

Notice that also the rule $(<)$ copies its principal formula $x<y$
in the conclusion; however, this rule will be applied only a
finite number of times. This is a consequence of the side
condition of the rule application and the fact that the number of
labels in a tableau is finite.

The terminating calculus $\calcoloRterminante$ is obtained by
replacing the $(\cond^{+})$ rule in Figure \ref{figura calcolo
rational} with the one presented in Figure \ref{figura calcolo
rational TERMINANTE}.

\begin{figure}[h]

\linea

\begin{footnotesize}
\[
  \begin{prooftree}
    \Gamma, u: A \cond B^L
    \justifies \Gamma, x: \nott A, u: A \cond B^{L,x}
    \quad\quad \Gamma, x: \nott \bbox \nott A, u: A \cond B^{L,x}
    \quad\quad \Gamma, x: B, u: A \cond B^{L,x}
    \using (\cond^{+})
  \end{prooftree}
\]

with $\ x \not\appartiene L$
\end{footnotesize}

\linea

 \caption{The rule $(\cond^{+})$ in the tableau system
$\calcoloRterminante$.}
 \label{figura calcolo rational TERMINANTE}
\end{figure}

\noindent It is easy to prove the following structural properties
of $\calcoloRterminante$:

\begin{lemma}\label{Estensione assiomi a formule qualsiasi R}
For any set of formulas $\Gamma$ and any world formula $x: F$,
there is a closed tableau for $\Gamma, x: F, x: \nott F$.
\end{lemma}

\begin{provaallettore}
By induction on the complexity of the formula $F$.
\end{provaallettore}

\begin{lemma}[Height-preserving admissibility of
weakening]\label{ammissibilità weakening} Given any set of
formulas $\Gamma$ and any formula $\alpha$, if $\Gamma$ has a
closed tableau of height $h$ then $\Gamma, \alpha$ has a closed
tableau whose height is no greater than $h$.
\end{lemma}

\begin{provaallettore}
  By induction on the height of the closed tableau for $\Gamma$.
\end{provaallettore}

\noindent Moreover, one can easily prove that all the rules of
$\calcoloRterminante$ are height-preserving invertible, that is to
say:

\begin{theorem}[Height-preserving invertibility of the rules of $\calcoloRterminante$]\label{invertibilità regole R}
  Given any rule $(${\bf R}$)$ of $\calcoloRterminante$, whose premise is $\Gamma$
  and whose conclusions are $\Gamma_i$, with $i \leq 3$, we have
  that if $\Gamma$ has a closed tableau of height $h$, then there is a closed tableau,
  of height no greater than $h$, for each $\Gamma_i$, i.e. the rules of $\calcoloRterminante$
are height-preserving invertible.
\end{theorem}


\noindent The proof is in the Appendix. Since all the rules are
invertible, we have that in $\calcoloRterminante$ the order of
application of the rules is not relevant. Hence, no backtracking
is required in the calculus, and we can assume without loss of
generality that a given set of formulas $\Gamma$ has a unique
tableau.

Let us now prove that $\calcoloRterminante$ is sound.

\begin{theorem}[Soundness]\label{soundness R}
  $\calcoloRterminante$ is sound w.r.t. rational models, i.e. if there is a closed tableau for a set of formulas $\Gamma$,
  then $\Gamma$ is unsatisfiable.
\end{theorem}

\begin{provaposu}
By induction on the height of the closed tableau for $\Gamma$.
\begin{rosso}If $\Gamma$ is an axiom, then we distinguish two different
cases. First case: $x: P \appartiene \Gamma$ and $x: \nott P
\appartiene \Gamma$, therefore there is no $I$ such that $I(x) \in
\WW$ and $\emme, I(x) \modello P$ and $\emme, I(x) \not\modello
P$, and $\Gamma$ is unsatisfiable. Second case: $x < y \in \Gamma$
and $y < x \in \Gamma$, therefore there is a cycle in the
preference relation, and the smoothness condition cannot be
satisfied. Indeed, since in any rational model $<$ is irreflexive
an transitive, it can be easily shown that there cannot be an $I$
such that $I(x)<I(y)$ and $I(y)<I(x)$, which makes $\Gamma$
unsatisfiable\end{rosso}.

For the inductive step, we have to show that, for each rule $r$,
if all the conclusions of $r$ are unsatisfiable, then the premise
is unsatisfiable too. As usual, we prove the contrapositive, i.e.
we prove for each rule that, if the premise is satisfiable, so is
(at least) one of the conclusions. In order to save space, we only
present the most interesting case of $(\bbox^{-})$. Since the
premise is satisfiable, then there is a model $\emme$ and a
mapping $I$ such that $\emme \modello_I \Gamma, x: \nott \bbox
\nott A$. Let $w \appartiene \WW$ such that $I(x)=w$; this means
that $\emme, w \not\modello \bbox \nott A$, hence there exists a
world $w'<w$ such that $\emme, w' \modello A$. By the strong
smoothness condition, we have that there exists a \emph{minimal}
such world, so we can assume that $w' \appartiene Min_{<}(A)$,
thus $\emme, w' \modello \bbox \nott A$. In order to prove that
the conclusion of the rule is satisfiable, we construct a mapping
$I'$ as follows: let $y$ be a new label, not occurring in the
current branch; we define $(1) \ I'(u)=I(u)$ for all $u \diverso
y$ and $(2) \ I'(y)=w'$. Since $y$ does not occur in $\Gamma$, it
follows that $\emme \modello_{I'} \Gamma$. By Definition
\ref{verità labelled formulas}, we have that $\emme \modello_{I'}
y < x$ since $w'<w$. Moreover, since $I'(y)=w'$, we have that
$\emme \modello_{I'} y: A$ and $\emme \modello_{I'} y: \bbox \nott
A$. Finally, $\emme \modello_{I'} \Gammam{x}{y}$ follows from the
fact that $I'(y)<I'(x)$ and from the transitivity of $<$. The only
conclusion of the rule is then satisfiable in $\emme$ via $I'$.
\end{provaposu}

\noindent In order to prove the completeness of the calculus, we
introduce the notion of saturated branch and we show that
$\calcoloRterminante$ ensures a terminating proof search. As a
consequence, we will observe that the calculus introduces a finite
number of labels in a tableau, and this result will be used to
prove the completeness of the calculus.

\begin{definition}[Saturated branch]\label{saturatedbranchR}
We say that a branch ${\bf B}=\Gamma_1, \Gamma_2,$ $\dots,
\Gamma_n, \dots$ of a tableau is \emph{saturated} if the following
conditions hold:
\begin{enumerate}
\item for the boolean connectives, the condition of saturation is
defined
    in the usual way. For instance, if $x: A \andd B \appartiene \Gamma_i$ in ${\bf B}$, then
    there exists $\Gamma_j$ in ${\bf B}$ such that $x: A
    \appartiene \Gamma_j$ and $x: B \appartiene \Gamma_j$;

    \item if $x: A \cond B \in \Gamma_i$, then for
any label $y$ in ${\bf B}$,
    there exists $\Gamma_j$ in ${\bf B}$ such that
    either $y: \nott A \in \Gamma_j$ or  $y: \neg\bbox \neg A \in \Gamma_j$
    or $y: B \appartiene \Gamma_j$.

    \item if $x:\neg(A \cond B)\in \Gamma_i$, then there is a
    $\Gamma_j$ in ${\bf B}$ such that, for some $y$,
    $y: A\in \Gamma_j$, $y:\bbox \neg A \in \Gamma_j$, and $y:\neg B \in \Gamma_j$.

    \item if $x:\neg\bbox \neg A \in \Gamma_i$, then there exists
    $\Gamma_j$ in ${\bf B}$ such that, for some $y$,
    $y< x \in \Gamma_j$, $y: A \in
    \Gamma_j$ and $y:\bbox \neg A \in \Gamma_j$.

    \item if $x < y \in \Gamma_i$, then for all labels $z$ in ${\bf B}$,
there exists $\Gamma_j$ in
    ${\bf B}$ such that either  $z < y \in \Gamma_j$ or $x < z \in \Gamma_j$.

\end{enumerate}

\end{definition}

\begin{blu}
\noindent We say that a branch ${\bf B}=\Gamma_1, \Gamma_2, \dots,
\Gamma_n, \dots$ is \emph{open} if there is no $\Gamma_i$ in ${\bf
B}$ such that $x: F \in \Gamma_i$ and $x: \nott F \in \Gamma_i$.
We can easily show the following Lemma:
\end{blu}

\begin{lemma}\label{irreflexivity}
Given a tableau starting with $x_0: F$, for any open, saturated
branch ${\bf B}=\Gamma_1, \Gamma_2, \dots, \Gamma_n, \dots$, we
have that:

\begin{enumerate}

  \item if $z < y \in \Gamma_i$ in ${\bf B}$ and  $y < x \in \Gamma_j$ in ${\bf B}$, then there
  exists $\Gamma_k$ in ${\bf B}$ such that $z < x
\in \Gamma_k$;

  \item if $x: \bbox \neg A \in \Gamma_i$ in ${\bf B}$ and $y < x \in \Gamma_j$ in ${\bf B}$,
then there exists $\Gamma_k$ in ${\bf B}$ such that $y: \neg A \in
    \Gamma_k$ and $y: \bbox \neg A \in \Gamma_k$;

  \item for no $\Gamma_i$ in ${\bf B}$, $x < x \in \Gamma_i$.
\end{enumerate}

\end{lemma}


\noindent The proof is in the Appendix. Also in
$\calcoloRterminante$ we introduce the restriction on the order of
application of the rules adopted for the other systems (see
Definition \ref{restrizione sull'ordine di applicazione delle
regole}), that is to say: the application of the $(\bbox^{-})$
rule is postponed to the application of all propositional rules
and to the test of whether $\Gamma$ is an instance of $(\bf AX)$
or not.

Let us now show that $\calcoloRterminante$ ensures a terminating
proof search. In order to prove this result in a rigorous manner,
we proceed as follows: first, we introduce the complexity measure
on a set of formulas $\Gamma$ of Definition \ref{definizione
misura di complessità R}, denoted by $m(\Gamma)$, which consists
of five measures $c_1, c_2, c_3, c_4$ and $c_5$ in a lexicographic
order, and the auxiliary Definition \ref{definizioni aggiuntive};
then, we prove that each application of $\calcoloRterminante$'s
rules reduces this measure, until the rules are no longer
applicable, or leads to a closed tableau.

The complexity of a formula $cp(F)$ is defined in Definition
\ref{complessità formule}; moreover, we use square brackets
$[\dots]$ to denote \emph{multisets}.

\begin{definition}\label{definizioni aggiuntive}
Given an initial set of formulas $\Gamma_0$, we define:
\begin{itemize}
  \item the set $\ellep{+}$ of boxed formulas $\bbox \nott A$ that
  can be generated in a tableau for $\Gamma_0$, i.e. $\ellep{+}=\{\bbox \nott A
  \tc A \cond B \appartiene_{+} \Gamma_0\} \unione \{\bbox \nott A
  \tc A \cond B \appartiene_{-} \Gamma_0\}$. We let  $n_0=\tc \ellep{+} \tc$;
  \item  the \emph{multiset} $\ellep{-}$ of negated boxed formulas
  that can be generated in a tableau for $\Gamma_0$, i.e.
$\ellep{-}=[\nott \bbox \nott A \tc A \cond B \appartiene_{+}
      \Gamma_0]$. We let $k_0=\tc \ellep{-} \tc$;
\end{itemize}

\noindent Given a label $x$ and a set of formulas $\Gamma$ in the
tableau for the initial set $\Gamma_0$, we define:

  \begin{itemize}
  \item the number $n_x$ of positive boxed formulas $\bbox \nott A$
  not labelled by $x$, i.e. $n_x=n_0-\tc \{\bbox \nott A \appartiene \ellep{+} \tc x: \bbox \nott A \appartiene \Gamma\}
  \tc$;
  \item the number $k_x$ of negated boxed formulas $\nott \bbox
  \nott A$ not yet expanded in a world $x$, i.e. $k_x=k_0-\tc [\nott \bbox \nott A \appartiene \ellep{-}
  \tc y: \bbox \nott A \in \Gamma \ \mbox{and} \ y < x \in \Gamma]
    \tc$\footnote{Notice that, in case there are two
    positive conditionals $A \cond B$ and $A \cond C$ with the same antecedent, then
    the multiset $\elle^{\Gamma_0}_{\bbox^{-}}$ contains two
    instances of $\nott \bbox \nott A$. Therefore, if the rule
    $(\bbox^{-})$ is applied to $x: \nott \bbox \nott A$ (for instance, generated
    by an application of $(\cond^{+})$ in $x$ on $A \cond B$), then
    $k_x$ decreases only by 1 unit, whereas the second instance of
    $\nott \bbox \nott A$, i.e. the one ``associated'' with $A
    \cond C$, is still considered to be \emph{not expanded} in
    $x$, thus it still ``contributes'' to $k_x$.

          }.
  \end{itemize}
\end{definition}

\begin{definition}[Lexicographic order]\label{definizione misura di complessità R}
  We define $m(\Gamma)=\sx c_1, c_2, c_3, c_4, c_5 \dx$ where:
  \begin{itemize}
    \item $c_1=\tc\{u: A \cond B \appartiene_{-} \Gamma\}\tc$
    \item $c_2$ is the multiset given by
    $[c_2^{x_1},c_2^{x_2},\dots,c_2^{x_n}]$, where
      $x_1, x_2, \dots, x_n$ are the labels occurring
        in $\Gamma$ and, given a label $x$, $c_2^x$ is a pair
        $(n_x,k_x)$ in a lexicographic order ($n_x$ and $k_x$ are defined as in Definition \ref{definizioni
        aggiuntive}). We consider the integer multiset ordering given by
        $c_2$
    \item $c_3=\tc \{\sx x,A \cond B \dx \tc u: A \cond B^L \appartiene \Gamma \ \mbox{and} \ x \not\appartiene L\} \tc$
    \item $c_4=\sum_{z} c_4^z$, where $z$ occurs in $\Gamma$ and
    $c_4^z=\tc \{x < y \appartiene \Gamma \tc \{x < z, z < y\} \intersezione \Gamma = \vuoto\} \tc$
    \item $c_5=\sum_{x: F \appartiene \Gamma} cp(F)$
  \end{itemize}
  \noindent We consider the lexicographic order given by $m(\Gamma)$.
\end{definition}

\noindent Roughly speaking, $c_1$ is the number of negated
conditionals that can still be expanded in the tableau. The
application of $(\cond^{-})$ reduces $c_1$. $c_2$ keeps track of
positive conditionals \emph{which can still create a new world}.
The application of $(\bbox^{-})$ reduces $c_2$. \hide{Indeed, if
$(\cond^{+})$ is applied to $u: A \cond B$, this application
introduces a branch containing $x: \nott \bbox \nott A$; when a
new world $y$ is generated by an application of $(\bbox^{-})$ on
$x: \nott \bbox \nott A$, $y: \bbox \nott A$ is added to the
current set of formulas. If $(\cond^{+})$ is applied to $u: A
\cond B$ by using the new world $y$, then the conclusion where $y:
\nott \bbox \nott A$ is introduced is closed, by the presence of
$y: \bbox \nott A$. }$c_3$ represents the number of positive
conditionals not yet expanded in a world $x$: the application of
$(\cond^{+})$ reduces this measure, since the rule is applied to
$u: A \cond B^L$ by using $x$ \emph{only if} $x$ does not belong
to $L$, i.e. $u: A \cond B$ has not yet been expanded in $x$.
$c_4$ represents the number of relations $x<y$ not yet added to
the current branch: the application of $(<)$ reduces $c_4$, since
it applies the modularity of $<$ in case $x < y$ and, given $z$,
neither $z < y$ nor $x < z$ belong to the current set of formulas.
$c_5$ is the sum of the complexities of the world formulas in
$\Gamma$: an application of the rules for boolean connectives
reduces $c_5$.

First of all, we prove that the application of any rule of
$\calcoloRterminante$ reduces $m(\Gamma)$ or leads to a closed
tableau, as stated by the following Lemma:

\begin{lemma}\label{lemma diminuzione misura R}
  Let $\Gamma'$ be a set of formulas obtained as a conclusion of
  an application of a rule of $\calcoloRterminante$ to a set of
  formulas $\Gamma$. We have that either the tableau for $\Gamma'$
  is closed or $m(\Gamma')<m(\Gamma)$.
\end{lemma}

\begin{provaposu}
  We consider each rule of the calculus:
  \begin{itemize}
    \item $(\cond^{-})$: an application of this rule reduces $c_1$,
  since it is applied to a negated conditional $u: \nott (A \cond
  B)$ belonging to its premise which is removed from the
  conclusion;

    \item $(\bbox^{-})$: first of all, observe that $c_1$ is not
    augmented in the conclusion, since no negated conditional is
    added by the rule. The application of $(\bbox^{-})$ reduces
    $c_2$ or leads to a closed tableau. We are considering the following rule application:
    \[
      \begin{prooftree}
        \Gamma, x: \nott \bbox \nott A
        \justifies \Gamma, y < x, \Gammam{x}{y}, y: A, y: \bbox
        \nott A \using (\bbox^{-})
      \end{prooftree}
    \]
    \noindent where $y$ is a new label. $c_2$ in the premise, say $c_{2_p}$,
    is a multiset $[\dots, c_{2_p}^x, \dots]$, whereas in the
    conclusion we have to consider a measure, called $c_{2_c}$, of the form
    $[\dots, c_{2_c}^x, c_{2_c}^y, \dots]$. By the standard
    definition of integer multiset ordering, we prove that either
    $c_{2_c} < c_{2_p}$ (by showing that $c_{2_c}^x < c_{2_p}^x$ and $c_{2_c}^y <
    c_{2_p}^x$, i.e. we replace an integer $c_{2_p}^x$ with two
    smaller integers, see \cite{multisetordering} for details on integer multiset orderings)
    or that the procedure leads to a closed tableau. We conclude the proof as follows:
    \begin{itemize}

    \begin{blu}
      \item let us consider $c_{2_c}^y$ and $c_{2_p}^x$. By definition,
      $c_{2_c}^y$ is a pair $(n_{y_c},k_{y_c})$ and $c_{2_p}^x$ is a pair
      $(n_{x_p},k_{x_p})$. We distinguish two cases: if $x: \bbox \nott A \not\in \Gamma$,
      then we easily prove that $n_{y_c} <
      n_{x_p}$, since  $y: \bbox \nott A$ belongs to the
      conclusion: therefore, the number of boxed formulas $\bbox
      \nott A$  not occurring with label $y$ is smaller than the number of
      boxed formulas not occurring with label $x$, and we are
      done (remember that all the positive boxed formulas labelled by $x$
      are also labelled by $y$ in the conclusion, by the presence of $\Gammam{x}{y}$);
      if $x: \bbox \nott A \in \Gamma$, then the application
      of $(\bbox^{-})$ leads to a node containing both $y: A$ and
      $y: \nott A$ and, by the restriction on the order of
      application of the rules (see Definition \ref{restrizione sull'ordine di applicazione delle
      regole}), the procedure terminates building a closed
      tableau;

      \item let us consider $c_{2_c}^x$ and $c_{2_p}^x$. It is easy to
      observe that $n_{x_p} = n_{x_c}$, since no formula $x: \bbox
      \nott A$ is added nor removed in the conclusion (the positive boxed formulas labelled by $x$
    are the same in both the premise and the conclusion). We conclude since $k_{x_c}<k_{x_p}$, since $x: \nott
    \bbox \nott A$ has been expanded in $x$, so it ``contributes'' to
    $k_{x_p}$ whereas it does not to $k_{x_c}$.

          \end{blu}

    \end{itemize}

    \item $(\cond^{+})$: first of all, notice that $c_1$ and $c_2$
    cannot be higher in the conclusions than in the premise.
    Notice that the addiction of $x: \nott \bbox \nott A$ in the inner conclusion does not
    increase $c_2$, since $\nott \bbox \nott A$ is a negated boxed
    formula still to be considered in both the premise and the
    conclusions.
    The application of $(\cond^{+})$ reduces $c_3$. Suppose that
    this rule is applied to $u: A \cond B^L$ by using label $x$ in
    the conclusions; by the restriction in Figure \ref{figura calcolo rational
    TERMINANTE}, this means that $x \not\appartiene L$, so $\sx x, A
    \cond B \dx$ belongs to the set whose cardinality determines
    $c_3$ in the premise. Obviously, since $x$ is added to $L$ for $u: A \cond B$ in
    the three conclusions of the rule, we can easily observe that $\sx x, A
    \cond B \dx$ does no longer belong to the set in the definition of
    $c_3$ in the conclusions: $c_3$ is then smaller in the
    conclusions than in the premise, and we are done;

    \item $(<)$: the application of $(<)$ reduces $c_4$, whereas
    $c_1$, $c_2$, and $c_3$ cannot be augmented (at most formulas
    $z: \bbox \nott A$ are added in the conclusions by $\Gammam{z}{x}$ and $\Gammam{y}{z}$, reducing
    $c_2$). To conclude the proof, just observe that, given a label $z$ and a formula $x<y$,
    $(<)$ is applied if $\{x<z,z<y\} \intersezione \Gamma =
    \vuoto$, i.e. $x < y$ belongs to the set used to define
    $c_4^z$ in the premise, say $c_{4_p}^z$. When the rule is applied, in the left premise $z<y$
    is added, and $x < y$ does no longer belong to the set used to
    define $c_4^z$  in the conclusion, say $c_{4_c}^z$. Therefore, $c_{4_c}^z <
    c_{4_p}^z$. The same for the right premise, and we are done;

    \item rules for the boolean connectives: these rules do not
    increase values of $c_1$, $c_2$, $c_3$, and $c_4$. Their
    application reduces $c_5$, since the (sum of) complexity of the
    subformula(s) introduced in the conclusion(s) is lower then
    the complexity of the principal formula to which the rule is
    applied.
  \end{itemize}
\end{provaposu}

\noindent Now we can prove that $\calcoloRterminante$ ensures a
terminating proof search:

\begin{theorem}[Termination of $\calcoloRterminante$]\label{terminazione del calcolo R}
  Let $\Gamma$ be a finite set of formulas, then any tableau generated
  by  $\calcoloRterminante$ is finite.
\end{theorem}

 \begin{provaposu} Let $\Gamma'$ be obtained by an application of
   a rule of $\calcoloRterminante$ to a
  premise $\Gamma$. By Lemma \ref{lemma diminuzione misura R} we have
  that either the procedure leads to a closed tableau (and in this
  case we are done) or we have that $m(\Gamma')<m(\Gamma)$.
  This means that, similarly to the case of \Pe, a finite number of applications of the rules
  leads either to close the branch or to a
  node whose measure is as follows: $\sx 0,[(0,0),(0,0),\dots,(0,0)],0,0,c_{5_{\mathit{min}}} \dx$,
  where $c_{5_{\mathit{min}}}$ is the minimal value that $c_5$ can assume for $\Gamma$. This
  means that no rule of $\calcoloRterminante$ is further
  applicable\hide{, with the exception of $(\bbox^{-})$ which is further
  applicable, but only a \emph{finite number of times}}. This is a consequence of the following facts:
  \begin{itemize}
    \item $(\cond^{-})$ is no longer applicable, since $c_1=0$;
    \item since $c_2=[(0,0),(0,0),\dots,(0,0)]$, given any label $x$,
    we have that $c_2^x=(0,0)$. \hide{Therefore, we can observe that $n_x=n_0-\{\bbox \nott A \appartiene \ellep{+} \tc x: \bbox \nott
A \appartiene
      \Gamma\}=0$, that is to say there is a formula $x: \bbox \nott A \appartiene \Gamma$ for each $\bbox \nott A$
      that can be generated. Therefore, the $(\bbox^{-})$ rule is
      applicable only one more time on each negated conditional.
      Indeed, if $(\bbox^{-})$ would be applicable to some $x: \nott \bbox \nott
      A \appartiene \Gamma$, then the procedure would lead to a node containing both $y: A$
      and $y: \nott A$ (and $y$ is a new label), since $x: \bbox \nott A \in \Gamma$; then, by the
      restriction on the order of application of the rules (see
      Definition \ref{restrizione sull'ordine di applicazione delle
      regole}), the procedure terminates building a closed tableau, and $(\bbox^{-})$ is not
      further applied;} Therefore, the $(\bbox^{-})$ rule is no longer applicable, since we can observe that
      there is no $x: \nott \bbox \nott A \in \Gamma$ (by the fact that $k_x=0$);
    \item the rule $(\cond^{+})$ is not further applicable since
    $c_3=0$; indeed, $c_3=0$ means that all formulas $A
    \cond B$ have already been expanded in each world $x$;
    \item $c_4=0$, i.e. for all formulas $x<y$, given any label
    $z$, we have that either $z<y \appartiene \Gamma$ or
    $x<z \appartiene \Gamma$, thus the $(<)$ rule is not further
    applicable;
    \item since $c_5$ assumes its \emph{minimal} value $c_{5_{\mathit{min}}}$, no rule for a boolean connective
    is further applicable. If a boolean rule is applicable, then
    its application reduces the value of $c_5$ in its
    conclusion(s) by Lemma \ref{lemma diminuzione misura R}, against the minimality of
    $c_{5_{\mathit{min}}}$in the premise.
  \end{itemize}
  \end{provaposu}

\begin{blu}
\noindent As a consequence of Theorem \ref{terminazione del
calcolo R}, we can observe that the tableau for a given set of
formulas $\Gamma_0$ contains a finite number of labels, since all
the branches in a tableau generated by $\calcoloRterminante$ are
finite.
\end{blu}

\begin{corollary}\label{finitezza delle labels}
Given a set of formulas $\Gamma$, the tableau generated by
$\calcoloRterminante$ for $\Gamma$ only contains a finite number
of labels.
\end{corollary}

\noindent Let us now show that $\calcoloRterminante$ is complete
with respect to the semantics:

\begin{theorem}[Completeness]\label{completeness R}
  $\calcoloRterminante$ is complete w.r.t. rational models, i.e.
  if a set of formulas $\Gamma$ is unsatisfiable, then it has a
  closed tableau in $\calcoloRterminante$.
\end{theorem}

\begin{provaposu}
We show the contrapositive, i.e. if there is no closed tableau for
$\Gamma$, then $\Gamma$ is satisfiable. Consider the tableau
starting with the set of formulas $\{x: F$ such that $F \in \Gamma
\}$ and any open, saturated branch ${\bf B}=\Gamma_1, \Gamma_2,
\dots, \Gamma_n$ in it. Starting from ${\bf B}$, we build a
canonical model $\emme=\langle \WW_B, <, V\rangle$ satisfying
$\Gamma$, where:
\begin{itemize}
  \item $\WW_B$ is the set of labels that appear in the branch ${\bf B}$;
  \item for each $ x, y \in \WW_B$, $x < y$ iff there exists $\Gamma_i$ in
${\bf B}$ such that $x < y \in \Gamma_i$;
  \item for each $ x \in \WW_B$,
$V(x) = \{P \in \mathit{ATM} \mid \ \mbox{there is} \ \Gamma_i \
\mbox{in} \ {\bf B} \ \mbox{such that} \ x: P \in \Gamma_i\}$.
\end{itemize}
\noindent We can easily prove that:

\noindent $(i)$ by Corollary \ref{finitezza delle labels}, we have
that $\WW_B$ is finite;

\noindent $(ii)$ $<$ is an irreflexive, transitive and modular
relation on $\WW_B$ satisfying the smoothness condition.
Irreflexivity, transitivity and modularity are obvious, given
Definition \ref{saturatedbranchR} and Lemma \ref{irreflexivity}
above. Since $<$ is irreflexive and transitive, it can be easily
shown that it is also acyclic. This property together with the
finiteness of $\WW_B$ entails that $<$ cannot have infinite
descending chains. In turn this last property together with the
transitivity of $<$ entails the smoothness condition.

\noindent $(iii)$ We show  that, for all formulas $F$ and for all
$\Gamma_i$ in ${\bf B}$, (i) if $x: F \in \Gamma_i$ then $\emme, x
\models F$ and (ii) if $x: \neg F \in \Gamma_i$ then $\emme, x
\not\models F$. The proof is by induction on the complexity of the
formulas. If $F \appartiene \mathit{ATM}$ this immediately follows
from definition of $V$. For the inductive step, we only present
the case of $F = A \cond B$. The other cases are similar and then
left to the reader. Let $x : A \cond B \in \Gamma_i$. By
Definition \ref{saturatedbranchR}, we have that, for all $y$,
there is $\Gamma_j$ in ${\bf B}$ such that either $y: \neg A \in
\Gamma_j$ or $y: B \in \Gamma_j$ or $y: \neg \bbox \neg A \in
\Gamma_j$. We show that for all $y \appartiene Min_{<}(A)$,
$\emme, y \modello B$. Let $y \in Min_<(A)$. This entails that
$\emme, y \models A$, hence $y: \neg A \not\in \Gamma_j$.
Similarly, we can show that $y: \neg \bbox \neg A \not\in
\Gamma_j$. It follows that $y : B \in \Gamma_j$, and by inductive
hypothesis $\emme, y \models B$. (ii) If $x : \neg (A \cond B) \in
\Gamma_i$, since ${\bf B}$ is saturated, there is a label $y$ in
some $\Gamma_j$ such that $y: A\in \Gamma_j$, $y:\bbox \neg A \in
\Gamma_j$, and $y:\neg B \in \Gamma_j$. By inductive hypothesis we
can easily show that $\emme, y \modello A$, $\emme, y \modello
\bbox \nott A$, hence $y \in Min_<(A)$, and $\emme, y \not\models
B$, hence $\emme, x \not\models A \cond B$.

Since $\calcoloRterminante$ makes use of the restriction in Figure
\ref{figura calcolo rational TERMINANTE}, we have to show that
this restriction preserves the completeness. We have only to show
that if $(\cond^{+})$ is applied twice on the same conditional $A
\cond B$, in the same branch, by using the same label $x$, then
the second application is useless. Since all the rules are
invertible (Theorem \ref{invertibilità regole R}), we can assume,
without loss of generality, that the two applications of
$(\cond^{+})$ are consecutive. We conclude that the second
application is useless, since each of the conclusions has already
been obtained after the first application, and can be removed.

\end{provaposu}

\noindent By Theorem \ref{soundness R} above and by the
construction of the model done in the proof of Theorem
\ref{completeness R} just above, we can show the following
Corollary.

\begin{corollary}[Finite model property]\label{corollario-finitezzar} \Ra \ has the finite model
property.
\end{corollary}

\hide{
\begin{corollary}\label{corollario-aciclicitàR} If $\Gamma$ is
satisfiable in \Ra, then it is satisfiable in an acyclic model.
\end{corollary}

\noindent From the two Corollaries above it immediately follows
that:
\begin{corollary}\label{infinite-chainsR}
If $\Gamma$ is satisfiable in \Ra , then it is satisfiable in a
model in which $<$ is transitive and does not have infinite
descending chains.
\end{corollary}
}

\subsection{Decision Procedure and Optimal Proof Search for \Ra}
In this section we define a systematic procedure which allows the
satisfiability problem for \Ra \ to be decided in
nondeterministically polynomial time, in accordance with the known
complexity results for this logic.

Let $n$ be the size of the starting set $\Gamma$ of which we want
to verify the satisfiability. The number of applications of the
rules is proportional to the number of labels introduced in the
tableau. In turn, this is $O(2^n)$ due to the interplay between
the rules $(\cond^{+})$ and $(\bbox^{-})$. Hence, the complexity
of the calculus $\calcoloRterminante$ is exponential in  $n$.\\
In order to obtain a better complexity bound for validity in \Ra \
we provide the following procedure. Intuitively, we do not apply
$(\bbox^{-})$ to all negated boxed formulas, but only to formulas
$y: \nott \bbox \nott A$ not already expanded, i.e. such that $z:
A, z: \bbox \nott A$ do not belong to the current branch. As a
result, we build a \emph{small} model for the initial set of
formulas in accordance with Theorem \ref{small model}. This is
made possible by the modularity of $<$ in \Ra.

Let us define a nondeterministic procedure
\texttt{CHECK}($\Gamma$) to decide whether a given set of formulas
$\Gamma$ is satisfiable. Let \texttt{EXPAND}($\Gamma$) be a
procedure that returns one saturated expansion of $\Gamma$ w.r.t.
all static rules. In case of a branching rule, \texttt{EXPAND}
nondeterministically selects (guesses) one conclusion of the rule.

\vspace{0.2cm}

\linea

\begin{footnotesize}
\noindent \texttt{CHECK}($\Gamma$)

\noindent 1. $\Gamma \longleftarrow$ \texttt{EXPAND}$(\Gamma)$;

\noindent 2. {\bf if} $\Gamma$ contains an axiom {\bf then return}
\texttt{UNSAT};

\noindent 3. $\Gamma \longleftarrow$ result of applying
$(\cond^{-})$ to each negated conditional in $\Gamma$;

\noindent 4. $\Gamma \longleftarrow$ \texttt{EXPAND}$(\Gamma)$;

\noindent 5. {\bf if} $\Gamma$ contains an axiom {\bf then return}
\texttt{UNSAT};

\noindent {\bf while} $\Gamma$ contains a $y: \nott \bbox \nott A$
not marked as \texttt{CONSIDERED} {\bf do}

\indent 6. select $y: \nott \bbox \nott A \appartiene \Gamma$ not
already marked as \texttt{CONSIDERED};

\indent\indent 6a. {\bf if} there is $z$ in $\Gamma$ such that $z:
A \appartiene \Gamma$ and $z: \bbox \nott A \appartiene \Gamma$

\indent\indent\indent {\bf then} 6a'. add $z<y$ and
$\Gammam{y}{z}$ to $\Gamma$;

\indent\indent\indent {\bf else} 6a''. $\Gamma \longleftarrow$
result of applying $(\bbox^{-})$ to $y: \nott \bbox \nott A$;

\indent\indent 6b. mark $y: \nott \bbox \nott A$ as
\texttt{CONSIDERED};

\indent 7. $\Gamma \longleftarrow$ \texttt{EXPAND}$(\Gamma)$;

\indent 8. {\bf if} $\Gamma$ contains an axiom {\bf then return}
\texttt{UNSAT};

\noindent {\bf endWhile}

\noindent 9. {\bf return} \texttt{SAT};

\end{footnotesize}

\linea \vspace{0.2cm}

\noindent Observe that the addition of the set of formulas $z<y,
\Gammam{y}{z}$ in step 6a' could be omitted and it has been added
mostly to enhance the understanding of the procedure. Indeed, the
rule ($<$), which is applied at each iteration to assure
modularity, already takes care of adding such formulas. The
procedure \texttt{CHECK} nondeterministically builds an open
branch for $\Gamma$.

\begin{theorem}[Soundness and completeness of the procedure]
The above procedure is sound and complete w.r.t. the semantics.
\end{theorem}

\begin{provaposu}
  (\emph{Soundness}). We prove that if the initial set of formulas
  $\Gamma$ is satisfiable, then the above procedure returns
  \texttt{SAT}. More precisely, we prove that each step of the
  procedure preserves the satisfiability of $\Gamma$. As far as
  \texttt{EXPAND} is concerned, notice that it only applies the
  static rules of $\calcoloRterminante$ and the soundness follows
  from the fact that these rules preserve satisfiability (see
  Theorem
  \ref{soundness R}). Consider now
  step 6. Let $y: \nott \bbox \nott A$ the formula selected in this step. If
  $(\bbox^{-})$ is applied to $y: \nott \bbox \nott A$ (step 6a'')
  we are done, since $(\bbox^{-})$ preserves satisfiability (see
  Theorem
  \ref{soundness R}). If $\Gamma$
  already contains $z: A, z: \bbox \nott A$, then step 6a' is
  executed, and the relation $z<y$ is added. In this case we
  reason as follows. Since $\Gamma$ is satisfiable, we have that there is a model $\emme$
    and a mapping $I$  such that $(1) \  \emme, I(y) \modello \nott \bbox
    \nott A$ and $(2) \ \emme, I(z) \modello A$ and $\emme, I(z) \modello \bbox
    \nott A$.
  We can observe that $I(z)<I(y)$ in $\emme$. Indeed, by the truth
  condition of $\nott \bbox \nott A$
  and by the strong smoothness
  condition, we have that there exists $w$ such that $w<I(y)$ and $\emme,
  w \modello A, \bbox \nott A$. By modularity of $<$, either
  $1. \ w<I(z)$ or $2. \ I(z)<I(y)$. $1$ is impossible, since
  otherwise we would have $\emme, w \modello \nott A$, which
  contradicts $\emme,
  w \modello A$. Hence, $2$ holds. Therefore, we can conclude that
  step 6a' preserves satisfiability.

  \noindent (\emph{Completeness}). It can be easily shown that in
  case the procedure  above returns \texttt{SAT}, then the branch
  built is saturated (see Definition \ref{saturatedbranchR}).
  Therefore, we can build a canonical model for the initial
  $\Gamma$, as done in the proof of Theorem \ref{completeness R}.
\end{provaposu}

\begin{theorem}[Complexity of the \texttt{CHECK}
procedure]\label{complessità check} By means of the procedure
\texttt{CHECK} the satisfiability of a set of formulas of logic
\Ra \ can be decided in nondeterministic polynomial time.
\end{theorem}

\begin{provaposu}
Observe that the procedure generates at most {\em O(n)} labels by
applying the rule $(\cond^{-})$ (step 3) and that the while loop
generates at most one new label for each $\nott \bbox \nott A$
formula. Indeed, the rule $(\bbox^{-})$ is applied to a labelled
formula $y: \nott \bbox \nott A$ to generate a new world only if
there is not a label $z$ such that $z: A \appartiene \Gamma$ and
$z: \bbox \nott A \appartiene \Gamma$ are already on the branch.
In essence, the procedure does not add a new minimal $A$-world on
the branch if there is already one. As the number of different
$\nott \bbox \nott A$ formulas is at most $O(n)$, then the while
loop can add at most $O(n)$ new labels on the branch. Moreover,
for each different label $x$, the expansion step can add at most
$O(n)$ formulas $x: \nott \bbox \nott A$ on the branch, one for
each positive conditional $A \cond B$ occurring in the set
$\Gamma$. We can therefore conclude that the while loop can be
executed at most $O(n^2)$ times.

As the number of generated labels is at most $O(n)$, by the
subformula property, the number of labelled formulas on the branch
is at most $O(n^2)$. Hence, the execution of step 6a has
complexity $O(n^2)$. The execution of the nondeterministic
procedure \texttt{EXPAND} has complexity $O(n^2)$, including a
guess of size $O(n^2)$, whereas to verify if $\Gamma$ contains an
axiom has complexity $O(n^4)$ (since it requires to check whether,
for each labelled formula $x: P \in \Gamma$,  the formula $x:\nott
P$ is also in
 $\Gamma$, and $\Gamma$ contains at most $O(n^2)$ labelled formulas). We can therefore conclude
that the execution of the \texttt{CHECK} procedure requires at
most $O(n^6)$ steps.
\end{provaposu}

By Theorem \ref{complessità check}, the validity problem for \Ra \
is in {\bf coNP}. {\bf coNP}-hardness is immediate, since \Ra \
includes classical propositional logic. Thus, we can conclude
that:

\begin{theorem}[Complexity of {\bf R}]\label{complessità di R}
  The problem of deciding the validity for rational logic {\bf R} is
  {\bf coNP}-complete.
\end{theorem}

\section{Conclusions}

In this paper, we have presented tableau calculi for all of the
KLM logical systems for default reasoning. Some preliminary
results have been presented in \cite{lpar2005} and
\cite{jelia2006}. We have given a tableau calculus for rational
logic \Ra, preferential logic \Pe, loop-cumulative logic \Cl, and
cumulative logic \Cu. The calculi presented give a decision
procedure for the respective logics. Moreover, for \Ra, \Pe \ and
\Cl \ we have shown that we can obtain {\bf coNP} decision
procedures by refining the rules of the respective calculi. In
case of \Cu, we obtain a decision procedure by adding a suitable
loop-checking mechanism. Our procedure gives an hyper exponential
upper bound. Further investigation is needed to get a more
efficient procedure. On the other hand, we are not aware of any
tighter complexity bound for this logic.

All the calculi presented in this paper have been implemented by a
theorem prover called KLMLean. KLMLean (not presented here) is a
SICStus Prolog implementation of the tableau calculi introduced in
this paper, and it is inspired to the ``lean'' methodology
\cite{leanTAP,leanTAP-Rev,lean2}, whose basic idea is to write
short programs and exploit the power of Prolog's engine as much as
possible. To the best of our knowledge, KLMLean is the first
theorem prover for KLM logics.

Artosi, Governatori, and Rotolo \cite{Governatori:02} develop a
labelled tableau calculus for {\bf C}. Their calculus is based on
the interpretation of \Cu \ as a conditional logic with a
selection function semantics. As a major difference from our
approach, their calculus makes use of labelled formulas, where the
labels represent possible worlds or sets of possible worlds. World
labels in turn are annotated by formulas to express minimality
assumptions, e.g. they represent by a label $w^A$ the fact that
$w$ is a minimal $A$-world, or in terms of the selection function,
belongs to $f(A,u)$. They use then a sophisticated unification
mechanism on the labels to match two annotated worlds, e.g. $w^A,
w^B$; observe that by CSO (which is equivalent to CUT+CM), the
equivalence of $A$ and $B$ might also be enforced by the
conditionals  contained in a tableau branch. Even if they do not
discuss decidability and complexity issues, their tableau calculus
should give a decision procedure for \Cu.

In \cite{GGOSTableaux2003} and \cite{Berlino} it is defined a
labelled tableau calculus for the logic {\bf CE} and some of its
extensions. The flat fragment of {\bf CE} corresponds to the
system {\bf P}. The similarity between the two calculi lies in the
fact that both approaches use a modal interpretation of
conditionals. The major difference is that the calculus presented
here does not use labels, whereas the one proposed in
\cite{GGOSTableaux2003} does. A further difference is that in
\cite{GGOSTableaux2003} the termination is obtained by means of a
loop-checking machinery, and it is not clear if it matches
complexity bounds and if it can be adapted in a simpler
way to \Cl \ and to \Cu. \\
\indent Lehmann and Magidor \cite{whatdoes} propose  a
non-deterministic algorithm that, given a finite set $K$ of
conditional assertions $C_i \cond D_i$ and a conditional assertion
$A \cond B$, checks if $A \cond B$ is not entailed by $K$ in the
logic \Pe. This is an abstract algorithm useful for theoretical
analysis, but practically unfeasible, as it requires to guess sets
of indexes and propositional evaluations. They conclude that
entailment in \Pe \ is {\bf coNP}, thus obtaining a complexity
result similar to ours. However, it is not easy to compare their
algorithm with our calculus, since the two approaches are
radically different. As far as the complexity result is concerned,
notice that our  result is more general than theirs, since  our
language is richer: we consider boolean combinations of
conditional assertions (and also combinations with propositional
formulas), whereas they do not. As remarked by Boutilier
\cite{Boutilier:94}, this more general result is not an obvious
consequence of  the more restricted one. Moreover, we prove the
{\bf coNP} result also for the system {\bf CL}.
At the best of our knowledge, this result was unknown up to now.\\
\indent We plan to extend our calculi to first order case. The
starting point will be the analysis of first order preferential
and rational logics by Friedman, Halpern and Koller in
\cite{halpern-first-order}.

\begin{acks}
This research has been partially supported by ``\emph{Progetto
Lagrange - Fondazione CRT}'' and by the projects ``\emph{MIUR
PRIN05: Specification and verification of agent interaction
protocols}'' and \emph{``GALILEO 2006: Interazione e coordinazione
nei sistemi multi-agenti''.}
\end{acks}

\bibliographystyle{acmtrans}
\bibliography{KLMToCL}

\begin{received}
  Received November 2006;
  revised ----;
  accepted ----
\end{received}

\elecappendix

\setcounter{section}{1}

\section*{Appendix with Proofs}

\vspace{0.2cm}


\begin{pf*}{\noindent Proof of Theorem \ref{small model}.}
\emph{For any $\Gamma \subseteq \elle$, if $\Gamma$ is satisfiable
in a rational model, then it is satisfiable in a rational model
containing at most $n$ worlds, where $n$ is the size of $\Gamma$,
i.e. the length of the string representing $\Gamma$.}

\vspace{0.2cm}

\noindent Let $\Gamma$ be satisfiable in a rational model
$\emme=\sx \WW, <, V \dx$, i.e. $\emme, x_0 \models \Gamma$ for
some $x_0 \appartiene \WW$. We build the model $\emme' = \langle
\WW', <', V'\rangle$ as follows:
\begin{itemize}
\item We build the set of worlds $\WW'$ by means of the following procedure:
\begin{enumerate}
    \item $\WW' \longleftarrow \{x_0\}$;

    \item {\bf for each} $A_i \cond B_i \in_- \Gamma$ {\bf do}

      \begin{itemize}
        \item choose  $x_i \in \WW$ s.t. $x_i \in Min_<(A_i)$ and $\emme, x_i \not\models B_i$;
        \item $\WW' \longleftarrow \WW' \cup \{x_i\}$;
      \end{itemize}

    \item {\bf for each} $A_i \cond B_i \in_+ \Gamma$ {\bf do}

     \quad {\bf if} $Min_<(A_i) \neq \emptyset$, and there is no $x_i$
    s.t. $x_i \in Min_<(A_i)$ and

    \qquad $x_i$ is already in $\WW'$ {\bf then}
     \begin{itemize}
     \item choose any  $x_i \in Min_<(A_i)$;
     \item $\WW' \longleftarrow \WW' \cup \{x_i\}$;
     \end{itemize}
\end{enumerate}

\item For all $x_i, x_j \in \WW'$, we let $x_i <' x_j$ if $x_i < x_j$;

\item For all $x_i \in \WW'$, we let $V'(x_i) = V(x_i)$.
\end{itemize}

In order to show that $\WW'$ is a rational model satisfying
$\Gamma$, we can show the following Facts:

\begin{fact}\label{fact0Vale}
$|\WW'| \leq n$.
\end{fact}

\begin{pf*}{Proof of Fact \ref{fact0Vale}.}
The proof immediately follows by construction of $\WW'$.
\provafatto{\ref{fact0Vale}}
\end{pf*}

\begin{fact}\label{fact1Vale}
$\emme'$ is a rational model, since  $<'$ is irreflexive,
transitive, modular and satisfies the Smoothness Condition.
\end{fact}

\begin{pf*}{Proof of Fact \ref{fact1Vale}.}
Irreflexivity, transitivity and modularity of $<'$ obviously
follow from the definition of $<'$. The Smoothness Condition  is
ensured by the fact that $<'$ does not have infinite descending
chains, since $<$ does not have. \provafatto{\ref{fact1Vale}}
\end{pf*}

\begin{fact}\label{fact2Vale}
For all $x_i \in \WW'$, for all propositional formulas $A$,
$\emme, x_i \models A$ iff $\emme', x_i \models A$.
\end{fact}

\begin{pf*}{Proof of Fact \ref{fact2Vale}.}
 By induction on the complexity of $A$.\provafatto{\ref{fact2Vale}}
\end{pf*}

\begin{fact}\label{fact3Vale}
For all $x_i \in \emme'$, for all formulas $A$ s.t. $A$ is the
antecedent of some conditional occurring in $\Gamma$, we have that
$x_i \in Min_{<'}(A)$ iff $x_i \in Min_<(A)$.
\end{fact}

\begin{pf*}{Proof of Fact \ref{fact3Vale}.}
First, we prove that if $x_i \in Min_{<'}(A)$, then $x_i \in
Min_<(A)$. Let $x_i \in Min_{<'}(A)$. Suppose that $x_i \not\in
Min_{<}(A)$. Since $A$ is the antecedent of a conditional in
$\Gamma$, by construction of $\WW'$, $\WW'$ contains $x_j$, $x_j
\diverso x_i$, s.t. $x_j \in Min_<(A)$  in $\emme$. Since $\emme,
x_i \modello A$ and $x_j \in Min_<(A)$, we have that $x_j < x_i$.
By Fact \ref{fact2Vale}, $\emme', x_j \models A$, and by the
definition of $<'$, $x_j <' x_i$, which contradicts the assumption
that $x_i \in Min_<'(A)$. We conclude that $x_i \in Min_{<}(A)$ in
$\emme$.

Now we prove that if $x_i \in Min_<(A)$, then $x_i \in
Min_{<'}(A)$. Let $x_i \in Min_<(A)$ in $\emme$. Suppose that $x_i
\not\in Min_{<'}(A)$. Then there is $x_j$ s.t. $\emme', x_j
\models A$ and $x_j <' x_i$. By Fact \ref{fact2Vale} (since $A$ is
a propositional formula), also $\emme, x_j \models A$, and by
definition of $<'$, $x_j < x_i$, which contradicts the assumption
that $x_i \in Min_<(A)$. Hence, $x_i \in Min_{<'}(A)$.
\provafatto{\ref{fact3Vale}}
\end{pf*}

\begin{fact}\label{fact4Vale}
For all conditional formulas $(\nott) A \cond B$ occurring in
$\Gamma$, if $\emme, x_0 \models (\neg) A \cond B$, then $\emme',
x_0 \models (\neg) A \cond B$.
\end{fact}

\begin{pf*}{Proof of Fact \ref{fact4Vale}.}
We distinguish the two cases:
\begin{itemize}
\item $\emme, x_0 \models \neg (A \cond B)$: by construction of $\WW'$,
there is $x_i \in \WW'$ s.t. $x_i \in Min_<(A)$ and $\emme, x_i
\not\models B$. By Facts \ref{fact2Vale} and \ref{fact3Vale}, $x_i
\in Min_{<'}(A)$ and $\emme', x_i \not\models B$, hence $\emme',
x_0 \models \neg (A \cond B)$.

\item  $\emme, x_0 \models A \cond B$:  consider any $x_i \in
Min_{<'}(A)$, by Fact \ref{fact3Vale} $x_i \in Min_{<}(A)$, hence
$\emme, x_i \models B$, and, by Fact \ref{fact2Vale}, $\emme', x_i
\models B$. We conclude that $\emme', x_0 \models A \cond B$.
\end{itemize}\provafatto{\ref{fact4Vale}}
\end{pf*}

\noindent By the Facts above, we have shown that $\Gamma$ is
satisfiable in a rational model containing at most $n$ worlds,
hence the Theorem follows.

\vspace{0.1cm}  \begin{flushright} $\blacksquare$
\end{flushright}

\end{pf*}



\begin{pf*}{Proof of Theorem
\ref{teorema nicola}.} \emph{Let $\Gamma$ be any set of formulas,
if $\Gamma$ is satisfiable then it has a multi-linear model.}

\vspace{0.2cm}

\noindent Let us make explicit the negated conditionals in
$\Gamma$ by rewriting it as
$$\Gamma = \Gamma', \nott (C_1\cond D_1), \ldots,
\nott (C_k \cond D_k).$$ Assume $\Gamma$ is satisfiable, then
there is a model $\emme = \sx \WW,<,V \dx$ and $x \in \WW$, such
that $\emme, x \models \Gamma$. We have that there are
$y_1,\ldots,y_k\in \WW$, such that for each $j=1,\ldots,k$,
$$y_j \in Min_{<}(C_j) \ \mbox{and } \emme,y_j \not\models D_j$$
We define for $x$ and each $y_j$:
$$\WW_x = \{z\in \WW \mid z < x\} \cup \{x\}$$
$$\WW_{y_j} = \{z\in \WW \mid z < y_j\} \cup \{y_j\}$$
Moreover, we consider for each $j=1,\ldots,k$ a renaming function
(i.e. a bijection) $f_j$ whose domain is $\WW_{y_j}$ that makes a
copy $\WW_{f_j(y_j)}$ of $\WW_{y_j}$ which is (i) disjoint from
$\WW_x$, (ii) disjoint from any $\WW_{y_l}$, and (iii) disjoint
from any other $\WW_{f_l(y_l)}$ with $l\not=j$. Observe that we
make $k$ disjoint sets $\WW_{f_j(y_j)}$ even if some $y_j$'s
coincide among themselves or coincide with $x$. We define a model
$\emme' =\sx \WW', <', V' \dx$ as follows:
$$\WW' = \WW_x \cup \WW_{f_1(y_1)} \ldots \WW_{f_k(y_k)}$$

\noindent The relation $<'$ is defined as follows:
\begin{center}
  $u <' v$ iff (i) $u,v \in \WW_x$ and $u < v$,\\ or (ii) $u, v \in
\WW_{f_j(y_j)}$ so that $u=f_j(z)$ and $v =f_j(w)$
\end{center}

\noindent where $z,w\in \WW_{y_j}$ and $z < w$. Observe that
elements in different components (i.e. $\WW_x$ or $\WW_{y_j}$) are
incomparable w.r.t. $<'$.

Finally, we let $V'(z) = V(z) $  for $z \in \WW_x$ and for $u\in
\WW_{f_j(y_j)}$ with $u=f_j(w)$, we let $V'(u) = V(w)$.

We prove that $\emme', x \models \Gamma$. The claim is obvious for
propositional formulas and for (negated) boxed formulas by
definition of $\WW_x$.

For any negated conditional $\nott (C_j \cond D_j)$, we have that
$y_j \in Min_{<}(C_j)$  and $\emme,y_j \not\models D_j$. By
definition of $\emme'$ we get that  $\emme',f_j(y_j) \models C_j$
and $\emme',f_j(y_j) \not\models D_j$; we have to show that there
is no $u\in \WW_{f_j(y_j)}$ such that $u <'f_j(y_j)$ and $\emme',
u \models C_j$. But if there were a such $u$, we would get that $u
= f_j(z)$ for some $z\in \WW_{y_j}$ with $z < y_j$ and $\emme,z
\models C_j$ against the minimality of $y_j$.

For any positive conditional in $\Gamma$, say $E \cond F$, let
$u\in Min_{<}(E)$: if $u \in \WW_x$ it must be $u\in Min_{<}(E)$
thus $\emme',u\models F$. If $u \in \WW_{f_j(y_j)}$, then $u =
f_j(z)$ for some $z\in \WW_{y_j}$; it must be $z \in Min_{<}(E)$,
for otherwise if it were $z' < z$ with $\emme,z' \models E$, since
$z'\in \WW_{y_j}$ we would have $f_j(z') < u$ and $\emme, f_j(z')
\models E$ against the minimality of $u$. Thus $z \in Min_{<}(E)$
and then $\emme,z \models F$, and this implies $\emme',u \models
F$.

We now define a multi-linear model $\emme_1 = \sx \WW', <_1, V'
\dx$ as follows: we let $<_1$ be  any total order on $\WW_x$ and
on each $\WW_{f_j(y_j)}$ which respects $<'$; the elements in
different components remain incomparable. More precisely $<_1$
satisfies:
\begin{itemize}
  \item if $u <' v$ then $u <_1 v$
  \item for each  $u,v \in \WW_x$ ($u,v \in \WW_{f_j(y_j)}$) with $u\not=v$, $u <_1 v$
  or $v <_1 u$
  \item for each $u \in \WW_x, \ v \in \WW_{f_j(y_j)}, u \not<_1 v$
  and $v \not<_1 u$
  \item for each $u \in \WW_{f_i(y_i)}, v \in \WW_{f_j(y_j)}$, with $i\not=j$ $u \not<_1 v$
  and $v \not<_1 u$
\end{itemize}

In Figure \ref{multi} we show an example of multi-linear model,
obtained by applying the above construction to the model
represented in Figure \ref{modellooriginale}.

\begin{figure}
{\centerline{\includegraphics[angle=0,width=2.8in]{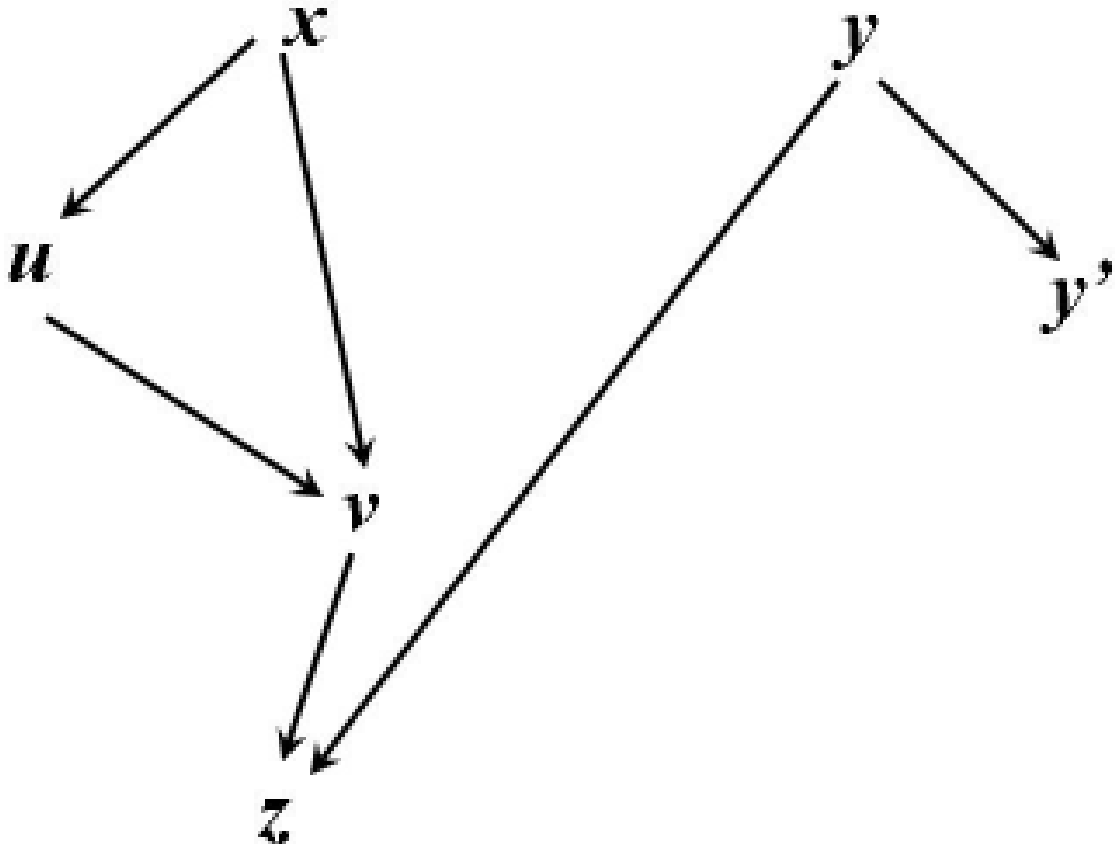}}}
 \caption{A preferential model satisfying a set of formulas $\Gamma$. Edges represent
 the preference relation $<$ ($u < x, v < u$, and so on).} \label{modellooriginale}
\end{figure}

\begin{figure}
{\centerline{\includegraphics[angle=0,width=2.8in]{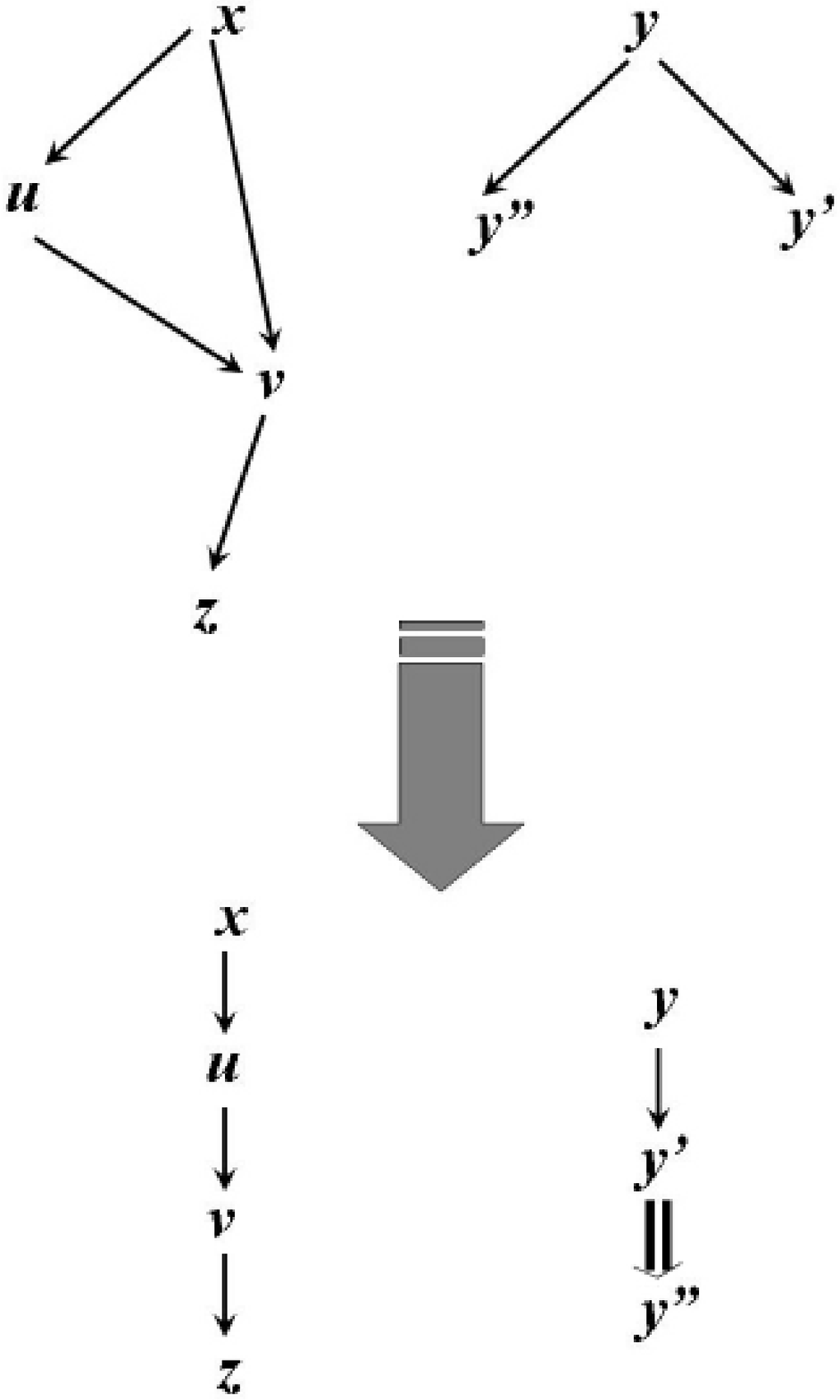}}}
 \caption{A multi-linear model obtained by means of the construction described in the proof of Theorem \ref{teorema nicola}.
 Edges represent the preference relation $<_1$. The empty edge has
 been added to let $<_1$ be a total order.
 In order to have that elements in different components are
incomparable w.r.t. $<_1$, in the rightmost component the world
$z$ has been renamed in $y''$.} \label{multi}
\end{figure}

\noindent We show that $\emme_1, x \models \Gamma$. For
propositional formulas the claim is obvious. For positive
boxed-formulas we have: if $\bbox\neg A\in \Gamma$, and $z <_1 x$,
then $z\in \WW_x$, thus $z <'x$, and the result follows by
$\emme',x\models \Gamma$. For negated boxed formulas, we similarly
have: $\nott\bbox\neg A \in \Gamma$, then $\emme',x \models
\nott\bbox\neg A$, thus there exists $z <'x$ such that $\emme',z
\models A$, but $z <' x$ implies $z <_1 x$ and we can conclude.

For negated conditionals, let $\nott (C_j \cond D_j)$, we know
that $\emme',x \models \nott (C_j \cond D_j)$, witnessed by the
$C_j$-minimal element $f_j(y_j)$. Since the propositional
evaluation is the same, we only have to check that $f_j(y_j)$ is
also minimal w.r.t. $<_1$. Suppose it is not, then there is $z \in
\WW_{f_j(y_j)}$ with $z <_1 f_j(y_j)$ such that $\emme_1, z
\models C_j$, but we would get $z <'f_j(y_j)$ against the
minimality of $f_j(y_j)$.

For positive conditionals in $\Gamma$, say $E \cond F$, let $u\in
Min_{<_1}(E)$. It must be also $u\in Min_{<_1}(E)$, for otherwise,
if there were $v <' u$, such that $\emme',v \models E$ then we
would get also $v <_1 u$ and $\emme_1, v \models E$, against the
minimality of $u$ in $\emme_1$.

\vspace{0.1cm} \begin{flushright}$\blacksquare$\end{flushright}
\end{pf*}



\noindent

\begin{pf*}{Proof of Proposition \ref{correspondence
CL}.} \emph{A boolean combination of conditional formulas is
satisfiable in a loop-cumulative model $\emme = \langle S, \WW ,
l, <, V \rangle$ iff it is satisfiable in a CL-preferential model
$\emme'= \langle \WW', R, <', V' \rangle$.}

\vspace{0.2cm}

\noindent The Proposition immediately follows from the following
Lemma:

\begin{lemma}
A set of conditional formulas $\{(\neg) A_1 \cond
B_1,\ldots,(\neg) A_n \cond B_n\}$ is satisfiable in a
loop-cumulative model $\emme = \langle S, \WW , l, <, V \rangle$
iff it is satisfiable in a CL-preferential model $\emme'= \langle
\WW', R, <', V' \rangle$.
\end{lemma}

\noindent First, we prove the \emph{only if} direction. Let $\emme
= \langle S, \WW, l, <, V \rangle$ be a loop-cumulative model, and
$s \in S$ s.t. $(\emme,s) \mid\hspace{-1pt}\equiv \{(\neg) A_i
\cond B_i \}$.

We build a CL-preferential model $\emme' = \langle \WW', R, <', V'
\rangle$ as follows:
\begin{itemize}
\item $\WW' = \{(s,w): s \in S$ and $w \in l(s)\}$;
\item $(s, w) R (s, w')$ for all $(s, w), (s, w') \in \WW'$;
\item $(s, w) <' (s', w')$ if $s < s'$;
\item $V'(s, w)$ = $V(w)$.
\end{itemize}

Observe that for each $s \in S$ there is at least one
corresponding $(s,w) \in \WW'$, since $l(s) \neq \emptyset$.  From
the fact that $<$ in $\emme$ is irreflexive and  transitive it
immediately follows by construction that also $<'$ in $\emme'$
satisfies the same properties. We show in Fact \ref{fatto 5
corrisp CL ->} below that $<'$ satisfies the smoothness condition
on $L$-formulas.

The relation $R$ is serial, since it is reflexive.

\begin{fact}\label{fatto 1 corrisp CL ->}
For every propositional formula $A$ we have that $(\emme, s)
\mid\hspace{-1pt}\equiv A$ iff $(\emme',(s, w)) \models LA$.
\end{fact}

\begin{pf*}{Proof of Fact \ref{fatto 1 corrisp CL ->}.}
 ($\Rightarrow$) Let $(\emme, s) \mid\hspace{-1pt}\equiv A$.
By definition, for all $w \in l(s)$, $(\emme,w) \models A$. By
induction on the complexity of $A$, we can easily  show that
$(\emme', (s, w)) \models A$. Since $R(s, w)= \{(s,w') \tc w'
\appartiene l(s)\}$, it follows that $(\emme',(s, w)) \models LA$.
\\($\Leftarrow$) Let $(\emme',(s, w)) \models LA$. Then, for all
$(s, w') \in R(s,w)$, $(\emme', (s, w')) \models A$. By definition
of $\emme'$ it follows that for all $w' \in l(s)$, $(\emme, w')
\models A$. Hence, $(\emme, s) \mid\hspace{-1pt}\equiv A$.
\provafatto{\ref{fatto 1 corrisp CL ->}}
\end{pf*}

\begin{fact}\label{fatto 2 corrisp CL ->}
$s \in Min_<(A)$ in $\emme$ iff $(s,w) \in Min_{<'}(LA)$ in
$\emme'$.
\end{fact}

\begin{pf*}{Proof of Fact \ref{fatto 2 corrisp CL ->}.}
($\Rightarrow$) Let $s \in Min_<(A)$ in $\emme$. Consider $(s, w)$
in $\emme'$. By Fact \ref{fatto 1 corrisp CL ->}, $(\emme',(s,w))
\models LA$. By absurd, suppose there exists a $(s',w')$ s.t.
$(\emme',(s',w')) \models LA$, and $(s',w') < (s, w)$. By Fact
\ref{fatto 1 corrisp CL ->}, $(\emme, s') \mid\hspace{-1pt}\equiv
A$, and  $s' < s$ by construction, which contradicts the fact that
$s \in Min_<(A)$ in $\emme$.
Hence $(s,w)\in Min_{<'}(LA)$ in $\emme'$.\\
($\Leftarrow$) Let $(s,w)\in Min_{<'}(LA)$ in $\emme'$. Consider
$s$ in $\emme$. By Fact \ref{fatto 1 corrisp CL ->}, $(\emme, s)
\mid\hspace{-1pt}\equiv A$. Furthermore there is no $s' < s$ s.t.
$(\emme, s') \mid\hspace{-1pt}\equiv A$. By absurd suppose there
was such a $s'$. By construction of $\emme'$, and by Fact
\ref{fatto 1 corrisp CL ->}, $(\emme',(s', w')) \models LA$, and
$(s', w') < (s,w)$, which is a contradiction. Hence $s \in
Min_<(A)$. \provafatto{\ref{fatto 2 corrisp CL ->}}
\end{pf*}

\begin{fact}\label{fatto 3 corrisp CL ->}
For every conditional formula $A \cond B$ we have that \\ $(\emme,
s) \mid\hspace{-1pt}\equiv A \cond B$ iff $(\emme',(s,w)) \modello
A \cond B$.
\end{fact}

\begin{pf*}{Proof of Fact \ref{fatto 3 corrisp CL ->}.}
($\Rightarrow$) Let $(\emme,s) \mid\hspace{-1pt}\equiv A \cond B$.
Then for all $s' \in Min_<(A)$, $(\emme, s')
\mid\hspace{-1pt}\equiv B$. By Facts \ref{fatto 1 corrisp CL ->}
and \ref{fatto 2 corrisp CL ->}, it follows that for all $(s',w')
\in Min_{<'}(LA)$, $(\emme', (s', w')) \models LB$, hence
$(\emme', (s, w)) \models A \cond B$. \\ ($\Leftarrow$) Let
$(\emme', (s, w)) \models A \cond B$. Then, for all $(s', w') \in
Min_<'(LA)$, $(\emme',(s', w')) \models LB$. By Facts \ref{fatto 1
corrisp CL ->} and \ref{fatto 2 corrisp CL ->} it follows that for
all $s' \in Min_<(A)$ in $\emme$, $(\emme,s')
\mid\hspace{-1pt}\equiv B$. Hence, $(\emme, s)
\mid\hspace{-1pt}\equiv A \cond B$. \provafatto{\ref{fatto 3
corrisp CL ->}}
\end{pf*}

\begin{fact}\label{fatto 4 corrisp CL ->}
For every negated conditional formula $\nott (A \cond B)$ we have
that $(\emme, s) \mid\hspace{-1pt}\equiv \nott (A \cond B)$ iff
$(\emme',(s,w)) \modello \nott (A \cond B)$.
\end{fact}

\begin{pf*}{Proof of Fact \ref{fatto 4 corrisp CL ->}.}
($\Rightarrow$) Let $(\emme,s) \mid\hspace{-1pt}\equiv \neg (A
\cond B)$. Then there is an $s' \in Min_< (A)$ s.t. $\emme, s'
\not \mid\hspace{-1pt}\equiv B$. Consider $(s', w')$ in $\emme'$.
By Facts \ref{fatto 1 corrisp CL ->} and \ref{fatto 2 corrisp CL
->} $(s',w') \in Min_{<'}(LA)$ and $(\emme',(s',
w'))  \not \models LB$. Hence, $(\emme', (s, w)) \models \neg (A \cond B)$.\\
($\Leftarrow$) Let $(\emme', (s, w)) \models \neg (A \cond B)$.
Then, there is a $(s', w') \in Min_<'(LA)$ s.t. $(\emme',(s',w'))
\not\models LB$. Consider $s'$ in $\emme$. From Facts \ref{fatto 1
corrisp CL ->} and \ref{fatto 2 corrisp CL ->} we conclude that
$s' \in Min_<(A)$, and $(\emme, s') \not \mid\hspace{-1pt}\equiv
B$. Hence $(\emme,s) \mid\hspace{-1pt}\equiv \neg (A \cond B)$.
\provafatto{\ref{fatto 4 corrisp CL ->}}
\end{pf*}

From Facts \ref{fatto 3 corrisp CL ->} and \ref{fatto 4 corrisp CL
->} we conclude that $(\emme',(s,w)) \models \{(\neg) A_i \cond
B_i \}$. Furthermore, we show that $<'$ satisfies the smoothness
condition on $L$-formulas.

\begin{fact}\label{fatto 5 corrisp CL ->}
$<'$ satisfies the smoothness condition on $L$-formulas.
\end{fact}

\begin{pf*}{Proof of Fact \ref{fatto 5 corrisp CL ->}.}
Let $(\emme',(s,w)) \models LA$, and $(s,w) \not\in Min_{<'}(LA)$.
By Fact \ref{fatto 1 corrisp CL ->} $(\emme, s)
\mid\hspace{-1pt}\equiv A$, and by Fact \ref{fatto 2 corrisp CL
->}, $s \not\in Min_<(A)$ in $\emme$. By the smoothness condition
in $\emme$ there is $s'$ such that $s' \in Min_<(A)$ and $s' < s$.
Consider any $(s',w') \in \emme'$. By Fact \ref{fatto 2 corrisp CL
->} $(s',w') \in Min_{<'}(LA)$, and by definition of $<'$,
$(s',w') <' (s,w)$. \provafatto{\ref{fatto 5 corrisp CL ->}}
\end{pf*}

\noindent Let us now consider the \emph{if} direction. Let the set
of conditionals $\{(\neg) A_i \cond B_i\}$ be satisfied in a
possible world $w$ in the CL-preferential model $\emme = \langle
\WW, R, <, V \rangle$. We build a Loop-Cumulative model $\emme' =
\langle S, \WW, l, <', V' \rangle$ as follows ($Rw$ is defined as
$Rw=\{w' \in \WW \tc (w,w') \in R \}$):

\begin{itemize}
\item $S = \{(w, Rw) \tc w \in \WW\}$;
\item $l((w, Rw)) = Rw$;
\item $(w, Rw) <' (w', Rw')$ if $w < w'$;
\item $V'(w)$ = $V(w)$.
\end{itemize}

\noindent From the fact that $<$ in $\emme$ is transitive and
irreflexive, it follows by construction that $<'$ in $\emme'$ is
transitive and irreflexive. As far as the  smoothness condition,
see Fact \ref{fatto 5 corrisp CL <-} below. Furthermore, for all
$(w, Rw) \in S$, $l(w, Rw) \neq \emptyset$, since $R$ is serial.

We now show that $(\emme', (w, Rw)) \mid\hspace{-1pt}\equiv
\{(\neg) A_i \cond B_i\}$.

\begin{fact}\label{fatto 1 corrisp CL <-}
For $A$ propositional, $(\emme, w) \models LA$ iff $(\emme', (w,
Rw)) \mid\hspace{-1pt}\equiv A$.
\end{fact}

\begin{pf*}{Proof of Fact \ref{fatto 1 corrisp CL <-}.}
($\Rightarrow$) Let $(\emme, w) \models LA$. Then for all $w' \in
Rw$, $\emme, w' \models A$. By definition of $\emme'$, it follows
that for all $w' \in l(w, Rw)$, $\emme', w' \models A$. Hence
$(\emme', (w, Rw)) \mid\hspace{-1pt}\equiv A$.\\ ($\Leftarrow$)
Let $(\emme', (w, Rw)) \mid\hspace{-1pt}\equiv A$. Then, for all
$w' \in l(w, Rw)$, $(\emme',w') \models A$. By induction on A, we
show that for all $w' \in Rw$, $(\emme, w') \models A$, hence
$(\emme, w) \models LA$.
\provafatto{\ref{fatto 1 corrisp CL <-}}
\end{pf*}

\begin{fact}\label{fatto 2 corrisp CL <-}
$w \in Min_<(LA)$ in $\emme$ iff $(w, Rw) \in Min_{<'}(A)$ in
$\emme'$.
\end{fact}

\begin{pf*}{Proof of Fact \ref{fatto 2 corrisp CL <-}.}
($\Rightarrow$) Let $w \in Min_<(LA)$ in $\emme$. Consider $(w,
Rw)$. By Fact \ref{fatto 1 corrisp CL <-} $(\emme', (w, Rw))
\mid\hspace{-1pt}\equiv A$. Furthermore, suppose by absurd there
was $(w', Rw')$ s.t. $(\emme', (w',Rw')) \mid\hspace{-1pt}\equiv
A$, and $(w', Rw') <' (w, Rw)$. Then in $\emme$, $\emme, w'
\models LA$ and $w' < w$, which contradicts the
fact that  $w \in Min_<(LA)$. It follows that in $\emme'$, $(w, Rw) \in Min_{<'}(A)$.\\
($\Leftarrow$) Let $(w, Rw) \in Min_{<'}(A)$ in $\emme'$. Consider
$w$ in $\emme$. By Fact \ref{fatto 1 corrisp CL <-}, $(\emme, w)
\models LA$. By absurd, suppose there was a $w'$ s.t. $\emme, w'
\models LA$ and $w' < w$. By Fact \ref{fatto 1 corrisp CL <-},
$(\emme',(w',Rw')) \mid\hspace{-1pt}\equiv A$, and $(w',Rw')<' (w,
Rw)$, which contradicts the fact that $(w, Rw) \in Min_{<'}(A)$.
It follows that $w \in Min_<(LA)$ in $\emme$.
\provafatto{\ref{fatto 2 corrisp CL <-}}
\end{pf*}

\noindent We can reason similarly to what done in Facts \ref{fatto
3 corrisp CL ->} and \ref{fatto 4 corrisp CL ->} above to prove
the following Fact:
\begin{fact}\label{fatti 3 e 4 corrisp CL  <-}
For every conditional formula $(\nott) A \cond B$ we have that
$(\emme, w)  \modello (\nott) A \cond B$ iff $(\emme',(w, Rw))
\mid\hspace{-1pt}\equiv (\nott) A \cond B$.
\end{fact}

We conclude that $(\emme', (w, Rw)) \mid\hspace{-1pt}\equiv
\{(\neg) A_i \cond B_i\}$.

Furthermore, we show that $<'$ satisfies the smoothness condition:

\begin{fact}\label{fatto 5 corrisp CL <-}
$<'$ satisfies the smoothness condition on $L$-formulas.
\end{fact}

\begin{pf*}{Proof of Fact \ref{fatto 5 corrisp CL <-}.}
Let $(\emme', (w, Rw)) \mid\hspace{-1pt}\equiv A$ and $(w, Rw)
\not\in Min_{<'}(A)$ in $\emme'$. By Facts \ref{fatto 1 corrisp CL
<-} and \ref{fatto 2 corrisp CL <-}, $(\emme, w) \models LA$ and
$w \not\in Min_<(LA)$ in $\emme$. By the smoothness condition on
$L$-formulas in $\emme$, it follows that in $\emme$ there is $w' <
w$ s.t. $(\emme, w') \models LA$ and $w' \in Min_<(LA)$. Consider
$(w', Rw')$ in $\emme'$. By Facts \ref{fatto 1 corrisp CL <-} and
\ref{fatto 2 corrisp CL <-} $(\emme', (w', Rw'))
\mid\hspace{-1pt}\equiv A$ and $(w', Rw') \in Min_{<'}(A)$ in
$\emme'$.
\provafatto{\ref{fatto 5 corrisp CL <-}}
\end{pf*}
\begin{flushright}$\blacksquare$\end{flushright}

\end{pf*}



\begin{pf*}{Proof of Lemma \ref{lemma saturazione}.} \emph{Given a consistent finite set
of formulas $\Gamma$, there is a consistent, finite, and
  saturated set $\Gamma' \supseteq \Gamma$.}

\vspace{0.2cm}

\noindent Consider the set $\Gamma^{\bigstar}$ of complex formulas
in $\Gamma$ such that there is a static rule that has not yet been
applied to that formula in $\Gamma$. For instance, if $\Gamma=\{P
\cond Q, \nott(R \imp S), R, \nott S, R \orr T, \nott \bbox \nott
Q\}$, then $\Gamma^{\bigstar}=\{P \cond Q\}$, since neither $\nott
P$ nor $\nott \bbox \nott P$ nor $Q$, resulting from an
application of the static rule $(\cond^{+})$ to $P \cond Q$,
belong to $\Gamma$. \hide{Notice that $\nott (R \imp S)$ does not
belong to $\Gamma^{\bigstar}$, since $R$ and $\nott S$, the
conclusion of an application of $(\imp^{-})$ to it, are in
$\Gamma$; the same for $R \orr T$, since $R \appartiene \Gamma$
($R$ is \emph{one} of the conclusions of an application of
$(\orr^{+})$). $\nott \bbox P$ is not in $\Gamma^{\bigstar}$ since
$\bbox^{-}$ is not a static rule.}

If $\Gamma^{\bigstar}$ is empty, we are done. Otherwise, we
construct the saturated set $\Gamma'$ as follows: 1. initialize
$\Gamma'$ with $\Gamma$; 2. choose a complex formula $F$ in
$\Gamma^{\bigstar}$ and apply the \emph{static} rule corresponding
to its principal operator; 3. add to $\Gamma'$ the formula(s) of
(one of) the \emph{consistent} conclusions obtained by applying
the static rule; 4. update $\Gamma^{\bigstar}$\footnote{The
complex formula analyzed at the current step must be removed from
$\Gamma^{\bigstar}$ and formulas obtained by the application of
the static rules that fulfill the definition of
$\Gamma^{\bigstar}$ must be added.} and repeat from 2. until
$\Gamma^{\bigstar}$ is empty. This procedure terminates, since in
all  static rules the conclusions have a lower complexity than the
premise; a brief discussion on the $(\cond^{+})$ rule: from the
premise $A \cond B$ we have the following possible conclusions:
$\nott A$, $\nott \bbox \nott A$ and $B$. If $\nott \bbox \nott A$
is introduced, then no other static rule will be applied to it.
Since  $A$ and $B$ are boolean combinations of formulas, then the
other applications of static rules to them will decrease the
complexity.

Furthermore, each step of the procedure preserves the consistency
of $\Gamma$. Indeed, each conclusion of the rule applied to
$\Gamma$ corresponds to a branch of a tableau for $\Gamma$. If all
the branches were inconsistent, $\Gamma$ would have a closed
tableau, hence would be inconsistent, against the hypothesis.

\begin{flushright}$\blacksquare$\end{flushright}
\end{pf*}


\begin{pf*}{Proof of Proposition \ref{proposizione nodi regolari}.} \emph{Given a pair $\Gamma_0;\vuoto$, where $\Gamma_0$ is a set of
formulas of $\elle$, all the tableaux obtained by applying
$\calcoloPterminante$'s rules only contain regular nodes.}

\vspace{0.2cm}

\noindent  Given a regular node $\Gamma;\Sigma$ and any rule of
  $\calcoloPterminante$, we have to show that each conclusion of
  the rule is still a regular node. The proof is immediate for all
  the propositional rules, for $(\bbox^{-})$ and for
  $(\cond^{-})$, since no negated box formula not belonging to
  their premise is introduced in their conclusion(s). Consider now
  an application of $(\cond^{+})$ to a regular node $\Gamma', A \cond B;\Sigma$:
  one of the conclusions is $\Gamma', \nott \bbox \nott A; \Sigma,
  A \cond B$, and it is a regular node since there is $A \cond B$
  in the auxiliary set of used conditional in correspondence of
  the negated boxed formula $\nott \bbox \nott A$.

\begin{flushright}$\blacksquare$\end{flushright}
\end{pf*}



\section*{Proof of Theorem \ref{eliminazione cut C} (Admissibility of (Weak-Cut) in
$\calcoloC$)}\label{eliminazione cut in appendice}

In order to prove in Theorem \ref{eliminazione cut C} below that
(Weak-Cut) is admissible in $\calcoloC$, we need to prove some
Lemmas.


First of all, we prove that weakening is height-preserving
admissible in our tableau calculi, i.e. if there is a closed
tableau for a set of formulas $\Gamma$, then there is also a
closed tableau for $\Gamma, F$  (for any formula $F$ of the
language) of height no greater than the height of a tableau for
$\Gamma$. Moreover, weakening is cut-preserving admissible, in the
sense that the closed tableau for $\Gamma, F$ does not add any
application of (Weak-Cut) to the closed tableau for $\Gamma$.
Furthermore, we prove that the rules for the boolean connectives
are height-preserving and cut-preserving invertible, i.e. if there
is a closed tableau for $\Gamma$, then there is a closed tableau
for any set of formulas that can be obtained from $\Gamma$ as a
conclusion of an application of a boolean rule.

\begin{lemma}[Height-preserving and cut-preserving admissibility of weakening]\label{weakening}
  Given a formula $F$, if there is a closed tableau of height $h$ for $\Gamma$, then there is also a
  closed tableau for $\Gamma, F$ of no greater height than $h$, i.e. weakening is
  height-preserving admissible. Moreover, the
closed tableau for $\Gamma, F$ does not add any application of
(Weak-Cut) to the closed tableau for $\Gamma$, i.e. weakening is
cut-preserving admissible.
\end{lemma}
\begin{provaallettore}
  By induction on the height $h$ of the closed tableau for
  $\Gamma$.
\end{provaallettore}

\begin{lemma}[Height-preserving and cut-preserving invertibility of boolean
rules]\label{invertibilità regole booleane C} The rules for the
boolean connectives are height-preserving invertible, i.e. given a
set of formulas $\Gamma$ and given any conclusion $\Gamma'$,
obtained by applying a boolean rule to $\Gamma$, if there is a
closed tableau of height $h$ for $\Gamma$, then there is a closed
tableau for $\Gamma'$ of height no greater than $h$. Moreover, the
closed tableau for $\Gamma'$ does not add any application of
(Weak-Cut) to the closed tableau for $\Gamma$, i.e. the boolean
rules are cut-preserving invertible.
\end{lemma}

\begin{provaposu}
  For each boolean rule (R), we proceed by induction on the height
  of the closed tableau for the premise. As an example, consider
  the $(\orr^{+})$ rule. We show that, if $\Gamma, F \orr G$ has a
  closed tableau, also $\Gamma, F$ and $\Gamma, G$ have. If $\Gamma, F \orr
  G$ is an instance of the axiom (AX), then there is an atom $P$
  such that $P \in \Gamma$ and $\nott P \in \Gamma$, since axioms
  are restricted to atomic formulas only. Therefore, $\Gamma$ has
  a closed tableau (it is an instance of (AX) too), and we
  conclude that both $\Gamma, F$ and $\Gamma, G$ have a closed
  tableau, since weakening is height-preserving admissible (Lemma \ref{weakening} above).
  For the inductive step, we consider the first rule in the
  tableau for $\Gamma, F \orr G$. If $(\orr^{+})$ is applied to $F \orr G$, then we are
  done, since we have closed tableaux for both $\Gamma, F$ and
  $\Gamma, G$ of a lower height than the premise's. If
  $(\cond^{-})$ is applied, then $F \orr G$ is removed from the
  conclusion, then $\Gamma$ has a closed tableau; we conclude since weakening
  is height-preserving admissible (Lemma \ref{weakening}). If a
  boolean rule is applied, then $F \orr G$ is copied into the
  conclusion(s); in these cases, we can apply the inductive hypothesis and
  then conclude by re-applying the same rule. As an example,
  consider the case in which $(\imp^{-})$ is applied to $\Gamma',
  \nott (H \imp I), F \orr G$ as follows:
  \[
    \begin{prooftree}
      \Gamma', \nott (H \imp I), F \orr G
      \justifies (*) \Gamma', H, \nott I, F \orr G \using (\imp^{-})
    \end{prooftree}
  \]
  By the inductive hypothesis, there is a closed tableau (of no
  greater height than the height of $(*)$) for $(**)\Gamma', H, \nott I, F$
  and for $(***)\Gamma', H, \nott I, G$. We conclude as follows:
  \[
    \begin{prooftree}
      \Gamma', \nott (H \imp I), F
      \justifies (**)\Gamma', H, \nott I, F \using (\imp^{-})
    \end{prooftree}
  \]
  \[
    \begin{prooftree}
      \Gamma', \nott (H \imp I), G
      \justifies (***)\Gamma', H, \nott I, G \using (\imp^{-})
    \end{prooftree}
  \]
  If the first rule of the closed tableau for $\Gamma, F \orr G$
  is $(\cond^{+})$ applied to $A \cond B \in \Gamma$, then we have the following situation:
  \[
    \begin{prooftree}
      \Gamma, F \orr G \justifies
      (1)\Gamma, F \orr G, \nott LA \quad\quad (2)\Gammas, LA, \bbox
      \nott LA \quad\quad (3)\Gamma, F \orr G, LA, \bbox \nott LA, LB
      \using (\cond^{+})
    \end{prooftree}
  \]
  By the inductive hypothesis on $(1)$, we have closed tableaux
  for $(1')\Gamma, F, \nott LA$ and for $(1'')\Gamma, G, \nott
  LA$. By the inductive hypothesis on $(3)$, we have closed tableaux
  for $(3')\Gamma, F, LA, \bbox \nott LA, LB$ and for $(3'')\Gamma, G, LA, \bbox \nott LA,
  LB$. We conclude as follows:
  \[
    \begin{prooftree}
      \Gamma, F \justifies
      (1')\Gamma, F, \nott LA \quad\quad (2)\Gammas, LA, \bbox
      \nott LA \quad\quad (3')\Gamma, F, LA, \bbox \nott LA, LB
      \using (\cond^{+})
    \end{prooftree}
  \]
  \[
    \begin{prooftree}
      \Gamma, G \justifies
      (1'')\Gamma, G, \nott LA \quad\quad (2)\Gammas, LA, \bbox
      \nott LA \quad\quad (3'')\Gamma, G, LA, \bbox \nott LA, LB
      \using (\cond^{+})
    \end{prooftree}
  \]
  If $F$ (resp. $G$) were a conditional formula (even negated), we
  conclude as above, replacing the inner conclusion with $\Gammas,
  F, LA, \bbox \nott LA$ (resp. $\Gammas, G, LA, \bbox \nott LA$),
  for which there is a closed tableau since weakening is
  height-preserving admissible (Lemma \ref{weakening}).

\end{provaposu}

\noindent Now we prove that we can assume, without loss of
generality, that the conclusions of (Weak-Cut) are never derived
by an application of $(\cond^{-})$, as stated by the following
Lemma:

\begin{lemma}\label{niente cond negativi}
  If $\Gamma$ has a closed tableau, then there is a closed tableau
  for $\Gamma$ in which all the conclusions of each application of
  \mbox{(Weak-Cut)} are derived by a rule different from
  $(\cond^{-})$.
\end{lemma}

\begin{provaposu}
 Consider an application of \mbox{(Weak-Cut)} in
    $\Gamma$ in which one of its conclusions is obtained by an
    application of $(\cond^{-})$. The application of
    \mbox{(Weak-Cut)} is useless, since the premise of the cut can
    be obtained by applying directly $(\cond^{-})$ without (Weak-Cut).
    For instance, consider the following derivation, in which the
    inner conclusion of \mbox{(Weak-Cut)} is obtained by an
    application of $(\cond^{-})$:
    \begin{footnotesize}
    \[
      \begin{prooftree}
        \[
        \Gamma', \nott (C \cond D)
        \justifies \Gamma', \nott (C \cond D), \nott LA \quad\quad
\Gamma'^{\cond{\pm}}, \Gamma'^{\bbox^{\freccia}}, \nott (C \cond
D), LA, \bbox \nott LA \quad\quad \Gamma', \nott (C \cond D),
\bbox \nott LA \using \mbox{(Weak-Cut)}
\]
  \justifies (*)\Gamma'^{\cond\pm}, LC, \bbox \nott LC, \nott LD \using
  (\cond^{-})
      \end{prooftree}
    \]
    \end{footnotesize}
    We can remove the application of \mbox{(Weak-Cut)}, obtaining
    the following closed tableau:

  \[
    \begin{prooftree}
    \Gamma', \nott (C \cond D)
  \justifies (*)\Gamma'^{\cond\pm}, LC, \bbox \nott LC, \nott LD \using
  (\cond^{-})
    \end{prooftree}
  \]
  Obviously, the proof can be concluded in the same way in the
  case the leftmost (resp. the rightmost) conclusion of
  \mbox{(Weak-Cut)} has a derivation starting with $(\cond^{-})$.

\end{provaposu}

\noindent Now we prove that cut is admissible on propositional
formulas and on formulas of the form $LA$. By cut we mean the
following rule:
\[
  \begin{prooftree}
    \Gamma
    \justifies \Gamma, F \quad\quad\quad \Gamma, \nott F \using (Cut)
  \end{prooftree}
\]

\noindent and we show that it can be derived if $F$ is a
propositional formula or a formula of kind $LA$.
\begin{lemma}\label{cut-prop} Given a set of propositional formulas $\Gamma$
and a propositional formula $A$, if there is a closed tableau for
both $(1)\Gamma, \neg A$  and $(2)\Gamma, A$ without (Weak-Cut),
then there is also a closed tableau for $\Gamma$ without
(Weak-Cut).
\end{lemma}

\begin{provaposu}
Since $\Gamma$ is propositional, the only applicable rules are the propositional rules.
The result follows by admissibility of cut in tableaux systems for
propositional logic.
\end{provaposu}

\begin{lemma}\label{cut-Lformula}
  If there is a closed tableau without (Weak-Cut) for $(1) \Gamma, \nott LA$
 and for $(2) \Gamma, LA$, then there is also a closed tableau without (Weak-Cut) for
 $\Gamma$.
\end{lemma}

\begin{provaposu}
Let $h1$ be the height of the tableau for $(1)$, and $h2$ the
height of the tableau for $(2)$. We proceed by induction on $h1 +
h2$.

 \noindent {\em Base Case:} $h1 + h2 = 0$. In this case,  $h1 = 0$ and $h2 = 0$. In this case, $(1)$ and $(2)$
contain an axiom, and since axioms are restricted to atomic
formulas, it can only be that $P, \nott P \appartiene \Gamma$,
where $P \appartiene \mathit{ATM}$. Therefore, we conclude that
there is a closed tableau for $\Gamma$ without (Weak-Cut).

For the inductive step, we show that if the property holds in case
$h1 + h2 = n-1$, then it also holds in case $h1 + h2 = n$. We
reason by cases according to which is the first rule applied to
$(1)$ or to $(2)$. If the first rule applied to $(1)$ is
$(\cond^{-})$, applied to a conditional $\neg(C \cond D) \in
\Gamma$, let $\Gamma_{\neg(C \cond D) }$ be the set of formulas so
obtained. It can be easily verified that the same set can be
obtained by applying the same rule to the same conditional in
$\Gamma$, hence $\Gamma$ has a closed tableau without (Weak-Cut).
If the first rule applied to $(1)$ is $(\cond^{+})$, applied to a
conditional $(C \cond D) \in \Gamma$, then we have that $(1a)
\Gamma, \neg LC, \neg LA$, $(1b) \Gammas, LC, \bbox \nott LC$, and
$(1c) \Gamma, LC, \bbox \neg LC, LD, \neg LA$ have a closed
tableau with height  smaller than $h1$. We can then apply
weakening and the inductive hypothesis first over (2) and $(1a)$
and then over (2) and $(1c)$, to obtain that (i)$\Gamma, \neg LC$,
and (ii) $\Gamma, LC, \bbox \neg LC, LD$ respectively have a
closed tableau without (Weak-Cut). Since (i),(1b),(ii) are
obtained from $\Gamma$ by applying $(\cond^+)$ on $C \cond D$, we
conclude that also
 $\Gamma$ has a closed tableau without (Weak-Cut). The case in which
the first rule applied is a propositional rule immediately follows
from the height-preserving invertibility of the boolean rules (see
Lemma \ref{invertibilità regole booleane C} above). For instance,
suppose $(1) \Gamma', F \andd G, \nott LA$ is derived by an
application of $(\andd^{+})$, i.e. $(1')\Gamma', F, G, \nott LA$
has a closed tableau of height smaller than $h1$. Since $(2)
\Gamma', F \andd G, LA$ has a closed tableau, and $(\andd^{+})$ is
height-preserving invertible, then also $(2')\Gamma', F, G, LA$
has a closed tableau of height no greater than $h2$, and we can
conclude as follows:
\[
  \begin{prooftree}
    \[
      \Gamma', F \andd G
      \justifies \Gamma', F, G \using (\andd^{+})
    \]
    \justifies
    (1')\Gamma', F, G, \nott LA \quad\quad (2')\Gamma', F, G, LA
    \using (cut)
  \end{prooftree}
\]

If the first rule applied to $(2)$ is either a propositional rule,
or $(\cond^{-})$, or $(\cond^{+})$, we can reason as for $(1)$. We
are left with the case in which the first rule applied both to
$(1)$ and to $(2)$ is $(L^{-})$.

 If the first rule applied
to $(1)$ is ($L^{-}$), applied to $\neg LB \in \Gamma$, let
$\Gamma_{\nott L B}$ be the set obtained. The same set can be
obtained by applying ($L^{-}$) to $\neg LB$  in $\Gamma$, hence
$\Gamma$ has a closed tableau without (Weak-Cut). If ($L^{-}$) is
applied to $\neg LA$ itself, then $(*)\Gamma^{L^{\freccia}}, \neg
A$ has a closed tableau. In this case, we have to consider the
tableau for (2). We distinguish two cases: in the first case
$(L^{-})$ in (2) has been applied to some $\neg LB \in \Gamma$, in
the second case $\Gamma$ does not contain any negated $L$ formula,
hence $(L^{-})$ has not been applied to any specific $\neg LB$.
First case: $(**)\Gamma^{L^{\freccia}}, \neg B, A$ has a closed
tableau. By weakening from $(*)$, also $(*')\Gamma^{L^{\freccia}},
\nott B, \neg A$  has  a closed tableau. By Lemma \ref{cut-prop}
applied to $(*')$ and $(**)$, also $\Gamma^{L^{\freccia}}, \neg B$
has a closed tableau, and since this set can be  obtained from
$\Gamma$ by applying $(L^{-})$ to $\neg LB$, it follows that
$\Gamma$ has a closed tableau, without (Weak-Cut). Second case:
$(***)\Gamma^{L^{\freccia}}, A$ has a closed tableau.  By Lemma
\ref{cut-prop} applied to $(*)$ and $(***)$, also
$\Gamma^{L^{\freccia}}$ has a closed tableau, and since this set
can be  obtained by applying $(L^{-})$ to $\Gamma$, it follows
that $\Gamma$ has a closed tableau, without (Weak-Cut).
\end{provaposu}

\begin{lemma}\label{induzBox}
Let $LA$ and $LB$ be such that there is a closed tableau without
(Weak-Cut) for  $\{LA, \neg LB \}$. Then for all sets of formulas
$\Gamma$, if $\Gamma, \bbox \neg LA, \bbox \neg LB$  has a closed
tableau without (Weak-Cut), also $\Gamma, \bbox \neg LB$ has.
\end{lemma}
\begin{provaposu}
By induction on the height $h$ of the tableau for $\Gamma, \bbox
\neg LA, \bbox \neg LB$. If $h = 0$, then $\Gamma, \bbox \neg LA,
\bbox \neg LB$ contains an axiom, hence (since axioms only concern
atoms), also $\Gamma$ does, and there is a tableau for $\Gamma,
\bbox \neg LB$ without (Weak-Cut). We prove that if the property
holds for all tableaux of height $h-1$, then it also holds for
tableaux of height $h$. We proceed by considering all possible
cases corresponding to the first rule applied to $\Gamma, \bbox
\neg LA, \bbox \neg LB$.

\noindent The case in which the first rule is boolean is easy and
left to the reader.

\noindent If the first rule is $(L^-)$, it can be easily verified
that the same set of formulas can be obtained by applying $(L^-)$
to $\Gamma,  \bbox \neg LB$ that hence has a closed tableau
without (Weak-Cut).

\noindent If the first rule is $(\cond^-)$, again it can be easily
verified that the same set of formulas can be obtained by applying
$(\cond^-)$ to $\Gamma,  \bbox \neg LB$, that hence has a closed
tableau without (Weak-Cut).

\noindent If the first rule is $(\cond^+)$ applied to a
conditional $C \cond D \in \Gamma$, then $(1) \Gamma, \bbox \neg
LA, \bbox \neg LB, \neg LC$; $(2) \Gammas, \neg LA, \neg LB, \bbox
\neg LC, LC$; $(3) \Gamma, \bbox \neg LA, \bbox \neg LB, \bbox
\neg LC, LC, LD$ have a closed tableau with height smaller than
$h$. By the inductive hypothesis from $(1)$ and $(3)$, we infer
that $(1') \Gamma, \bbox \neg LB, \neg LC$ and $(3') \Gamma, \bbox
\neg LB, \bbox \neg LC, LC, LD$ have a closed tableau without
(Weak-Cut). Furthermore, since by hypothesis $\{LA, \neg LB \}$
has a closed tableau without (Weak-Cut), from $(2)$,  weakening
(Lemma \ref{weakening}), and  Lemma \ref{cut-Lformula}, we infer
that also $(2') \Gammas, LC, \bbox \nott LC, \neg LB$ has a closed
tableau without (Weak-Cut). Since $(1'), (2'), (3')$ can be
obtained from $\Gamma, \bbox \neg LB$ by applying $\cond^+$ to $C
\cond D$ in $\Gamma$, we conclude that $\Gamma, \bbox \neg LB$ has
a closed tableau without (Weak-Cut).

\noindent There are no other cases, hence the result follows.
\end{provaposu}


\noindent Now we are able to prove that the \mbox{(Weak-Cut)} is
admissible in $\calcoloC$, as stated by Theorem \ref{eliminazione
cut C}:\\

\noindent \textsc{Theorem \ref{eliminazione cut C}.} \emph{Given a
set of formulas $\Gamma$ and a propositional formula $A$, if there
is a closed tableau for each of the following sets of
  formulas:
  \begin{itemize}
    \item[$(1)$] $\Gamma, \nott LA$
    \item[$(2)$] $\Gammas, LA, \bbox \nott LA$
    \item[$(3)$] $\Gamma, \bbox \nott LA$
  \end{itemize}
\noindent   then  there is also a closed tableau for $\Gamma$,
i.e. the \emph{(Weak-Cut)} rule is admissible.}\\


\begin{provaposu}
We prove that for all sets of formulas $\Gamma$, if  there is a
closed tableau without \mbox{(Weak-Cut)} for each of the following
sets of
  formulas:
  \begin{itemize}
    \item[$(1)$] $\Gamma, \nott LA$
    \item[$(2)$] $\Gammas, LA, \bbox \nott LA$
    \item[$(3)$] $\Gamma, \bbox \nott LA$
  \end{itemize}
\noindent   then  there is also a closed tableau  without
\mbox{(Weak-Cut)} for $\Gamma$. By this property, given a closed
tableau for a starting set of formulas $\Gamma_0$, a closed
tableau without (Weak-Cut) for $\Gamma_0$ can be obtained by
eliminating all applications of \mbox{(Weak-Cut)}, starting from
the leafs towards the root of the tableau.

\noindent First of all, notice that in general there can be
several closed tableaux for $\Gamma_0$. We only consider closed
tableaux for $\Gamma_0$ that are {\em minimal} (in the number of
nodes). Moreover, by Lemma \ref{niente cond negativi} we can
restrict our concern to applications of \mbox{(Weak-Cut)} whose
conclusions are not obtained by applications of $(\cond^{-})$.

\noindent Let $h1, h2, h3$ be the heights of the tableaux for
$(1)$, $(2)$, and $(3)$ respectively. We proceed by  induction on
$h1 + h2 + h3$. The reader might be surprised by the fact that the
proof is carried on by single induction on the heights of the
premises of (Weak-Cut). Indeed, the proof relies on the proof of
admissibility of ordinary cut at the propositional level (Lemma
\ref{cut-prop}), that is proved as usual by double induction on
the complexity of the cut formula and on the heights of the
derivation of the two conclusions.

For the base case, notice that always $h2 > 0$, since axioms are
restricted to atomic formulas, and $\Gammas$ by definition does
not contain any atomic formula. Our base case will hence be the
case in which $h1 = 0$ or $h3= 0$, i.e. (1) or (3) are axioms, and
$h2$ is minimal.

\noindent {\em Base Case:}  $h1 = 0$,  $h3 = 0$, and $h2$ is
minimal.  If $h1 = 0$, then $(1)$ is an axiom. In this case, since
axioms restricted to atomic formulas, it must be that $P, \nott P
\appartiene \Gamma$ with $P \appartiene \mathit{ATM}$. Therefore
we can conclude that there is a closed tableau for $\Gamma$
without \mbox{(Weak-Cut)}.

  \noindent For the inductive step, we distinguish the two
  following cases:

  \begin{enumerate}
\item one of the conclusions of \mbox{(Weak-Cut)} is obtained
    by an application of a rule for the boolean connectives;
\item all the conclusions of \mbox{(Weak-Cut)} are obtained by
   $(\cond^{+})$ or by $(L^{-})$.
    \end{enumerate}

\noindent The list is exhaustive; indeed, by Lemma \ref{niente
cond negativi}, we can consider, without loss of generality, a
closed tableau in which all the conclusions of each application of
(Weak-Cut) are obtained by a rule different from $(\cond^{-})$.

We consider the two cases above:

  \begin{enumerate}

    \item rules for the boolean connectives: first, notice that
    a boolean rule cannot be applied to
    the inner conclusion of (Weak-Cut), since it only contains conditional formulas (even negated),
    $L$- formulas (even negated), and a positive box formula. In
    these cases, we conclude by permuting the (Weak-Cut) rule over the boolean
    rule, i.e. we first cut the conclusion(s) of the boolean rule
    with the other conclusions of (Weak-Cut), then we conclude by
    applying the boolean rule on the sets of formulas obtained. As
    an example, consider the following closed tableau, where the
    rightmost conclusion of (Weak-Cut) is obtained by an
    application of $(\orr^{+})$, and where $F$ and $G$ are conditional formulas:

\begin{footnotesize}
    \[
      \begin{prooftree}
       \[
        \Gamma', F \orr G
        \justifies (1)\Gamma', F \orr G, \nott LA \quad\quad
          (2)\Gamma'^{\cond\pm}, \Gamma'^{\bbox^{\freccia}}, LA, \bbox
          \nott LA \quad\quad (3) \Gamma', F \orr G, \bbox \nott LA
          \using \mbox{(Weak-Cut)}
        \]
        \justifies \quad\quad\quad\quad\quad\quad\quad\quad\quad\quad\quad\quad\quad\quad\quad\quad\quad\quad\quad\quad\quad (3a)\Gamma', F,
        \bbox \nott LA \quad\quad\quad (3b) \Gamma', G, \bbox \nott LA \using
        (\orr^{+})
      \end{prooftree}
    \]
\end{footnotesize}

\noindent     Since $(\orr^{+})$ is height-preserving and
cut-preserving
    invertible (see Theorem \ref{invertibilità regole booleane
    C}), there is a closed tableau of no greater height than $(1)$,
    having no applications of (Weak-Cut) (since the closed tableau
    starting with $(1)$ does not contain any application of it), of $(1')
    \Gamma', F,$ $\nott LA$ and $(1'') \Gamma', G, \nott LA$. By the
    height-preserving admissibility of weakening (see Lemma
    \ref{weakening}), we have also a closed tableau, of no greater
    height than $(2)$, for $(2') \Gamma'^{\cond\pm}, \Gamma'^{\bbox^{\freccia}}, F, LA, \bbox
          \nott LA$ and for $(2'') \Gamma'^{\cond\pm}, \Gamma'^{\bbox^{\freccia}}, G, LA, \bbox
          \nott LA$. We can apply the inductive hypothesis to
          $(1')$, $(2')$, and $(3a)$ to obtain a closed tableau,
          without applications of (Weak-Cut), of $(*)\Gamma', F$.
          Notice that we can apply the inductive hypothesis since
          the tableaux for
          $(1')$ and $(2')$ do not contain applications of (Weak-Cut) and
          have no greater height than $(1)$ and $(2)$,
          respectively; however, the closed tableau for $(3a)$,
          not containing any application of (Weak-Cut), has
          obviously a smaller height than the one for $(3)$.
    In the same way, we can apply the inductive hypothesis to
    $(1'')$, $(2'')$, and $(3b)$ to obtain a closed tableau for
    $(**)\Gamma', G$. We can conclude by applying $(\orr^{+})$ to
    $(*)$ and $(**)$ to obtain a proof without (Weak-Cut) for
    $\Gamma', F \orr G$.

   \item $(\cond^{+})$ or $(L^{-})$: let us denote with a triple
   $<R_1,R_2,R_3>$ the rules applied, respectively, to the
   three conclusions $(1)$, $(2)$, and $(3)$ of (Weak-Cut), where
   $R_i \in \{(L^{-}),(\cond^{+}),(\mbox{ANY})\}$. We use
   $(\mbox{ANY})$ to represent either $(L^{-})$ and $(\cond^{+})$.
   As an example, $<(L^{-}),(\cond^{+}),(\mbox{ANY})>$ is used to
   represent the case in which the leftmost conclusion of
   (Weak-Cut) $(1)$ is obtained by an application of $(L^{-})$,
   whereas the inner conclusion $(2)$ is obtained by an
   application of $(\cond^{+})$; the rightmost conclusion $(3)$
   can be either obtained by an application of $(L^{-})$ or by an
   application of $(\cond^{+})$.

   We distinguish the following cases: $(\dag) \ <(\mbox{ANY})(\mbox{ANY})(L^{-})>$,
   $(\dag\dag) \ <(\mbox{ANY}),(L^{-}),(\cond^{+})>$, and $(\dag\dag\dag) \ <(\mbox{ANY}),(\cond^{+}),(\cond^{+})>$. The list is exhaustive.

  \begin{itemize}

\item $(\dag) \ <(\mbox{ANY})(\mbox{ANY})(L^{-})>$: the tableau
for $(3)$ is started with an application of
    $(L^{-})$ to a formula $\nott LB \appartiene \Gamma$ ($\Gamma=\Gamma', \nott LB$):

    \[
      \begin{prooftree}
        (3) \Gamma', \nott LB, \bbox \nott LA
        \justifies \Gamma^{'L^{\freccia}}, \nott B \using (L^{-})
      \end{prooftree}
    \]

    \noindent We can conclude since $\Gamma^{'L^{\freccia}}, \nott B$ can be obtained by applying
    $(L^{-})$ to $\Gamma=\Gamma'. \nott LB$ (the formula $\bbox \nott LA$ is
    removed from the conclusion in an application of $(L^{-})$). In case the tableau for
    $(3)$ is started with an application of $(L^{-})$ on a formula
    $LB \in \Gamma$, and there is no $\nott LB_i \in \Gamma$, we
    can obviously conclude in the same manner;

    \item $(\dag\dag) \ <(\mbox{ANY}),(L^{-}),(\cond^{+})>$: the proof of $(3)$ is started with an application of
    $(\cond^{+})$ on a formula $C \cond D \appartiene \Gamma$; we have that the following sets of formulas have three closed
    tableaux without (Weak-Cut):

    \begin{itemize}
    \item (3a) $\Gamma, \bbox \neg LA, \neg LC $
    \item (3b) $\Gammas, \neg LA, \bbox \neg LC, LC$
    \item (3c) $\Gamma, \bbox \neg LA, \bbox \neg LC, LC, LD$
    \end{itemize}

\noindent Moreover, since the first rule applied to $(2)$ is
$(L^-)$, we have that $\{LA, \Gamma^{\bbox^{\freccia}}\}$ has a
closed tableau without (Weak-Cut).

We want to show  that also the three following sets of formulas:

\begin{itemize}
  \item (I) $\Gamma, \neg LC$
  \item (II) $\Gammas, \bbox \neg LC, LC$
  \item (III) $\Gamma,  \bbox \neg LC, LC, LD$
\end{itemize}

\noindent have a closed tableau without (Weak-Cut), from which we
conclude by an application of $(\cond^{+})$. We thus want to show
that:

\noindent (I). By (3a) we know that $\Gamma, \neg LC, \bbox \neg
LA$ has a closed tableau without (Weak-Cut); by weakening from (1)
we also know that $\Gamma, \neg LC, \neg LA$ has a closed tableau
without (Weak-Cut); last, by (2) we know that $\Gammas, \bbox \neg
LA, LA$ has a closed tableau without (Weak-Cut). Since the sum of
the heights of (1), (2) and (3a) is smaller than h1 + h2 + h3
(because the height of (3a) is smaller than h3), we can apply the
inductive hypothesis and conclude that (I) $\Gamma, \neg LC$ has a
closed tableau without (Weak-Cut).

\noindent (II) Since $\{LA, \Gamma^{\bbox^{\freccia}}\}$ has a
closed tableau without (Weak-Cut), by weakening, also  $\Gammas,
LA, \bbox \neg LC, LC$ has a closed tableau without (Weak-Cut),
and from (3b) and  Lemma \ref{cut-Lformula}, also (II) $\Gammas,
\bbox \neg LC, LC$ has.

\noindent (III). By weakening from (1), we know that $\Gamma, LC,
\bbox \neg LC, LD,$ $\neg LA$ has a closed tableau without
(Weak-Cut); by $(3c)$ we know that $\Gamma, LC, \bbox \neg LC, LD,
\bbox \neg LA$ has a closed tableau without (Weak-Cut), and by
weakening from (2) we know that $\Gammas,$ $\neg LC, \bbox \neg
LA, LA$ has a closed tableau without (Weak-Cut). Furthermore, the
sum of the heights of the tableaux for (1), (2) and (3c) is
smaller than h1+h2+h3. We can then apply the inductive hypothesis
to conclude that (III): $\Gamma, \bbox \neg LC, LC, LD$ has a
closed tableau without (Weak-Cut).

\item $(\dag\dag\dag) \ <(\mbox{ANY}),(\cond^{+}),(\cond^{+})>$:
as in the previous case, the proof of $(3)$ is started with an
application of $(\cond^{+})$ on a formula $C \cond D \appartiene
\Gamma$; we have that the following sets of formulas have three
closed
    tableaux without (Weak-Cut):

    \begin{itemize}
    \item (3a) $\Gamma, \bbox \neg LA, \neg LC $
    \item (3b) $\Gammas, \neg LA, \bbox \neg LC, LC$
    \item (3c) $\Gamma, \bbox \neg LA, \bbox \neg LC, LC, LD$
    \end{itemize}

\noindent Moreover, we have that the first rule applied in (2) is
$(\cond^+)$ applied to some conditional $C_1 \cond D_1 \in
\Gamma^{\cond^{\pm}}$. We show that there is a conditional $C_i
\cond D_i \in \Gamma$ such that:
\begin{itemize}
  \item $(i)$ $\Gamma, \neg LC_i$
  \item $(ii)$ $\Gammas, \bbox \neg LC_i, LC_i$
  \item $(iii)$ $\Gamma, \bbox \neg LC_i, LC_i, LD_i$
\end{itemize}
\noindent have a closed tableau without (Weak-Cut), hence also
$\Gamma$ has, since $(i)$, $(ii)$, and $(iii)$ can be obtained
from $\Gamma$ by applying $(\cond^+)$ to $C_i \cond D_i$.

The tableau for $(2)$ can contain a sequence $s$ of applications
of $(\cond^+)$. By considering only the leftmost branch introduced
by the applications of $(\cond^+)$ in $s$, consider the last
application of $(\cond^+)$ to a conditional $C_i \cond D_i$. We
will have that (2a) $\Gammas, LA, \bbox \nott LA, \neg LC_1, \dots
,\neg LC_i$ has a closed tableau; (2b) $\Gamma^{\cond^{\pm}}, \neg
LA, \bbox \neg LC_i, LC_i$ has a closed tableau; (2c) $\Gammas,$
$LA, \bbox \nott LA, \neg LC_1, \dots ,\neg LC_{i-1}, \bbox \neg
LC_i, LC_i, LD_i$ has a closed tableau. The situation can be
represented as follows:

\lineacorta

\vspace{-0.5cm}

\begin{footnotesize}
\[
  \begin{prooftree}
    \[
    \Gamma
    \justifies (1) \Gamma, \nott LA \quad\quad (2) \Gammas, LA, \bbox \nott
    LA \quad\quad (3) \Gamma, \bbox \nott LA \using (\mbox{Weak-Cut})
    \]
    \justifies \shortstack{$\Gammas, LA, \bbox \nott LA, \nott LC_1 \quad \dots$ \\ $\vdots$ \\
    $(2a)\Gammas, LA, \bbox \nott LA, \nott LC_1, \dots, \nott LC_i \quad \dots (2b) \dots (2c) \dots$}
    \using (\cond^{+})
  \end{prooftree}
\]
\end{footnotesize}

\vspace{-0.3cm}

\lineacorta

 \noindent Furthermore, since by hypothesis the tableau for
$(2)$ does not contain any application of (Weak-Cut), also the
tableaux for (2a), (2b), and (2c) do not contain any application
of (Weak-Cut).

Observe that (2a) $\Gammas, LA, \bbox \nott LA, \neg LC_1, \dots
,\neg LC_i$ cannot be an instance of (AX), since axioms are
restricted to atomic formulas only, and (2a) only contains
conditionals, $L$-formulas, and boxed formulas. Therefore, we can
observe that the rule applied to (2a) is $(L^{-})$: indeed, if the
rule was $(\cond^{-})$ applied to some $\nott (C_k \cond C_j) \in
\Gamma^{\cond\pm}$, then we could find a shorter derivation,
obtained by applying $(\cond^{-})$ to $(2) \Gammas, LA,$ $\bbox
\nott LA$, against the minimality of the closed tableau we are
considering. This situation would be as follows:

\lineacorta

\begin{footnotesize}
\[
  \begin{prooftree}
    \[
    \[
    \Gamma
    \justifies (1) \Gamma, \nott LA \quad (2) \Gammas, LA, \bbox \nott
    LA \quad (3) \Gamma, \bbox \nott LA \using (\mbox{Weak-Cut})
    \]
    \justifies \shortstack{$\Gammas, LA, \bbox \nott LA, \nott LC_1 \quad \dots$ \\ $\vdots$ \\
    $\Gammas, LA, \bbox \nott LA, \nott LC_1, \dots, \nott LC_i \quad \dots$}
    \using (\cond^{+})
    \]
      \justifies (\Gamma^{\cond\pm} - \{\nott (C_k \cond C_j)\}),
    LC_k, \bbox \nott LC_k, \nott LC_j \using (\cond^{-})
  \end{prooftree}
\]
\end{footnotesize}

\tratteggio

\begin{footnotesize}
\[
  \begin{prooftree}
\[
    \Gamma
    \justifies (1) \Gamma, \nott LA \quad\quad (2) \Gammas, LA, \bbox \nott
    LA \quad (3) \Gamma, \bbox \nott LA \using (\mbox{Weak-Cut})
\]
    \justifies \dots \quad\quad (\Gamma^{\cond\pm} - \{\nott (C_k \cond C_j)\}),
    LC_k, \bbox \nott LC_k, \nott LC_j  \quad \dots
\using (\cond^{-})
  \end{prooftree}
\]
\end{footnotesize}

\lineacorta

\noindent More precisely, we can observe that the rule applied to
(2a) is $(L^-)$ applied to $\neg LC_i$: indeed, if $(L^-)$ was
applied to a previously generated negated formula, there would be
a shorter tableau obtained by immediately applying $(L^-)$ to that
formula, as represented by the following derivations:

\lineacorta

\begin{footnotesize}
\[
  \begin{prooftree}
    \[
      \shortstack{$\vdots$ \\ $\Gammas, LA, \bbox \nott LA, \nott LC_1, \dots, \nott LC_h$}
      \justifies
        \shortstack{$\Gammas, LA, \bbox \nott LA, \nott LC_1, \dots, \nott LC_h, \nott LC_{h+1} \quad \dots$ \\ $\vdots$ \\
        $\Gammas, LA, \bbox \nott LA, \nott LC_1, \dots, \nott LC_h, \dots, \nott LC_i \quad\quad \dots$} \using (\cond^{+})
    \]
    \justifies A, \nott C_h \using (L^{-})
  \end{prooftree}
\]
\end{footnotesize}

\tratteggio

\begin{footnotesize}
\[
  \begin{prooftree}
    \shortstack{$\vdots$ \\ $\Gammas, LA, \bbox \nott LA, \nott LC_1, \dots, \nott LC_h$}
    \justifies A, \nott C_h \using (L^{-})
  \end{prooftree}
\]
\end{footnotesize}

\lineacorta

\noindent Hence, $A, \neg C_i $ has a closed tableau without
(Weak-Cut), and hence also $(*) LA, \neg LC_i $ has a closed
tableau without (Weak-Cut). From $(*)$ and $(1)$, with opportune
weakenings, by Lemma \ref{cut-Lformula} we derive that $(i):
\Gamma, \neg LC_i$ has a closed tableau without (Weak-Cut). By
weakening from $(2b)$, we have that $\Gammas,$ $\neg LA, \bbox
\neg LC_i, LC_i$ has a closed tableau without (Weak-Cut). From
this set of formulas, $(2)$, weakening, and Lemma
\ref{cut-Lformula}, we also know that: $\Gammas,$ $\bbox \neg LA,
\bbox \neg LC_i, LC_i$ has a closed tableau without (Weak-Cut).
From this set of formulas, (*) and Lemma \ref{induzBox}, we
conclude that also $(ii) \Gammas, \bbox \neg LC_i, LC_i$ has a
closed tableau without (Weak-Cut).

\noindent Consider now $(iii)$: we can show that $\Gamma, \bbox
\neg LC_i, LC_i, LD_i$ has a closed tableau without (Weak-Cut).
Indeed, we can repeat the same proofs of case $(\dag\dag)$ in
order to show that (I) $\Gamma \neg LC$ and (III) $\Gamma, \bbox
\neg LC, LC, LD$ have a closed tableau without (Weak-Cut) (in
$(\dag\dag)$ there was no assumption on the first rule applied to
$(2)$ to show that (I) and (III) have a closed tableau). In order
to show that $(iii)$ has a closed tableau without (Weak-Cut), we
observe that: by weakening from (I), (I')$\Gamma, \bbox \neg LC_i,
LC_i,$ $LD_i, \neg LC$ has a closed tableau without (Weak-Cut); by
weakening from (III), (III')$\Gamma, \bbox \neg LC_i, LC_i, LD_i,
\bbox \neg LC,$ $LC, LD$ has a closed tableau without (Weak-Cut);
since (3b) and (*) have closed tableaux without (Weak-Cut), by
Lemma \ref{cut-Lformula} we have that (II')$\Gammas,$ $\neg LC_i,
\bbox \neg LC, LC$ has a closed tableau without (Weak-Cut). We
conclude that $(iii) \Gamma, LC_i, \bbox \nott LC_i, LD_i$ has a
closed tableau without (Weak-Cut), obtained by applying
$(\cond^{+})$ to (I'), (II'), and (III').

 We have hence proven that $(i), (ii)$, and $(iii)$ have a closed tableau
without (Weak-Cut), then we can conclude by applying $(\cond^{+})$
to them to obtain a closed tableau for $\Gamma$.

\end{itemize}

\end{enumerate}

\end{provaposu}



\begin{pf*}{\noindent Proof of Theorem \ref{invertibilità regole R}.}
\emph{Given any rule ({\bf R}) of $\calcoloRterminante$, whose
premise is $\Gamma$ and whose conclusions are $\Gamma_i$, with
$i=1,2,3$, we have that if $\Gamma$ has a closed tableau of height
$h$, then there is a closed tableau, of height no greater than
$h$, for each $\Gamma_i$, i.e. the rules of $\calcoloRterminante$
are height-preserving invertible.}

\vspace{0.2cm}

\noindent  We consider each rule of the calculus, then we proceed
by induction on the height of the closed tableau for the premise.

  \begin{itemize}
    \item $(\cond^{+})$: given a closed tableau for $\Gamma, u: A \cond B$,
    then we can immediately conclude that there is also a closed tableau for
    $\Gamma, u: A \cond B, x: \nott A$, for $\Gamma, u: A \cond B, x: \nott \bbox \nott A$,
    and for $\Gamma, u: A \cond B, x: B$,
    since weakening is height-preserving
    admissible (see Lemma \ref{ammissibilità weakening});

    \item $(<)$: as in the previous case, a closed tableau for the
    two conclusions of the rule can be obtained by weakening from
    the premise;

    \item $(\bbox^{-})$: given a closed tableau for $(1)\Gamma, x:
    \nott \bbox \nott A$ and a label $y$ not occurring in $\Gamma$, we have to show that there is a closed
    tableau for $\Gamma, y < x, \Gammam{x}{y}, y: A, y: \bbox
    \nott A$. By induction on the height of the proof of $(1)$, we
    distinguish the following cases:
    \begin{itemize}
      \item the first rule applied is $(\bbox^{-})$ on $x: \nott
      \bbox \nott A$: in this case, we are done, since we have a
      closed tableau for $\Gamma, y < x, \Gammam{x}{y}, y: A, y: \bbox
    \nott A$;
      \item otherwise, i.e. another rule ({\bf R}) of $\calcoloRterminante$ is applied
      to $\Gamma, x: \nott \bbox \nott A$, we can apply the inductive hypothesis on
      the conclusion(s) of ({\bf R}), since no rule removes side
      formulas in a rule application. In detail, we have that $x: \nott \bbox \nott A$ belongs to all the
    conclusions, then we can apply the inductive hypothesis and
    then conclude by re-applying ({\bf R}). As an example, suppose
    the derivation starts with an application of $(\cond^{-})$ as
    follows:
    \[
      \begin{prooftree}
        (1)\Gamma', u: C \cond D, x: \nott \bbox \nott A
        \justifies (2)\Gamma', v: C, v: \bbox \nott C, v: \nott D, x:
        \nott \bbox \nott A \using (\cond^{-})
      \end{prooftree}
    \]
    \noindent We can apply the inductive hypothesis on the closed tableau
    for $(2)$, concluding that there is a closed tableau for
    $(2')\Gamma', v: C, v: \bbox \nott C, v: \nott D, y < x, \Gammam{x}{y}, y: A, y: \bbox \nott
    A$, from which we can conclude obtaining the following closed
    tableau:
    \[
      \begin{prooftree}
        \Gamma', u: C \cond D, y < x, \Gammam{x}{y}, y: A, y: \bbox \nott
    A
        \justifies (2')\Gamma', v: C, v: \bbox \nott C, v: \nott D, y < x, \Gammam{x}{y}, y: A, y: \bbox \nott
    A \using (\cond^{-})
      \end{prooftree}
    \]
    \end{itemize}
    \noindent Notice that the proof has (at most) the same height of the
    closed tableau for $(1)$.

    \item other rules: the proof is similar to the one for
    $(\bbox^{-})$ and then left to the reader.
  \end{itemize}

\begin{flushright}$\blacksquare$\end{flushright}
\end{pf*}



\begin{pf*}{\noindent Proof of Lemma \ref{irreflexivity}.}
\emph{Given a tableau starting with $x_0: F$, for any open,
saturated branch ${\bf B}=\Gamma_1, \Gamma_2, \dots, \Gamma_n,
\dots$, we have that:}

\begin{enumerate}

  \item \emph{if $z < y \in \Gamma_i$ in ${\bf B}$ and  $y < x \in \Gamma_j$ in ${\bf B}$, then there
  exists $\Gamma_k$ in ${\bf B}$ such that $z < x
\in \Gamma_k$};

  \item \emph{if $x: \bbox \neg A \in \Gamma_i$ in ${\bf B}$ and $y < x \in \Gamma_j$ in ${\bf B}$,
then there exists $\Gamma_k$ in ${\bf B}$ such that $y: \neg A \in
    \Gamma_k$ and $y: \bbox \neg A \in \Gamma_k$};

  \item \emph{for no $\Gamma_i$ in ${\bf B}$, $x < x \in \Gamma_i$}.
\end{enumerate}

\vspace{0.2cm}

\noindent Let us consider an open, saturated branch ${\bf
B}=\Gamma_1, \Gamma_2, \dots, \Gamma_n, \dots$. We consider
separately the three claims in the definition of the Lemma:
\begin{enumerate}
\item We are considering the case when $z < y \in \Gamma_i$ in ${\bf B}$
  and  $y < x \in \Gamma_j$ in ${\bf B}$. Since ${\bf B}$ is
  saturated, then there exists $\Gamma_k$ in ${\bf B}$ such that
  either $z < x \in \Gamma_k$ or $x < y \in \Gamma_k$. If $x < y
  \in \Gamma_k$, then the branch is closed
  \begin{rosso}($\Gamma_k$ is an instance of {\bf(AX)})\end{rosso}.
   Thus, we conclude that $z < x \in \Gamma_k$.
\item \begin{rosso} A relation formula $y < x$ can only be introduced by an
  application of either $(\bbox^{-})$ or $(<)$; in both cases,
  $\Gammam{x}{y}$ is added to the current branch of the tableau.
  Consider any $x: \bbox \nott A \in \Gamma_i$; if $j > i$, i.e.
  $y < x$ is introduced in the branch \emph{after} $x: \bbox \nott
  A$, then we are done, since $y: \nott A \in \Gammam{x}{y}$ and
  $y: \bbox \nott A \in \Gammam{x}{y}$. Otherwise, if $y < x$ is introduced in the branch
  \emph{before} $x: \bbox \nott A$, then we are considering the
  case such that $x: \bbox \nott A$ is introduced by an
  application of $(<)$, i.e. $x: \bbox \nott A \in \Gammam{k}{x}$
  (by the presence of some $k: \bbox \nott A$ in the branch)
  for some $k$ and $x < k$ is also introduced in the branch. Since
  the branch is saturated, then either $(1) \ x < y$ or $(2) \ y < k$ are
  introduced in the branch: $(1)$ cannot be, otherwise the branch
  would be closed.
  If $(2)$ is
  introduced after $k: \bbox \nott A$, then we are done since $y:
  \nott A \in \Gammam{k}{y}$ and $y: \bbox \nott A \in
  \Gammam{k}{y}$; otherwise, $(2)$ has also been introduced by an
  application of $(<)$, and we can reason in the same way
  described here above. This process terminates: indeed, we can
  easily observe the following facts: - a boxed formula $u: \bbox
  \nott A$ is initially introduced in the tableau by an application of
  either $(\bbox^{-})$ or $(\cond^{-})$, whereas
  $(<)$ and $(\bbox^{-})$ only ``propagate'' it to other worlds; - in both cases
  $(\bbox^{-})$ and $(\cond^{-})$, $u$ is a new label not occurring
  in the branch, therefore all the relation formulas $v < u$ will
  be introduced \emph{later} in the branch, i.e. when $u: \bbox
  \nott A$ already belongs to the branch. In conclusion of our
  proof, a relation formula $y < u$ will be introduced in the
  branch \emph{not before} of $u: \bbox \nott A$: by this fact and by
  the saturation of the branch, we conclude that also $y < u$ and,
  then, $y: \nott A$ and $y: \bbox \nott A$ belong to the branch.
 \end{rosso}
\item A relation $x < x$ cannot be introduced by rule $(\bbox^-)$, since this
rule establishes a relation between $x$ in ${\bf B}$ and a label
distinct from $x$. On the other hand, it cannot be introduced by
modularity. Indeed, for rule $(<)$ to introduce a relation $x <
x$, there must be in ${\bf B}$ some relation $y < x$ (resp. $x <
y$) for some $y$. But in this case the side condition of the rule
would not be fulfilled, and the rule could not be applied.

\end{enumerate}

\begin{flushright}$\blacksquare$\end{flushright}
\end{pf*}


\end{document}